\documentclass[a4paper,12pt]{article}
\usepackage{jheppub,esint,shuffle,psfrag}
\usepackage[utf8]{inputenc}
\usepackage{simpler-wick}

\usepackage{epsfig,amssymb,amsmath,psfrag,subfigure,rotate,color,wasysym,xcolor}
\usepackage[bbgreekl]{mathbbol}
\usepackage{empheq}
\usepackage{array}
\usepackage[normalem]{ulem}

\allowdisplaybreaks
\newcommand{\insertfig}[2]{\includegraphics[width=#1mm]{#2}}

\DeclareSymbolFontAlphabet{\mathbbm}{bbold}
\DeclareSymbolFontAlphabet{\mathbb}{AMSb}%

\def \ba  {\begin{eqnarray}}
\def \ea  {\end{eqnarray}}
\def \baa {\begin{eqnarray*}}
\def \eaa {\end{eqnarray*}}
\def \lab #1 {\label{#1}}

\newcommand\re[1]{(\ref{#1})}
\def\d{\hbox{{d}\kern-.20em\hbox{l}}}
\def \qqquad {\qquad\quad}
\def \qqqquad {\qquad\qquad}
\def \matrix #1 {\left(\begin{array}{cc} #1 \end{array}\right)}

\def \e  {\mathop{\rm e}\nolimits}
\newcommand\lr[1]{{\left({#1}\right)}}

\newcommand \vev [1] {\langle{#1}\rangle}

\newcommand \ket [1] {|{#1}\rangle}
\newcommand \bra [1] {\langle {#1}|}
\newcommand{\bit}[1]{\mbox{\boldmath$#1$}}
\def\1{\hbox{{1}\kern-.25em\hbox{l}}}

\def\Xint#1{\mathchoice
   {\XXint\displaystyle\textstyle{#1}}%
   {\XXint\textstyle\scriptstyle{#1}}%
   {\XXint\scriptstyle\scriptscriptstyle{#1}}%
   {\XXint\scriptscriptstyle\scriptscriptstyle{#1}}%
   \!\int}
\def\XXint#1#2#3{{\setbox0=\hbox{$#1{#2#3}{\int}$}
     \vcenter{\hbox{$#2#3$}}\kern-.5\wd0}}

\def\dashint{\Xint-}

\newcommand\SZ[1]{{\color{violet}{\bf SZ: #1}}}
\newcommand\GK[1]{{\color{blue}{\bf GK: #1}}}

\newcommand{\om}{\omega}
\newcommand{\eps}{\epsilon}

\def\be#1\ee{\begin{align}#1\end{align}}
\def \nn {\nonumber}  
 
\begin{document}

\begin{flushleft}
 \hfill \parbox[c]{40mm}{CERN-TH-2021-135\\
  {IPhT--T21/061} }
\end{flushleft}

\author{Gregory P. Korchemsky$^{a,b}$ and Alexander Zhiboedov$^c$ }

\affiliation{
$\null$
$^a${Institut de Physique Th\'eorique\footnote{Unit\'e Mixte de Recherche 3681 du CNRS}, Universit\'e Paris Saclay, CNRS, 91191 Gif-sur-Yvette, France}  \\
$\null$
$^b${Institut des Hautes \'Etudes Scientifiques, 91440 Bures-sur-Yvette, France}  \\
$\null$
$^c$CERN, Theoretical Physics Department, CH-1211 Geneva 23, Switzerland}

\title{ 
On the light-ray algebra  
in conformal field theories
}
 
 \abstract
{
We analyze the commutation relations of light-ray operators in conformal field theories.
We first establish the algebra of light-ray operators built out of higher spin currents in free CFTs and 
 find explicit expressions for the corresponding structure constants.
 The resulting algebras are remarkably similar to the generalized Zamolodchikov's $W_\infty$ algebra in a two-dimensional conformal field theory. 
We then compute the commutator of generalized energy flow operators in a generic, interacting CFTs in $d>2$. We show that it
receives contribution from the energy flow operator itself, as well as from the light-ray operators built out of scalar primary operators of dimension $\Delta \leq d-2$, that are present in the OPE of two stress-energy tensors. 
Commutators of light-ray operators considered in the present paper lead to CFT sum rules which
generalize the superconvergence relations and naturally connect to the dispersive sum rules, both of which have been studied recently.
}

\maketitle
\flushbottom
\setcounter{footnote} 0

\section{Introduction and summary of results}

In this paper we investigate the commutation relations of a special class of light-ray operators in conformal field theories (CFTs). These operators represent a natural generalization of the so-called energy flow operator which plays an important role in the study of weighted cross sections in QCD~\cite{Sveshnikov:1995vi,Korchemsky:1997sy,Korchemsky:1999kt} and more recently in CFTs~\cite{Hofman:2008ar}.\,\footnote{A recent discussion of the energy flow correlators applications in QCD 
can be found in Ref.~\cite{Chen:2021gdk} and references therein.} 

They take the following general form~\cite{Korchemsky:2021okt} 
\begin{align}\label{J-def}
\mathcal J_{\omega,S}(n) = (n\bar n)^{1-S} \lim_{r\to\infty} r^{\Delta - S} \int_{-\infty}^\infty dt \e^{ -i t \omega (n\bar n)}\, O_{\mu_1\dots\mu_S} (r n+ t \bar n)
\bar n^{\mu_1} \dots  \bar n^{\mu_S} \,,
\end{align}
where a local conformal operator $O_{\mu_1\dots \mu_S}$ with scaling dimension $\Delta$ and Lorentz spin $S$
is sent to null infinity in the direction specified by the null vector $n=(1,\vec n)$, with $\vec n$ being a unit vector on the sphere $S^{d-2}$, in a $d-$dimensional CFT. 
All Lorentz indices on the right-hand side of \re{J-def} are contracted with the auxiliary null vector $\bar n$ (with $\bar n^2=0$ and $(n\bar n)\neq 0$). The light-ray operator $\mathcal J_{\omega,S}(n)$ does not depend on the choice of this vector, see Ref.~\cite{Korchemsky:2021okt} for details.
The Fourier integral in \re{J-def} selects a mode with the total momentum $\omega n^\mu$
\be
\label{eq:momentum}
[P^\mu, \mathcal J_{\omega,S}(n)] = \omega n^\mu  \mathcal J_{\omega,S}(n) \,, 
\ee
where $P^\mu$ is the generator of translation. As a consequence, the operator $\mathcal J_{\omega,S}(n)$ annihilates the left or right vacuum for positive or negative $\omega$ respectively, e.g. $\mathcal J_{\omega,S}(n) | 0 \rangle=0$ for $\omega\le 0$. The operator $\mathcal J_{\omega=0,S}(n)$ annihilates both the left and the right vacuum \cite{Kravchuk:2018htv}. 

The energy flow operator ${\cal E}_{\omega}(n)$ corresponds to the special case of \re{J-def} when the local operator is given by the stress-energy tensor $T_{\mu_1\mu_2}$ with $\Delta=d$ and $S=2$
\begin{align}\label{E-def}
\mathcal E_{\omega}(n) = (n\bar n)^{-1} \lim_{r\to\infty} r^{d-2} \int_{-\infty}^\infty dt \e^{ -i t \omega (n\bar n)}\, T_{\mu_1\mu_2} (r n+ t \bar n)
\bar n^{\mu_1}   \bar n^{\mu_2}\,.
\end{align}
For $\omega=0$ it coincides with the ANEC operator \cite{Hartman:2016lgu}. The operator ${\cal E}_{\omega}(n)$ has a clear physical meaning in terms of the flow of the energy in the final state created by a source. Inserted inside a matrix element $\vev{\Psi | \mathcal E_{\omega_1}(n_1)\dots \mathcal E_{\omega_k}(n_k)  |\Psi }$, the energy flow operators measure the energy flux that reaches detectors located on the celestial sphere in the directions defined by unit vectors $\vec n_1$, $\dots$, $\vec n_k$, in the state $|\Psi \rangle$. The integration over $t$ in \re{E-def} has the meaning of an average over the  ``working'' time of the detector, dependence on $\omega$ encodes the ``time resolution" of the detector.

For $\vec n \neq \vec n'$ the detectors are space-like separated on the celestial sphere. Therefore, one should expect that the flow operators commute, and, as a consequence, the above matrix element should not depend on the ordering of the flow operators. Indeed, one can show, following  Ref.~\cite{Kologlu:2019bco}, 
that the light-ray operators \eqref{J-def} commute at non-coincident points $n=(1,\vec n)$ and $n'=(1,\vec n')$
\be
\label{eq:gencommutator}
[\mathcal J_{\omega,S}(n) , \mathcal J_{{\omega'},S'}(n') ] = 0 \,, \qquad \text{for $n \neq n'$} \,, 
\ee
given certain simple conditions on the quantum numbers of local operators out of which the light-ray operators are built are satisfied. The most important one being $S + S' - 1 > J_0$, where $J_0$ is the nonperturbative Regge intercept in the underlying CFT.

The  purpose of the present note is to generalize the commutation relations \re{eq:gencommutator} to the case of coincident points $n = n'$
 and to study their algebraic properties. 
The previous works addressing some aspects of this problem can be found in \cite{Cordova:2018ygx,Belin:2020lsr,Besken:2020snx}.

In what follows we consider various CFTs:

\subsection*{CFT${}_{\boldsymbol{d=2}}$}

In $d=2$ dimensions, the celestial sphere consists of two points, say $n=(1,1)$ and $\bar n=(1,-1)$. Applying \re{E-def} and \re{J-def}, we can define the corresponding flow operators $\mathcal E_{\omega}$ and $\mathcal J_{\omega,S}$. The commutator of the energy flow operators is related to the Virasoro algebra and takes the following form  
\begin{align}\label{cintro}
[\mathcal E_{\omega},\mathcal E_{{\omega'}}] = ({\omega'}-\omega) \mathcal E_{\omega+{\omega'}} - {c\over 12} \omega^3 \delta(\omega +{\omega'})\,,
\end{align}
where $c$ is the central charge of CFT${}_2$.
Due to holomorphy of the stress-energy tensor and the fact that the light-ray operators are co-dimension one, the relation \eqref{cintro} has well-known far-reaching consequences \cite{Belavin:1984vu}.

In the presence of an additional symmetry, CFT${}_2$ has (an infinite number of) conserved currents of higher spin $O_{\mu_1\dots\mu_S}$. A well-known example is an extended conformal algebra known as 
the generalized Zamolodchikov's
$W-$algebra, see Refs.~\cite{Zamolodchikov:1985wn,Bouwknegt:1992wg}. Defining the flow operators
\re{J-def}, we show that their commutation relations look as
\begin{align}  \label{w-intro}
{[\mathcal J_{\omega,S} , \mathcal J_{\omega',S'}  ]} & =  \sum_{S''=1}^{S+S'-2}
W_{SS'}^{S''}(\omega,\omega') \mathcal J_{\omega+\omega',S''} - c_S\, \omega^{2S-1} \delta_{SS'} \delta(\omega+\omega')
 \,,
\end{align}
where the sum runs over spins $S''$ of the same parity as $S+S'$. The structure constants $W_{SS'}^{S''}(\omega, \omega')$ are homogenous polynomials in $\omega$ and $\omega'$ of degree $S+S'-S''-1\ge 1$ whose expansion coefficients are given by the structure constants of the $W-$algebra. The central charge  $c_S$ is proportional to the central charge of the Virasoro algebra \re{cintro}. The latter algebra arises as the special case of \re{w-intro} for $S=S'=2$.

\subsection*{Free fermion and  gauge field}

In the free theory of fermions and gauge fields in $d-$dimension,~\footnote{For $d=4$ the gauge fields are described by the Maxwell theory, whereas for even $d>4$ they are defined  in terms of $(d-2)/2$ form, see Appendix B of Ref.~\cite{Buchel:2009sk} for the definition. The reason for this is that the Maxwell theory is scale-invariant but not conformal-invariant in $d \neq 4$, see Ref.~\cite{El-Showk:2011xbs}. \label{form}} there exist infinitely many higher spin conserved currents 
 $O_{\mu_1\dots\mu_S}$ with scaling dimension $\Delta=d-2+S$ and spin $S$. 
 
 Restricting our consideration to the light-ray operators \re{J-def} and \re{E-def} built out of these currents, we obtain the following algebra of the energy flow operators \re{E-def} for $d>2$
\begin{align}\label{commE} 
& [\mathcal E_{\omega}(n),\mathcal E_{{\omega'}}(n')]  = \delta^{(d-2)}(n,n') \Big[ ({\omega'} - \omega) {\cal E}_{\omega + {\omega'}}(n) - C \, \omega^3 \delta(\omega+{\omega'}) \Big] ,
\end{align}
and its generalization to the light-ray operators of an arbitrary spin \re{J-def}
\begin{align} \label{comm} 
& {[\mathcal J_{\omega,S} (n) , \mathcal J_{\omega',S'} (n') ]}  =\delta^{(d-2)}(n,n')  \Big[  \sum_{S''=1}^{S+S'-2}
C_{SS'}^{S''}(\omega,\omega') \mathcal J_{\omega+\omega',S''} (n)  
  - C_S\, \omega^{2S-1} \delta_{SS'} \delta(\omega+\omega') \Big].
\end{align}
Here the sum over the spins is similar to that in \re{w-intro} but the explicit expressions for the structure constants 
are different (see Eqs.~\re{W1-int} -- \re{C-phi-4} below). 
 
For $n\neq n'$ the light-ray operators commute, in agreement with \re{eq:gencommutator}. In addition, for $\omega=\omega'=0$ the 
expression on the right-hand side of \re{comm} vanishes  and, therefore, the flow operators $\mathcal J_{\omega=0,S} (n)$ and $\mathcal J_{\omega'=0,S'} (n')$ commute for arbitrary $n$ and $n'$. 

According to \re{commE} and \re{comm}, the commutator of the light-ray operators is given by a linear combination of the light-ray operators and  
 the identity operator. The contribution of the identity operator is localized at $\omega=-{\omega'}$ and it involves the central charges $C$ and $C_S$. They are proportional to a volume of $(d-2)-$dimensional Euclidean subspace and take infinite values for $d>2$. We will discuss in the concluding remarks how one can meaningfully treat this infinity.

The relations \re{commE} and \re{comm} are remarkably similar to the commutation relations in $d=2$, see Eqs.~\re{cintro} and \re{w-intro}.
Indeed, for $d=2$ the two algebras coincide after we replace $\delta^{(0)}(n,n')\to 1$. For $d>2$ the structure constants in \re{comm} are different from those at $d=2$. Nevertheless, we show below that for $d=4$ (and, more generally, for any even $d$) they are related to each other by a linear finite difference transformation. 

\subsection*{Free scalar field}

For free complex scalars in $d-$dimensions, the commutation relations of the energy flow operators take the form
\be
\label{eq:freescalar}
&[\mathcal E_{\omega}(n),\mathcal E_{{\omega'}}(n')] = \delta^{(d-2)}(n,n') \Big[ ({\omega'} - \omega) {\cal E}_{\omega + {\omega'}}(n) -C  \,\omega^3 \delta(\omega+{\omega'}) + \omega^{2}  {\omega'}^{2} {\cal O}^-_{\omega, {\omega'}}(n)  \Big]. 
\ee
As compared to \re{commE}, the commutator contains an additional contribution of the light-ray operator ${\cal O}^-_{\omega, {\omega'}}(n)$. It  is built out of two scalar fields and is defined as follows
\be\label{nonloc}
{\cal O}^\mp _{\omega, \omega'}(n) = - {i(d-2)^2 \over 32(d-1)^2} (n\bar n)^{2} \lim_{r\to\infty} r^{d-2}  \int_{-\infty}^\infty dt dt'  \,  \e^{-i( t \omega + t' \omega')(n \bar n)}
{\rm sign}(t-t') \nn \\
\times \Big\{\phi( r n + t \bar n)\bar\phi( r n + t'\bar n) \pm  \bar\phi( r n + t \bar n) \phi( r n + t' \bar n) \Big\} ,
\ee
where 
${\cal O}_{\omega, {\omega'}}^{\pm}(n)=\pm {\cal O}_{{\omega'}, \omega}^{\pm}(n)$.\footnote{For ${\cal O}_{\omega, {\omega'}}^{+}(n)$ we implicitly mean only the connected part of the product of the fields (namely we subtract the contribution of the identity operator from their OPE).}
Note that ${\cal O}^\pm_{\omega, {\omega'}}(n)$ is different from \re{J-def} in that it is not given by the light transform of a local operator.
Instead, expanding the product of scalar fields in \re{nonloc} in powers of $t-t'$, we can express ${\cal O}^\mp _{\omega, {\omega'}}(n)$ as an infinite sum over the light-ray operators \re{J-def} of the leading twist $\tau=\Delta-S=d-2$ and arbitrary spin $S\ge 0$, see Appendix \ref{sec:extralightray}.

For the light-ray operators \re{J-def} of an  arbitrary spin, the commutation relations also receive an additional contribution of the form 
\begin{align}\label{odd}
{[\mathcal J_{\omega,S} (n) , \mathcal J_{{\omega'},S'} (n') ]}\sim   \delta^{(d-2)}(n,n') \left[ \omega^{S}  {\omega'}^{S'} {\cal O}^{\,\epsilon}_{\omega, {\omega'}}(n) + \dots\right],
\end{align}
where the signature of the operator is $\epsilon=(-1)^{S+S'+1}$ and the dots denote the remaining `universal' contribution given by \re{comm}.
We show below that the additional contribution to the commutators \re{eq:freescalar} and \re{odd} involving the operators \re{nonloc} arises due to interference between the detectors induced by an exchange of scalar particles with zero energy, the so-called cross-talk phenomenon~\cite{Belitsky:2013bja}.

Already these simple examples of free fields illustrate a general lesson regarding the commutation relations of the light-ray operators for $d>2$:
their algebra is not universal and it depends on the details of the theory.  

\subsection*{Generic interacting CFT}

In a generic, interacting CFT we expect that the commutator $[{\cal E}_{\omega}(n) , {\cal E}_{{\omega'}}(n')]$ is localized at $n=n'$ and is given by the sum of light-ray operators \re{J-def}. The contribution of a given light-ray operator can be identified by studying three-point functions
\be
 \label{eq:threepointalgebra}
 \langle [{\cal E}_{\omega}(n) , {\cal E}_{{\omega'}}(n')]O_{\mu_1 ... \mu_{S''}}(x)  \rangle\,.
\ee
Restricting our attention to the light-ray operators built out of symmetric-traceless operators $O_{\mu_1 ... \mu_{S''}}(x)$ with twist $\tau=\Delta-S$, we found that the operators with twist $\tau > d-2$ do not contribute to the commutator, whereas operators with twist $\tau < d-2$ produce divergent contributions. Finally, the operators with twist $\tau = d-2$ contribute to the algebra with the coefficient fixed by the corresponding three-point function \re{eq:threepointalgebra}. 

This leads to the following result
\be\notag
\label{eq:algebrahigherd}
 [{\cal E}_{\omega}(n) , {\cal E}_{{\omega'}}(n')] =\delta^{(d-2)}(n,n') \Big( & ({\omega'} - \omega + \tilde n_\phi \om^2 \om'^2  f_\phi(\omega,{\omega'})){\cal E}_{\omega + {\omega'}}(n)
\\
&  - C \omega^3 \delta(\omega+{\omega'})+ [S=0,\Delta\le d-2 ] \Big) ,    
\ee 
where the coefficient function $f_\phi(\omega,{\omega'})$ and the structure constant $\tilde n_\phi >0$ are defined in \re{n-phi} below. The second term inside the brackets in \re{eq:algebrahigherd} contains a divergent central charge. The last term $[S=0,\Delta\le d-2]$ denotes the contribution of light-ray operators built out of scalar primary operators present in the OPE of the stress-energy tensors.~\footnote{For $d \geq 4$ the OPE of two stress-energy tensors contains conformal operators belonging to more complicated representations of Lorentz group. These are sometimes called mixed-symmetry tensors  \cite{Costa:2016hju}.
We checked explicitly in $d=4$ that only symmetric traceless operators can contribute to the algebra. For $d>4$ we have not analyzed the contribution of mixed-symmetry tensors to the light-ray algebra.}  The contribution of scalars with dimension ${d-2 \over 2} \le \Delta < d-2$ is divergent.

The appearance of $({\omega'} - \omega) {\cal E}_{\omega + {\omega'}}(n)$ on the right-hand side of \eqref{eq:algebrahigherd} can be traced back to the conformal Ward identities \cite{Cordova:2018ygx}. 
The presence of $\tilde n_\phi \om^2 \om'^2 f_\phi(\omega,{\omega'}) {\cal E}_{\omega + {\omega'}}(n)$ in the commutator \eqref{eq:algebrahigherd} is related to the fact  that the three-point function of stress-energy tensors in a generic interacting CFT contains a conformal tensor structure\footnote{For the explicit form see for example Appendix C in \cite{Zhiboedov:2012bm}.}  which coincides with the correlation function  in the free complex scalar theory $\langle TTT\rangle_\text{CFT} = n_\phi \langle TTT\rangle_{\phi}+ \dots $, see Refs. \cite{Zhiboedov:2013opa,Meltzer:2017rtf} and Eq.~\eqref{eq:stresstensorthree} below. As a consequence, the function $f_\phi(\omega,{\omega'})$ can be found by computing the three-point function \re{eq:threepointalgebra}   in the free scalar theory with $O_{\mu_1 ... \mu_{S''}}(x)$ replaced with the stress-energy tensor (see Appendix~\ref{sec:extralightray}).  The explicit expressions for $f_\phi(\omega,{\omega'})$ in $d=3$ and $d=4$ dimensions are
\begin{align}\label{eq:fs3and4}\notag
& f_\phi(\omega,{\omega'})\big|_{d=3}=\frac{16 \left(4 (\omega' - \omega)+ 
   \left( \omega^2-6 \omega\omega'+\omega'^2\right) (-\omega \omega')^{-1/2} [\theta(\om') - \theta(\om) ] \right)}{\left(\omega
   +\omega'\right){}^4} \, , 
\\
 & f_\phi(\omega,{\omega'}) \big|_{d=4}=\frac{30 \left(3 (\omega'^2 - \omega^2)+\left(\omega^2-4 \omega' \omega
   +\omega'^2\right) \log |\frac{\omega}{\omega'}|\right)}{\left(\omega+\omega'\right){}^5}  \,.
\end{align}
 
\subsection*{Holographic CFTs, $\mathcal N=4$ SYM, $O(N)$ critical and 3d Ising models} 
  
Based on the discussion above we conclude that in holographic CFTs, ${\cal N}=4$ SYM at non-zero coupling,  $O(N)$ critical and 3d Ising models the energy flow operators satisfy the following relation
\be\label{eq:energyalgebra}
&[{\cal E}_{\omega}(n) , {\cal E}_{{\omega'}}(n')] =\delta^{(d-2)}(n,n') \Big[  ({\omega'} - \omega + \tilde n_\phi \om^2 \om'^2 f_\phi (\omega,{\omega'})) {\cal E}_{\omega + {\omega'}}(n) -C \omega^3 \delta(\omega+{\omega'}) \Big] .
\ee 
This follows from the fact that the OPE of stress-energy tensors in these theories
does not contain light conformal operators with zero spin, as well as due to the absence of higher spin conserved currents.  
 
 \subsection*{Relation to the previous work}
 
 The light-ray operators \re{J-def} are closely related to the so-called generalized ANEC operators studied in Refs.~\cite{Casini:2017roe,Cordova:2018ygx,Kologlu:2019bco,Huang:2020ycs,Belin:2020lsr,Besken:2020snx}. Applying a conformal transformation (see Appendix~\ref{App:conv} for the definition) we can obtain another equivalent representation of the operator \re{J-def}
 \begin{align}\label{J-z}
 {\mathcal J}_{\omega,S}(n) = \int_{-\infty}^\infty dz^- \e^{-i z^- \omega}\, O_{\underbrace{\scriptstyle -\ldots-}_{S}} (0,z^-,\bit z)\,, 
\end{align}
where  $n^\mu=(1,\bit z^2,\sqrt{2}\bit z)$ in the light-cone coordinates~\footnote{The $(d-2)-$dimensional vector $\bit z$ defines a unit vector $\vec n$ on the sphere $S^{d-2}$ via a stereographic projection.
The scalar product of null vectors is given by the distance on the transverse plane $(nn')=(\bit z-\bit z')^2$.} and the operator $O_{-\ldots-}(z)$ is integrated along the $z^--$direction for $z^+=0$ and arbitrary $\bit z=(z^1,\dots,z^{d-2})$. 
The operators \re{J-z} satisfy the algebra \re{comm} with $\delta^{(d-2)}(n,n')$ replaced with the delta-function  $\delta^{(d-2)}(\bit z-\bit z')$ on the transverse $(d-2)-$dimensional plane.
 
The generalized ANEC operators arise from expanding \re{J-z} at small $\omega$
 \be\label{z-bad}
 (i \partial_\omega)^k \mathcal J_{\omega,S}(n) \Big|_{\omega=0}  = \int_{-\infty}^\infty dz^-(z^-)^k\, O_{\scriptstyle -\ldots-} (0,z^-,\bit z)\,.
 \ee
It is however clear on general grounds that this operation should be done with great care when studying the matrix elements $\langle \mathcal J_{\omega_1,S_1}(n) \dots \mathcal J_{\omega_k,S_k}(n_k) \rangle$ (or generalized event shapes).  The reason for this is that expanding $\mathcal J_{\omega_i,S_i}(n_i)$ in $\omega_i$ brings extra powers of $z^-$ under the integral \re{z-bad}. This worsen the convergence of the integral over $z^-$, eventually making the matrix elements of interest divergent. Another manifestation of the same phenomenon is that the generalized event shapes are not analytic around $\omega_i=0$, see for example the coefficient functions \eqref{eq:fs3and4} in a generic CFT. This problem does not arise for $\omega_i \neq 0$ and for this reason we prefer to work with the light-ray operators \eqref{J-def} and \re{J-z}.

A proposal for defining generalized ANEC operators was put forward in Ref.~\cite{Besken:2020snx}. In distinction to \re{z-bad}, the resulting operators  
 are well-defined for arbitrary $k$ but they
 do not form, however, a closed algebra in $d=4$ dimensions. 
 
The paper is organized as follows. In Section~\ref{sec:freeCFTs}, we present expressions for the higher spin conserved currents  in the free CFTs in $d-$dimensions. We compute three- and four-point correlation functions involving the corresponding light-ray operators and
use the obtained expressions to establish the commutation relations of these operators. Their properties and relation to the  Virasoro and $W$ algebras are discussed in section~\ref{sect:4}. 

In Section \ref{sec:CFTlightrayalgebra} we use the three-point functions \eqref{eq:threepointalgebra} to identify the contribution of a given primary operator in the commutator of two energy flow operators. Our analysis is complete in $d=3$ and $d=4$. In $d>4$ we only analyze the contribution of the symmetric traceless operators into the algebra. We briefly discuss closure of the light-ray algebra, or the Jacobi identities, for the light-ray commutators in interacting CFTs.
 
In Section \ref{sec:Mellin} we explain how the discussion of the present paper can be implemented in Mellin space,   extending the analysis of our previous work \cite{Belitsky:2013xxa,Belitsky:2013bja,Korchemsky:2021okt} to coincident points between detectors.

Section~\ref{sec:conclusions} contains concluding remarks. Some technical details are presented in Appendices.

\section{The light-ray algebra in free CFTs} 
\label{sec:freeCFTs}

In this section, we establish the commutation relations of the light-ray operators \re{J-def} in three different families of free CFTs describing scalars, fermion and gauge fields in $d$ dimensions.
 
\subsection{Leading twist operators}\label{sect:1}

We start with recalling the definition of higher spin conserved currents 
\be
O_S(x)\equiv O_{\mu_1\dots\mu_S} (x)\bar n^{\mu_1} \dots  \bar n^{\mu_S}
\ee
in the free CFTs.  
These operators are built out of two fundamental fields (scalars, fermions and gauge field strength tensor) and their derivatives. They carry the leading twist
$\tau=\Delta-S=d-2$ and  have the following schematic form
\begin{align}\label{O-ex}\notag
& O^{(\phi)}_{S} \sim \bar\phi (i\partial_-)^S \phi + \dots\,,\quad 
\\\notag
& O^{(\psi)}_{S} \sim \bar\psi \gamma_- (i\partial_-)^{S-1} \psi + \dots\,,\quad
\\
& O^{(F)}_{S} \sim F_{-\boldsymbol \mu} (i\partial_-)^{S-2} F_{-}{}^{\boldsymbol \mu}+ \dots\,,
\end{align}
where all Lorentz indices are contracted with the auxiliary null vector $\bar n^\mu$,  in particular, $\partial_-\equiv (\bar n\partial_x)$, and dots denote terms with derivatives distributed between the two fields. 
 The operator $O^{(F)}_{S}$ is well-defined for even $d\ge 4$ in which case $\boldsymbol \mu=(\mu_2,\dots,\mu_{d/2})$ denotes a set of Lorentz indices (see footnote~\ref{form}).
 In the free theory, the operators \re{O-ex} have the same quantum numbers. In an interacting theory, they start to mix under the conformal transformations and develop anomalous dimensions. In this case, the conformal operators are given by linear combinations of \re{O-ex}.

In the free theory,  operators \re{O-ex} can be constructed by employing a collinear $SL(2,\mathbb R)$ subgroup of the conformal group in $d-$dimensions, see e.g. Ref.~\cite{Braun:2003rp}. It leaves the light-ray $x^\mu = z \bar n^\mu$ invariant and generates projective transformations of the light-cone coordinates, $z\to (\alpha z+\beta)/(\gamma z+\delta)$ with $\alpha \delta-\beta \gamma=1$.
The conformal operators with spin $S$ and scaling dimension $\Delta$ are transformed under these transformations as
\begin{align}\label{trans-O}
O_S(z\bar n) \to (\gamma z+\delta)^{-2j_O} O_S\lr{{\alpha z+\beta\over \gamma z+\delta}\bar n}\,.
\end{align}
The operator $O_S(0)$ defines the highest weight of the $SL(2)$ representation labelled by its conformal spin $j_O$ and twist $\tau_O$ 
\begin{align}
j_O=\frac12(\Delta_O+S)\,, \qqqquad \tau_O=\Delta_O-S\,.
\end{align}
The leading twist operators \re{O-ex} have the following general form 
\begin{align}\label{O-sum-der}
O_S(x) = \sum_{k=0}^\ell c_k  \lr{i \partial_-}^k \bar\Phi (x) \lr{i \partial_-}^{\ell-k} \Phi(x) \,, 
\end{align}
where nonnegative integer $\ell$ and expansion coefficients $c_k$ depend on the spin $S$.
For $d\ge 3$, $\bar\Phi(x)$ and $\Phi(x)$ can be either complex scalar field, or the so-called `good' (or leading twist) components of fermions and gauge field strength tensor (for even $d$)
\begin{align}\label{fields}
\Phi = \{ \phi\,,   \psi\,,    F_{-\boldsymbol\alpha}\}\,,
\end{align}
where $F_{-\boldsymbol\alpha}=F_{\mu \alpha_2\dots \alpha_{d/2}} \bar n^\mu$ is antisymmetric in $\alpha$'s. To separate independent components of the  strength tensor, we impose the condition on its Lorentz indices $1\le \alpha_2 < \dots < \alpha_{d/2}\le d-2$. In $d=2$ dimensions, $\Phi$ can be either scalar or fermion field.

The fields \re{fields} carry the scaling dimension $\Delta_\phi=d/2-1$, $\Delta_\psi=(d-1)/2$ and $\Delta_F=d/2$ and diagonalize 
the operator of angular momentum  $i[M_{+-},\Phi(0)]=s_\Phi\Phi(0)$ with 
\be
s_\phi=0\,,\qqqquad s_\psi=1/2\,,\qqqquad s_F=1\,.
\ee

As a consequence, the fields \re{fields} have the same twist  $\tau_\Phi=\Delta_\Phi-s_\Phi=(d-2)/2$ and their conformal spins are given by
\begin{align}\label{js}
j_\phi={d-2\over 4}\,,\qqqquad j_\psi= {d\over 4}\,,\qqqquad j_F={d+2\over 4}\,.
\end{align}
Note that $j_\Phi=(d-2)/ 4+ s_\Phi$. The remaining field components that are not present in \re{fields} have higher twist.   

In general, the fields $\bar\Phi$ and $\Phi$ can be different. For the sake of simplicity, we shall take them to be conjugated 
to each other, $\bar\Phi=\Phi^\dagger$.
Each derivative in \re{O-sum-der} increases the dimension and the spin of the operator by one unit, leading to
\begin{align}\label{ell}
S=\ell+2s_\Phi\,,\qqqquad \Delta=S+ d-2\,,
\end{align}
where $\ell\ge 0$. The conformal spin of the operator \re{O-sum-der}  and its twist  are
\begin{align}\label{jO}
j_O= S+{d\over 2}-1=\ell+2j_\Phi \,,\qqqquad \tau_O=d-2\,.
\end{align}
In what follows we use $S$ and $\ell$ interchangeably. The relation \re{ell} establishes the relation between the two variables.

To define the expansion coefficients in \re{O-sum-der}, it is convenient to introduce homogenous polynomials
\begin{align}\label{P-c}
P_\ell(p_1,p_2) = \sum_{k=0}^\ell c_{k\ell} \, p_1^k \,p_2^{\ell-k}\,.
\end{align}
Replacing $p_i$ with light-cone derivatives $i\partial_-$, we can rewrite the operators \re{O-sum-der} in a compact form
\begin{align}\label{O-P}
O_S(x)= \bar \Phi (x) P_{\ell} (\stackrel{\rightarrow}{i\partial_-},\stackrel{\leftarrow}{i\partial_-})\Phi(x)\,, 
\end{align}
where arrows indicate the fields the derivative acts on. The polynomials $P_\ell$ define unambiguously the operators $O_S(x)$.
Properties of these polynomials play an important role in our analysis.

To determine the polynomials $P_\ell$, we use the well-known fact that a general form of two- and three-point correlation functions is fixed by the conformal symmetry. For the operators located at $x_i=(x_i^+,x_i^-,\bit{0})$ the corresponding Wightman functions are
\begin{align}\label{phiOphi}\notag
& \vev{\Phi(x_1) \bar \Phi(x_2)} = {i^{1-d/2-2j}  \over (2\pi)^{d/2}}
 {\Gamma(2j)\over (x_{12}^+-i0)^{(d-2)/2} (x_{12}^--i0)^{2j}} \,,
\\
& \vev{\Phi(x_1) O_S(x_2) \bar\Phi(x_3)} \sim {1\over (x_{12}^+ x_{23}^+)^{t_O/2}} {1\over (x_{12}^- x_{23}^-)^{j_O} (x_{13}^-)^{2j-j_O}} \,,
\end{align}
where  we used a shorthand notation for $x_{ij}^\pm =x_i^\pm-x_j^\pm$ and
$j=j_\Phi$ is the conformal spin of fields $\Phi$ and $\bar\Phi$.  All poles in the denominator of \re{phiOphi} have `$\pm i0$' prescription. The sign is dictated by the ordering of the operators on the left-hand side, e.g.
$x_{ij}^\pm \to x_{ij}^\pm -i0$ for $i<j$.  
Replacing $O_S(x_2)$ in \re{phiOphi} with its expression \re{O-P} and comparing expressions on both sides of \re{phiOphi}, we obtain a relation that fixes the expansion coefficients in \re{P-c} up to an overall normalization constant~\cite{Makeenko:1980bh}. 

The resulting expression for the polynomial is~\cite{Makeenko:1980bh,Ohrndorf:1981qv,Braun:2003rp}
\begin{align}\label{Geg}
P_\ell(p_1,p_2) =\kappa_\ell \sum_{k=0}^\ell\binom \ell k
{(-p_1)^k p_2^{\ell-k}\over \Gamma(2j+k)\Gamma(2j+\ell-k) } =(p_1+p_2)^\ell C_\ell^{2j-\frac12}\lr{p_1-p_2\over p_1+p_2}\,,
\end{align}
where $C_\ell^{\lambda}(x)$ is the Gegenbauer polynomial and $\kappa_\ell$ is a normalization constant whose form is not important for our purposes. 
Here the conformal spin $j$ takes the values \re{js} and nonnegative integer $\ell$ is defined in \re{ell}.~\footnote{Note that for scalar field, $j_{\phi}=(d-2) /4$, the polynomial \re{Geg} is proportional to $(d-3)$ for $\ell\ge 1$.\label{foot}} The polynomial \re{Geg} satisfies
\begin{align}\label{P-prop}
P_{\ell}(p_1,p_2)= (-1)^\ell P_{\ell}(p_2,p_1) = P_\ell(-p_2,-p_1)\,.
\end{align}

The appearance of Gegenbauer polynomials in \re{Geg} is not accidental.  
 By virtue of conformal symmetry, the two-point function of the conformal operators \re{O-P} should be diagonal in spins
\begin{align}\label{OO}
\vev{O_S(x) O_{S'}(x')} =\delta_{SS'} \mathcal N_{\ell} {(2i\bar n(x-x'))^{2S}\over [-(x-x')^2]^{2S+d-2} }
{\Gamma(2S+d-2)\over (2\pi)^d}
\,,
\end{align}
where $\ell=S-2s_\Phi$ and  the last factor was introduced for convenience.
Replacing operators with \re{O-P} and repeating the above calculation, one can translate \re{OO} to the orthogonality condition of the Gegenbauer polynomials (see Eq.~\re{CC} in Appendix~\ref{App:Geg})
\begin{align}\label{ortho}
\int_0^1 dt\,  (t(1-t))^{2j-1} P_{\ell}(t,1-t)P_{\ell'}(t,1-t) = \mathcal N_{\ell}\,\delta_{\ell \ell'}\,,
\end{align}
where the normalization factor is given by
\begin{align}\label{norm}
\mathcal N_\ell=  \frac{\pi\,2^{4 (1-2 j)}\, \Gamma (\ell+4 j-1)}{\ell! \, (2 \ell+4 j-1)
   \Gamma^2 \left(2 j-\frac{1}{2}\right)} \,.
\end{align}

Notice that for $d=2$ and $j_\phi=0$, the normalization factor looks as 
$\mathcal N_{\ell}\sim \Gamma(\ell-1)$ and it diverges for $\ell=0$ and $\ell=1$. 
The reason for this is that 
the corresponding scalar operators, $O^{(\phi)}_0=\bar\phi\phi$ and $O^{(\phi)}_1=-i\bar\phi(\stackrel{\rightarrow}{\partial_-}-\stackrel{\leftarrow}{\partial_-})\phi$, are not conformal primary operators in $d=2$. At the same time, for $\ell\ge 2$ the polynomial $P_\ell(p_1,p_2)$ is proportional to $p_1p_2$ (see Eq.~\re{P-sc}) and, as a consequence, the operator $O^{(\phi)}_S$ is built out of $i\partial_-\phi$ which is a conformal primary operator, in distinction to $\phi$.

Thus, the scalar operators $O_S^{(\phi)}$ in $d=2$ have  the spin $S\ge 2$. 
Additional restrictions on the possible values of the spin $S$ come from the condition for the degree of polynomial \re{Geg} to be nonnegative, $\ell\ge 0$. Together with \re{ell} this leads to $S \ge 2s_\Phi$, or equivalently $S_\phi \ge 0$, $S_\psi \ge 1$ and $S_F\ge 2$, where the subscript refers to the choice of $\Phi$.
The relation $S_\psi \ge 1$ holds for arbitrary $d$ and the remaining two, $S_\phi \ge 0$ and $S_F\ge 2$, only for $d>2$.~\footnote{We recall that the leading twist operators in $d=2$ only involve scalar and fermion fields.}

For $S=2$, the operators \eqref{O-P} are proportional to the stress-energy tensor  $T_{--} = T_{\mu\nu} \bar n^\mu \bar n^\nu$ in the corresponding free CFTs. In a free theory of a single complex scalar, Dirac fermion, and gauge field we have~\cite{Osborn:1993cr,Buchel:2009sk}
\begin{align}\notag\label{T}
T^{(\phi)}_{--} &= \partial_-\bar\phi \partial_-\phi -{d-2\over 4(d-1)} \partial_-^2(\bar\phi\phi)  ={1\over 2(d-1)(d-3)} O^{(\phi)}_{S=2}  \,,
\\\notag
T^{(\psi)}_{--} &=  \frac{i}2\bar\psi \gamma_- \!\!\stackrel{\leftrightarrow}{\partial_-} \!\! \psi =  {1\over 2(d-1)} O^{(\psi)}_{S=2} \,,
\\
T^{(F)}_{--} &= {1\over \Gamma(d/2)}F_{-\alpha_2\dots \alpha_{d/2}}F_{- \alpha_2\dots \alpha_{d/2}} = O^{(F)}_{S=2}\,,
\end{align}
where $\stackrel{\leftrightarrow}{\partial}_- =\stackrel{\rightarrow}{\partial}_--\stackrel{\leftarrow}{\partial}_-$ and  $\alpha_i=1,\dots,d-2$. Here the operators $O^{(\Phi)}_{S=2}$ are given by \re{O-P} and \re{Geg} with the corresponding values of $j$ and $\ell$ defined in \re{js} and \re{ell}. The pole at $d=3$
in the first relation in \re{T} is fictitious since $O^{(\phi)}_{S=2}$ is proportional to $(d-3)$ (see footnote~\ref{foot}). 

\subsection{Correlation functions}\label{sect:2}

Next we compute the correlation functions $\vev{\Phi(x_1) \mathcal J_{\omega,S}(n)\bar\Phi(x_2)}$ and $\vev{\Phi(x_1) \mathcal J_{\omega,S} (n) \mathcal J_{\omega',S'} (n') \bar\Phi(x_2)}$. Their explicit expressions are given by Eqs.~\re{3pt1} and \eqref{4pt-Q} below.
Later in this section we establish the commutation relations \re{comm}, by examining the difference of four-point correlation functions
\begin{align}\label{diff}
\vev{\Phi(x_1) \mathcal J_{\omega,S} (n) \mathcal J_{\omega',S'} (n') \bar\Phi(x_2)}
-\vev{\Phi(x_1)\mathcal J_{\omega',S'} (n') \mathcal J_{\omega,S} (n)\bar\Phi(x_2)} \ .
\end{align}
and by comparing it with the three-point function. 

\subsubsection*{Three-point function}

Let us start with the three-point Wightman function $\vev{\Phi(x_1) \mathcal J_{\omega,S}(n)\bar\Phi(x_2)}$.  
Replacing the flow operator $\mathcal J_{\omega,S}(n)$ with its definition \re{J-def} we encounter the correlation function
\begin{align}\label{phiOphi1} 
\vev{\Phi(x_1) O_S(x_3) \bar\Phi(x_2)} =\wick{\c1{\Phi}(x_1) \c1{\bar\Phi}(x_3)}  P_{\ell} (\stackrel{\rightarrow}{i\partial}_{3-},\stackrel{\leftarrow}{i\partial}_{3-})\wick{\c1{\Phi}(x_3)\c1{\bar\Phi}(x_2)}\,,
\end{align}
where 
we applied \re{O-P} and performed Wick contractions.
The two-point function in this relation is given by
\begin{align}\label{PhiPhi}
\vev{\Phi(x_1)\bar\Phi(x_3)} = {\Gamma(2j) \over  2\pi^{d/2}}
 {(2i (\bar nx_{13}))^{2s} \over (-x_{13}^2+i0 x_{13}^0)^{2j}} \,.
\end{align}
Here and in the rest of this section we use a shorthand notation for the conformal and Lorentz spins of the fields, $j \equiv j_\Phi$ and $s \equiv s_\Phi$, respectively.
One can check that in the special kinematics, for $x_i^\mu=(x_i^+,x_i^-,\bit{0})$ and $\bar n^\mu=(0,1,\bit{0})$, 
the last two relations coincide with \re{phiOphi}. 

To obtain $\vev{\Phi(x_1) \mathcal J_{\omega,S}(n)\bar\Phi(x_2)}$ from \re{phiOphi1}, we have to replace $x_3=r n+ t \bar n$ and apply the transformations specified in \re{J-def}. In this way, we find from \re{PhiPhi} for $r\to\infty$
\begin{align}\notag\label{prop1}
\vev{\Phi(x_1)\bar\Phi(x_3)} &\sim 
{(-i)^{2s}\over (2\pi)^{d/2}}
 {(n\bar n)^{2s}\over r^{(d-2)/2}} {\Gamma(2j)\over ((n x_{1})-t(n\bar n)-i0)^{2j}} 
 \\
&={i^{d/2-1}\over (2\pi)^{d/2}}
 {(n\bar n)^{2s}\over r^{(d-2)/2}} \int_0^\infty d\alpha\, \alpha^{2j-1} \e^{-i\alpha ((n x_{1})-t(n\bar n))}\,,
\end{align}
where the minus sign in the exponent on the second line is dictated by `$-i0$' prescription. The propagator $\vev{\Phi(x_3)\bar\Phi(x_2)}$ is given by a complex conjugated expression
\begin{align}\label{prop2}
\vev{\Phi(x_3)\bar\Phi(x_2)} \sim{(-i)^{d/2-1}\over (2\pi)^{d/2}}
 {(n\bar n)^{2s}\over r^{(d-2)/2}} \int_0^\infty d\alpha\, \alpha^{2j-1} \e^{-i\alpha (t(n\bar n)- (nx_{2}))}\,.
\end{align}
At the next step, we have to substitute the last two relations to \re{phiOphi1}, apply the derivatives $i\partial_{3-} = i(\bar n\partial_{x_3})=i(n\bar n)\partial_t$ and, then, perform integration over $t$. Due to a growing number of terms in the polynomial  \re{Geg}, the evaluation of \re{phiOphi1} becomes very cumbersome at high spin $S$ if one uses the expressions for the two-point functions  given on the first line of \re{prop1}.
As we show below, an integral representation \re{prop1} and \re{prop2} is advantageous in carrying out the calculation.

In particular, the action of $i(n\bar n)\partial_t$ on the propagator \re{prop1} amounts to inserting the factor of $-\alpha(n\bar n)$ inside the integral on the right-hand side of \re{prop1}. In this way, substituting \re{prop1} into \re{phiOphi1} 
we find for $r\to\infty$
\begin{align}\label{phiOphi2}\notag
& \vev{\Phi(x_1) O_S(x_3) \bar\Phi(x_2)} \sim
 {(n\bar n)^{S+2s}\over (2\pi)^{d} \,r^{d-2}}
 \\
&\qquad \times \int_0^\infty d\alpha_1d\alpha_2\, (\alpha_1\alpha_2)^{2j-1} P_\ell(\alpha_2,-\alpha_1)
\e^{ -i\alpha_1(n x_{1})+ i\alpha_2(nx_{2})+it(n\bar n)(\alpha_1-\alpha_2)}\,,
\end{align}
where we took into account that $P_\ell$ is a homogenous polynomial of degree $\ell$.
 Note that for $d=2$ the 
expression \re{phiOphi2} approaches a finite value as $r\to\infty$. For $d>2$, it is proportional to $1/r^{d-2}$ but this factor is compensated by the analogous factor coming from the definition of the flow operator \re{J-def}. 

\begin{figure}[h!t]
\psfrag{x1}{$x_1$}\psfrag{x2}{$x_2$}  
\psfrag{w}{$\omega$} 
\psfrag{s1}{$\scriptstyle \alpha_1$}\psfrag{s2}{$\scriptstyle \alpha_2$} 
\centerline{\parbox[c]{60mm}{\insertfig{60}{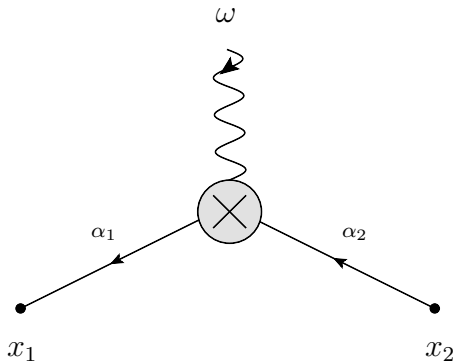}}}
\caption{Diagrammatic representation of the correlation function \re{3pt}.  Gray blob with a wavy line attached represents the flow operator $\mathcal J_{\omega,S} (n)$. The arrow indicates the direction in which the particle propagates, index $\alpha_i$ defines its energy.  }\label{three}
\end{figure}

Finally, we perform a Fourier transform of \re{phiOphi2} with respect to $t$ to get
\begin{align}\notag\label{3pt}
\vev{\Phi(x_1) \mathcal J_{\omega,S}(n)\bar\Phi(x_2)} =  {(n\bar n)^{2j+1-d/2}\over (2\pi)^{d-1} } \int_0^\infty d\alpha_1d\alpha_2\, \e^{-i\alpha_1(n x_{1})+ i\alpha_2(nx_{2})}
\\
\times  (\alpha_1\alpha_2)^{2j-1} P_\ell(\alpha_1,-\alpha_2)\delta(\alpha_1-\alpha_2-\omega)\,,
\end{align}
where we used the identity $P_\ell(\alpha_2,-\alpha_1)=P_\ell(\alpha_1,-\alpha_2)$.
This relation has a simple interpretation as shown in Figure~\ref{three}. 
The correlation function \re{3pt} defines a transition of a spinning particle from point $x_2$ to $x_1$ in the presence of the detector described by the flow operator $\mathcal J_{\omega,S}(n)$. The detector selects a particle propagating in the direction of the null vector $n$ and transfers it the momentum $n\omega$.
Thus, the particle enters the detector with the momentum $p=n\alpha_2$ and leaves it with the momentum $p'=n\alpha_1=n(\alpha_2+\omega)$. The integration variables $\alpha_i$ have the meaning of the energy and $(\alpha_1\alpha_2)^{2j-1} P_\ell(\alpha_2,-\alpha_1)$ defines the detector weight. It depends on the energy of incoming and outgoing particles.

Integration over $\alpha_2$ in \re{3pt} yields 
\begin{align}\notag\label{3pt1}
& \vev{\Phi(x_1) \mathcal J_{\omega,S}(n)\bar\Phi(x_2)} =  {(n\bar n)^{2s}\over (2\pi)^{d-1} }\e^{-i\omega (nx_2)}
\\
& \qquad \times  \int_0^\infty d\alpha_1\,(\alpha_1(\alpha_1-\omega))^{2j-1}\theta(\alpha_1-\omega) \e^{-i\alpha_1(n x_{12})}  P_\ell(\alpha_1,-\alpha_1+\omega) \,,
\end{align}
where a $\theta-$function arises due to $\alpha_2=\alpha_1-\omega\ge 0$. Replacing the polynomial $P_\ell$ with its expression \re{Geg}, we obtain from \re{3pt1} a concise representation of the correlation function 
$\vev{\Phi(x_1) \mathcal J_{\omega,S}(n)\bar\Phi(x_2)}$ for any flow operator. 

The relation \re{3pt1} holds for arbitrary $\omega$. For positive $\omega$, the lower limit of the integration in \re{3pt1} changes from $0$ to $\omega$. This leads to a nonanalyticity of \re{3pt1} at $\omega=0$.
To show this we rewrite the relation \re{3pt1} as
\begin{align}\label{fpm}
\vev{\Phi(x_1) \mathcal J_{\omega,S}(n)\bar\Phi(x_2)} &=
{(n\bar n)^{2s}\over (2\pi)^{d-1} }\e^{-i\omega (nx_1)}\left[
\theta(\omega) f_+(\omega) + \theta(-\omega) f_-(\omega)\right]\,,
\end{align}
where $f_+(\omega)$ and $f_-(\omega)$ are given by the integral on the second line of \re{3pt1} with the $\theta-$function removed and 
 the lower integration limit put to $\omega$ and $0$, respectively. Continuing $f_+(\omega)$ to negative $\omega$, we examine the difference
\begin{align}\notag\label{f-diff}
f_+(\omega)-f_-(\omega) &= - \int_0^{\omega} d\alpha_1 \e^{-i\alpha_1(n x_{12})} (\alpha_1(\alpha_1-\omega))^{2j-1} P_\ell(\alpha_1,-\alpha_1+\omega)
\\
&=(-1)^{2j}\omega^{4j+\ell-1} \int_0^1 dt \, (t(1-t))^{2j-1} P_\ell(t,1-t)\e^{-it \omega (nx_{12})}\,,
\end{align}
where in the second relation we replaced $\alpha_1=\omega t$. The integral in the last relation can be expressed in terms of a Bessel function. Its leading leading asymptotics at small $\omega$ can be found by expanding $\e^{-it\omega (nx_{12})}$ in powers of $\omega t$. Due to orthogonality \re{ortho} of the polynomials $P_\ell$, the first few terms of the expansion vanish after integration. The leading contribution comes from the $O((\omega t)^\ell)$ term, 
\begin{align}\label{f-disc}
 f_+(\omega)-f_-(\omega) = O(\omega^{2j_O-1}) \,,
\end{align}
where $j_O=\ell+2j$ is the conformal spin of the operator $O_S$ defined in \re{js}.
Combining together the relations \re{fpm} and \re{f-disc}, we observe that the function $\vev{\Phi(x_1) \mathcal J_{\omega,S}(n)\bar\Phi(x_2)}$ is continuous at the origin but its derivative of order $2j_O-1$ has a jump at $\omega=0$. Its value can be found from \re{f-diff}.

Invoking interpretation of \re{fpm} as a particle transition amplitude, the discontinuity \re{f-disc} arises because the 
functions $f_+(\omega)$ and $f_-(\omega)$ describe two different processes: for positive (or negative) $\omega$ the detector increases (or decreases) the energy of the particle. As a consequence, the energy of the incoming particle can be arbitrary small for $\omega>0$ but it should be larger than $(-\omega)$ otherwise. The discontinuity \re{f-disc} appears due to this mismatch. 

Let us expand both sides of the relation \re{fpm} at small $\omega$. 
Using representation \re{J-z}, we find that the coefficient in front of $(i\omega)^k$ in the formal expansion of the three-point function on the left-hand side of \re{fpm} involves the operator
\begin{align}\label{V-def}
\mathcal J^{(k)}_{S}(n)  = \int_{-\infty}^\infty dz^- (z^-)^{k}\, O_S (0^+,z^-,\bit z) \,.
\end{align}
We can show that this operator produces a finite contribution to
\re{fpm} for $0\le k\le 2(j_O-1)$ with the conformal spin $j_O$ defined in \re{jO}. According to \re{J-def},  behaviour of \re{fpm} at small $\omega$ is related to asymptotics of the three-point correlation function \re{phiOphi2} at large $t\sim 1/\omega$. Changing the integration variables in \re{phiOphi2} as $\alpha_i\to \alpha_i/t$, one finds that $\vev{\Phi(x_1) O_S(x_3)\bar\Phi(x_2)}\sim 1/t^{2j_O}$ at large $t$ leading to $\vev{\Phi(x_1)\mathcal J^{(k)}_{S}(n) \bar\Phi(x_2)}\sim\int^\infty dt \, t^{k-2j_O}$.  Obviously, the integral is finite for $k\le 2(j_O-1)$.

\subsubsection*{Four-point function}\label{sect:4pt}

The four-point Wightman function $\vev{\Phi(x_1) \mathcal J_{\omega,S}(n)\mathcal J_{\omega,S'}(n')\bar\Phi(x_2)}$ can be split into the sum of disconnected and connected parts. The former factorises into the product of two-point functions, whereas the latter is given by the sum of two diagrams shown in Figure~\ref{four} below.

The calculation of the connected part of  $\vev{\Phi(x_1) \mathcal J_{\omega,S}(n)\mathcal J_{\omega',S'}(n')\bar\Phi(x_2)}$ goes along the same lines as before. 
We start with the four-point function 
$G_4=\vev{\Phi(x_1) O_S(x_3) O_{S'}(x_4) \bar\Phi(x_2)}$, replace the operators with their expressions \re{O-P} 
 and perform Wick contractions. This gives $G_4=G_{4,a}+G_{4,b}$, where
\begin{align}\label{part1}\notag
& G_{4,a} = \vev{\Phi(x_1)\bar\Phi(x_3)}P_\ell(\stackrel{\leftarrow}{\partial}_3,\stackrel{\rightarrow}{\partial}_3) \vev{\Phi(x_3)\bar\Phi(x_4)}P_{\ell'}(\stackrel{\leftarrow}{\partial}_4,\stackrel{\rightarrow}{\partial}_4)\vev{\Phi(x_4)\bar\Phi(x_2)}\,,
\\[2mm]
& G_{4,b}= (-1)^{2s_\Phi}\vev{\Phi(x_1)\bar\Phi(x_4)}P_{\ell'}(\stackrel{\rightarrow}{\partial}_4,\stackrel{\leftarrow}{\partial}_4)\vev{\bar\Phi(x_3)\Phi(x_4)}
P_\ell(\stackrel{\rightarrow}{\partial}_3,\stackrel{\leftarrow}{\partial}_3) 
\vev{\Phi(x_3)\bar\Phi(x_2)}\,.
\end{align}
Here $(-1)^{2s_\Phi}$ takes into account Bose-Fermi statistics of the field $\Phi$.  
As before, the derivatives $\partial_i=(\bar n\partial_{x_i}) \equiv \partial_{i-}$ act in the direction indicated by the arrow.

\begin{figure}[h!t]
\psfrag{x3}{$x_1$}\psfrag{x4}{$x_2$}  
\psfrag{n1w1}[cc][rr]{$\omega$}\psfrag{n2w2}{$\omega'$}
\psfrag{s1}{$\scriptstyle \alpha_1$}\psfrag{s3}{$\scriptstyle \alpha_2$}
\psfrag{s2}{$\scriptstyle \alpha_3$} 
\psfrag{(a)}{(a)}\psfrag{(b)}{(b)} 
\centerline{\parbox[c]{150mm}{\insertfig{150}{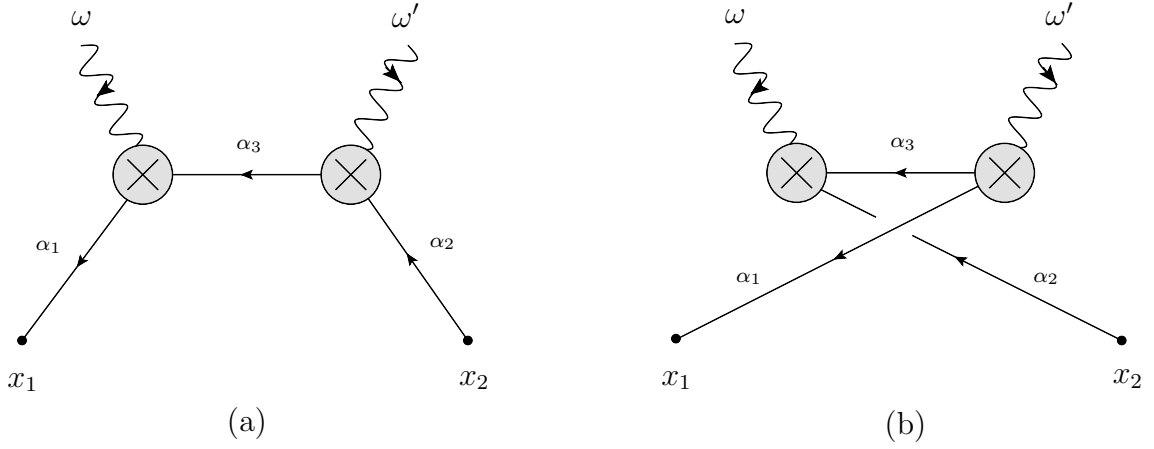}}}
\caption{Connected part of the correlation function $\vev{\Phi(x_1) \mathcal J_{\omega,S} (n) \mathcal J_{\omega',S'} (n') \bar\Phi(x_2)}$.  }
\label{four}
\end{figure}
 
 The two relations in \re{part1} correspond to diagrams shown in Figure~\ref{four}(a) and (b), respectively.
They involve two propagators attached to the external points, $x_1$ and $x_2$, and internal one connecting the points $x_3$ and $x_4$. 
Replacing $x_3=rn+t\bar n$ and $x_4=r'n'+t'\bar n$ we find that for $r,r'\to\infty$ the former propagators are given by \re{prop1} and \re{prop2}. The internal propagator is given by \re{PhiPhi}
\begin{align}\label{inter0}
\vev{\Phi(x_3)\bar\Phi(x_4)} \sim (r r')^{1-d/2} {\Gamma(2j) \over (2\pi)^{d/2}} i^{1-d/2-2j}
 {\epsilon^{1-d/2} \over [-(nn')/\epsilon +t-t'-i0]^{2j}} \,,
\end{align}
where the notation was introduced for $\epsilon=(n\bar n)/r'-(n'\bar n)/r$. 
Going to the limit $\epsilon\to 0$, we notice that the relation \re{inter0} has different behaviour for $d=2$ and $d>2$. 
 
For $d=2$ the null vectors $n$ and $n'$ coincide  and, therefore, $(nn')=0$.~\footnote{In $d=2$ dimensions we have $\bar n \sim e_-$ and $n\,, n' \sim e_+$ where $e_\pm^\mu=(1,\pm 1)$ are basis null vectors.  } 
Then, the relation \re{inter0} takes the form
\begin{align}\label{prop-2d}
\vev{\Phi(x_3)\bar\Phi(x_4)}\big|_{d=2} \sim {(-i)^{2j}\over (2\pi)^{d/2}} 
 {\Gamma(2j) \over (t-t'-i0)^{2j}}\,.
\end{align}

For $d>2$  the expression \re{inter0} vanishes for $(nn')\neq 0$ and diverges for $(nn')=0$. This suggests that it should be proportional to a delta-function. Indeed, carefully taking the limit we arrive at 
 (see \re{limit} in Appendix~\ref{App:conv})
\begin{align}\label{inter}
\vev{\Phi(x_3)\bar\Phi(x_4)} \sim  {1 \over 2\pi}(r r')^{1-d/2} \delta^{(d-2)}(n,n') 
\int_0^\infty d\alpha_3 \, \alpha_3^{2s-1} \e^{-i\alpha_3(t-t')}
\,,
\end{align}
where $s=j-(d-2)/4$ is the spin of the field $\Phi$ and $ \delta^{(d-2)}(n,n')$ is the delta-function on the sphere $S^{d-2}$ defined in \re{delta}. For $d=2$ this relation reduces to \re{prop-2d} after one replaces $\delta^{(d-2)}(n,n')\to 1$.
As before, the integral representation \re{inter} is advantageous in carrying out the calculations. 

The integral in \re{inter} is well defined for $s>0$. This condition is satisfied for fermion and gauge fields but it is violated for scalars in which case the integral \re{inter} diverges logarithmically as $\alpha_3\to 0$. To obtain the two-point function for scalars, we
replace $j=(d-2)/4$ in \re{inter0} and take the limit $\epsilon\to 0$  
\begin{align}\label{phiphi}
\vev{\phi(x_3)\bar\phi(x_4)} \sim (r r')^{1-d/2} {\Gamma(d/2-1) \over (2\pi)^{d/2}}  
 {1 \over (nn')^{d/2-1}}\,.
\end{align}
In distinction to \re{inter}, it remains nonzero for $(nn')\neq 0$ and, in addition, it is independent on $t$ and $t'$.
Such a difference between \re{inter} and \re{phiphi} is a manifestation of the cross-talk between the two detectors ~\cite{Belitsky:2013bja}. We recall that the momentum of a particle entering and leaving the detector is aligned with the null vector defining its orientation. In diagrams shown in Figure~\ref{four}, 
the particle with the energy $\alpha_3$ is exchanged between the two detectors and, therefore, its momentum should be aligned with $n$ and $n'$ simultaneously. This is only possible if $n=n'$ or $\alpha_3=0$. The former corresponds to \re{inter} and the latter to the exchange of a zero-energy scalar between the detectors. This explains why \re{phiphi} does not depend on the time difference $t-t'$.

We are now ready to compute the contribution of the diagrams shown in Figure~\ref{four}. 
We start with \re{part1} and replace the two-point functions by their expressions \re{prop1}, \re{prop2} and \re{inter}.
Then, we apply to $G_{4,a}$ and $G_{4,b}$ the transformation \re{J-def} 
to convert the operators at the points $x_3$ and $x_4$ to the flow operators $\mathcal J_{\omega,S}(n)$ and $\mathcal J_{\omega',S'}(n')$, respectively. Denoting the resulting expressions as $\mathcal G_{4,a}$ and $\mathcal G_{4,b}$
we get
\begin{align}\notag\label{Ga}
\mathcal G_{4,a}= {1\over (2\pi)^{d-1}}  \delta^{(d-2)}(n,n') \int_0^\infty d\alpha_1 d\alpha_2 d\alpha_3 \, (\alpha_1\alpha_2)^{2j-1} \alpha_3^{2s-1} \e^{-i\alpha_1 (nx_{1})+i\alpha_2 (n'x_{2})}
\\[2mm]
 \times P_{\ell}(\alpha_1,-\alpha_3)P_{\ell'}(\alpha_3,-\alpha_2)  \delta(\alpha_1-\alpha_3-\omega)\delta(\alpha_2-\alpha_3+\omega')\,,
\\[2mm]\label{Gb}
\notag
\mathcal G_{4,b} = {1\over (2\pi)^{d-1}}  \delta^{(d-2)}(n,n')  \int_0^\infty d\alpha_1 d\alpha_2 d\alpha_3 \, (\alpha_1\alpha_2)^{2j-1} \alpha_3^{2s-1} \e^{-i\alpha_1 (nx_{1})+i\alpha_2 (n'x_{2})}
\\[2mm]
 \times (-1)^{2s}P_{\ell'}(\alpha_1,\alpha_3)P_{\ell}(-\alpha_3,-\alpha_2)\delta(\alpha_1+\alpha_3-\omega')\delta(\alpha_2+\alpha_3+\omega)\,.
\end{align}
A general form of these relations can be read off from the diagrams shown in Figure~\ref{four}. The integration variables  $\alpha_i$ parameterize energy of the particles, whereas $\omega$ and $\omega'$ define energy transferred to them by the detectors. The delta-functions express the energy conservation and each detector vertex comes with a polynomial weight that depends on the energy of particles attached to it.~\footnote{The additional minus signs in the arguments of the polynomials correspond to the incoming particles.} Note that the two diagrams in Figure~\ref{four} can be obtained one from another by exchanging the detectors and flipping the direction of the line with index $\alpha_3$. As a consequence, the second line in the expression for $\mathcal G_{4,b}$ can be obtained from the analogous expression for $\mathcal G_{4,a}$ by exchanging $(\omega,\ell)$ with $(\omega',\ell')$  and replacing $\alpha_3\to -\alpha_3$.

The product of $\delta-$functions in \re{Ga} and \re{Gb} localizes integrals over $\alpha_2$ and $\alpha_3$. It also imposes constraints on the possible values of $\omega$'s. In particular, it follows from \re{Gb} that $\mathcal G_{4,b}$ is different from zero only for $\omega<0$ and $\omega'>0$. Combining together \re{Ga} and \re{Gb}, we obtain 
\begin{align}\label{4pt}
& \vev{\Phi(x_1) \mathcal J_{\omega,S}(n)\mathcal J_{\omega',S'}(n')\bar\Phi(x_2)}_c= {1\over (2\pi)^{d-1}}  \delta^{(d-2)}(n,n') \, \mathcal I_\Phi\,,
\end{align}
where the subscript `$c$' refers to the connected part and the notation was introduced for 
\begin{align}\notag\label{I-gen-j}
\mathcal I_\Phi = & \e^{-i\omega'' (n x_{2})}  \int_0^\infty d\alpha_1 \, (\alpha_1(\alpha_1-\omega''))^{2j-1}  \e^{-i\alpha_1 (nx_{12})}\theta(\alpha_1-\omega'')
\\\notag
&\times \Big[\theta(\alpha_1-\omega)(\alpha_1-\omega)^{2s-1} P_{\ell}(\alpha_1,-\alpha_1+\omega)P_{\ell'}(\alpha_1-\omega,-\alpha_1+\omega'')
\\
&{} \ \ -\theta(\omega'-\alpha_1)(\alpha_1-\omega')^{2s-1} P_{\ell'}(\alpha_1,-\alpha_1+\omega')P_{\ell}(\alpha_1-\omega',-\alpha_1+\omega'')\Big]\,,
\end{align}
with $\omega''=\omega+\omega'$ being the total energy transferred by the detectors. We recall that the polynomials 
$P_\ell$ and $P_{\ell'}$ are given by \re{Geg} and their degree is related to the spin of the flow operators through \re{ell}.
 
To obtain the correlation function $\vev{\Phi(x_1)\mathcal J_{\omega',S'}(n') \mathcal J_{\omega,S}(n)\bar\Phi(x_2)}_c$ from \re{4pt} it suffices to exchange $(\omega, \ell)$ and $(\omega', \ell')$ in \re{I-gen-j}. Combining the two expressions we obtain
\begin{align}\notag\label{4pt-Q}
\vev{\Phi(x_1) [\mathcal J_{\omega,S} (n), \mathcal J_{\omega',S'} (n')] \bar\Phi(x_2)}_c={1\over (2\pi)^{d-1}}  \delta^{(d-2)}(n,n')    \e^{-i\omega'' (n x_{2})}
\\
\qquad
\times \int_0^\infty d\alpha_1 \, (\alpha_1(\alpha_1-\omega''))^{2j-1} \theta(\alpha_1-\omega'')  \e^{-i\alpha_1 (nx_{12})}Q(\alpha_1)\,,
\end{align}
where $\omega''=\omega+\omega'$ and $Q(\alpha_1)$ is given by
\begin{align}\notag\label{Q}
Q(\alpha_1)= & (\alpha_1-\omega)^{2s-1} P_{\ell}(\alpha_1,-\alpha_1+\omega)P_{\ell'}(\alpha_1-\omega,-\alpha_1+\omega'')
\\[2mm]
{} -&(\alpha_1-\omega')^{2s-1} P_{\ell'}(\alpha_1,-\alpha_1+\omega')P_{\ell}(\alpha_1-\omega',-\alpha_1+\omega'')\,.
\end{align}
Here nonnegative integer $\ell$ and $\ell'$ are related to Lorentz spins of the  operators as follows
\begin{align}
\ell=S-2s\,,\qquad \ell'=S'-2s\,.
\end{align}
As compared with  \re{I-gen-j}, the relation \re{Q} does not involve theta--functions.

Notice that the relation \re{4pt-Q} is only well-defined for $s>0$, that is for the flow operators built out of fermion and gauge fields.
For scalars, $s=0$, the integral in \re{I-gen-j} and \re{4pt-Q} diverges logarithmically for $\alpha_1\to \omega$ and $\alpha_1\to\omega'$. This singularity arises from the exchange of a scalar with zero energy between the two detectors (see Figure~\ref{four}).
As explained above, the propagator of this particle takes the form \re{phiphi}. If does not vanish for $n'\neq n$ and scales as $1/(nn')^{d/2-1}$.
Repeating the calculation of the diagrams shown in Figure~\ref{four}, we find (see Appendix~\ref{app:scalar})
\begin{align}\label{4pt-s} 
\vev{\phi(x_1) \mathcal J_{\omega,S}(n)\mathcal J_{\omega',S'}(n')\bar\phi(x_2)}_c=
 {1\over (2\pi)^{d-1}} \left[{\Gamma(d/2-1)\over (2\pi)^{d/2-1}} {X_\phi \over (nn')^{d/2-1}} +\dots \right],
\end{align} 
where dots denote terms localized at $n'=n$ and the function $X_\phi$ describes the cross-talk between detectors
\begin{align}\notag\label{X}
X_\phi = (-\omega \omega')^{d/2-2} & \Big[\e^{-i\omega (nx_1)-i\omega' (n'x_2)}
P_{\ell}(\omega,0)P_{\ell'}(0,\omega') \theta(\omega)\theta(-\omega') 
\\[2mm] 
&
+ \e^{-i\omega' (n' x_1)-i\omega(n x_2)} P_{\ell'}(\omega',0) P_{\ell}(0,\omega)\theta(\omega')\theta(-\omega)\Big]\,.
\end{align}
The polynomials have vanishing argument because the particle exchanged between the two detectors has zero energy. Notice that $X_\phi$ is symmetric under the exchange of the detectors and, therefore, it cancels in the difference of the correlation functions \re{diff} for $n'\neq n$. 
Carefully examining this difference we find  (see Appendix~\ref{app:scalar} for details) 
that $\vev{\phi(x_1) [\mathcal J_{\omega,S}(n),\mathcal J_{\omega',S'}(n')]\bar\phi(x_2)}_c$ is given by \re{4pt-Q} and \re{Q} evaluated at $s=0$ and $j=(d-2)/4$. Most importantly, the poles of $Q(\alpha_1)$ at $\alpha_1=\omega$ and $\alpha_1=\omega'$ are integrated using the principal value prescription.

\subsubsection*{Disconnected part}
 
Disconnected part of the four-point function is given by $\vev{\Phi(x_1)  \bar\Phi(x_2)}\vev{\mathcal J_{\omega,S} (n) \mathcal J_{\omega',S'} (n')}$.
To find $\vev{\mathcal J_{\omega,S} (n) \mathcal J_{\omega',S'} (n')}$ we start with the two-point function
$\vev{O_S(x) O_{S'}(x')}$ defined in \re{OO} and transform $O_S(x)$ and $O_{S'}(x')$  into the flow operators \re{J-def}. 

Replacing $x=n r+ \bar n t$ in \re{OO} we obtain for $r\to\infty$
\begin{align}
\vev{O_S(x) O_{S'}(x')} \sim\delta_{SS'} (2r)^{2-d}\mathcal N_{\ell} {i^{2S}(n \bar n)^{2S}\over (2\pi)^d }
{\Gamma(2S+d-2)\over (-(n \bar n) t+(nx')+i0)^{2S+d-2}}
\,.
\end{align}
Then, taking Fourier transform with respect to $t$ we arrive at
\begin{align} \label{JO}
\vev{\mathcal J_{\omega,S}(n)  O_{S'}(x') } = 
 \delta_{SS'} {2i^{2-d}\over (4\pi)^{d-1}}\mathcal N_{\ell}(n\bar n)^{S}   \theta(-\omega)(-\omega) ^{2S+d-3}\e^{-i\omega (nx')} \,.
\end{align}
This relation has the following interpretation. The operator $O_{S'}(x')$ creates out of vacuum two particles which propagate
to the final state and  
get absorbed by the detector $\mathcal J_{\omega,S}(n)$. The total energy of the particles is 
$(-\omega)>0$.
This explains why \re{JO} vanishes for positive $\omega$, leading to 
$\bra{0}\mathcal J_{\omega,S}(n)=0$ for $\omega>0$. Taking into account the relation $(\mathcal J_{\omega,S}(n))^\dagger = \mathcal J_{-\omega,S}(n)$, we deduce that $\mathcal J_{\omega,S}(n)\ket{0}=0$ for $\omega<0$.

To convert the operator $O_{S'}(x')$ into the flow operator \re{J-def}, we replace $x'=n' r' + t' \bar n$ in \re{JO},  take the limit $r'\to\infty$ and perform a Fourier transform with respect to $t'$. This gives
\begin{align} \label{JJ}
\vev{\mathcal J_{\omega,S}(n)\mathcal J_{\omega',S'}(n') } &= \delta_{SS'}
\theta(-\omega) \delta(\omega+\omega') (-\omega)^{2S-1}\Omega_S\lr{n,n'}\,,
\end{align}
where the notation was introduced for 
\begin{align}\label{Omega}
\Omega_S\lr{n,n'}= {\mathcal N_{\ell}\over (4\pi)^{d-2}} (n\bar n)^{S} (n'\bar n)^{-S} \lim_{r'\to\infty} (-i\omega r')^{d-2}  \,\e^{-i\omega (n n') r' } \,.
\end{align}
It depends on two null vectors $n^\mu=(1,\vec n)$ and $n'{}^\mu=(1,\vec n')$ where $\vec n$ and $\vec n'$ are unit vectors on  the sphere $S^{d-2}$. The function $\Omega_d\lr{n,n'}$ has different properties at $d=2$ and for $d>2$.

For $d=2$ we have $n =n'$ and, as a consequence, the function \re{Omega} takes a finite value for the scalar ($j_\phi=0$) and fermionic ($j_\psi=1/2$) flow operators
\begin{align}\label{Omega2}\notag
& \Omega_{S}^{(\phi)} \lr{n,n}\Big|_{d=2}= {4\over S(S-1)(2S-1)}\,,
\\ 
& \Omega_{S}^{(\psi)} \lr{n,n}\Big|_{d=2}= {1\over 2S-1}\,.
\end{align}

For $d>2$ the function \re{Omega} vanishes for $(nn')\neq 0$ and diverges for $(nn')=0$. Replacing $\omega r'=1/\epsilon$, we
find that the expression on the right-hand side of \re{Omega} scales as $1/\epsilon^{d-2}\e^{i(nn')/\epsilon}$ for $\epsilon\to 0$. As compared with the analogous expression for $\delta^{(d-2)} (n,n')$, Eq.\,\re{delta}, it contains an additional factor of $1/\epsilon^{d/2-1}$. 
As a consequence, the integral of $\Omega_S\lr{n,n'}$ with a test function on the sphere $S^{d-2}$ diverges as $\epsilon\to 0$, and the limit \eqref{Omega} does not produce a well-defined distribution.

This divergence is not surprising. In the  correlation function $\vev{\mathcal J_{\omega,S}(n)\mathcal J_{\omega',S'}(n') }$, the flow operator $\mathcal J_{\omega',S'}(n')$ creates a pair of particles which propagate in the direction of the null vector $n$ and carry the total energy $\omega'$. These particles are absorbed by the detector oriented along the direction $n'$. Each particle brings in the factor of $\delta^{(d-2)}(n,n')$ (see Eq.\,\re{inter}), so that the transition amplitude is proportional to the square of the delta-function, schematically,
\begin{align}\label{Omegad}
\Omega_S(n,n') = C_S \,\delta^{(d-2)}(n,n')\,,
\end{align}
where $C_S \sim 1/\epsilon^{d/2-1}$ diverges as $\epsilon\to 0$.
Indeed, it is easy to see using \re{delta} that the additional factor of $1/\epsilon^{d/2-1}$  comes from $\delta^{(d-2)}(n,n)$. 

The fact that the correlation function \re{JJ} is divergent for $d>2$ is in agreement with the findings of 
Refs.~\cite{Belin:2020lsr,Besken:2020snx} where similar correlation functions were studied for $d=4$ and $S=2$.

\subsection{Light-ray algebra}\label{sect:3}

In this subsection we find the commutation relations of the flow operators by comparing \re{4pt-Q} with the analogous expression for the three-point function \re{3pt1}. 

\subsubsection*{Free fermion and gauge field}

It is easy to see from \re{Q} that  $Q(\alpha_1)$ is a polynomial in $\alpha_1$  for fermion ($s_\psi=1/2$) and gauge field ($s_F=1$) operators. Due to antisymmetry under the exchange of $(\omega,\ell)$ and $(\omega',\ell')$, its degree is  
 $\ell+\ell'+2(s-1)$.
Aside from the delta-function $\delta^{(d-2)}(n,n')$, the relation \re{4pt-Q} is remarkably similar to the expression for the three-point function \re{3pt1}. To match the two relations, we take advantage of the properties of Gegenbauer polynomials and expand \re{Q} over the orthogonal basis of polynomials $P_{\ell''}(\alpha_1,-\alpha_1+\omega'')$ defined in \re{Geg}
\begin{align}\label{Q-sum}
Q(\alpha_1) = \sum_{\ell''=0}^{\ell+\ell'+2(s-1)} C_{\ell\ell'}{}^{\ell''} (\omega,\omega') P_{\ell''}(\alpha_1,-\alpha_1+\omega'')\,,
\end{align}
where $s$ is the spin of the field $\Phi$ and the maximal value of $\ell''$ coincides with the degree of $Q(\alpha_1)$.

Substituting \re{Q-sum} into \re{4pt-Q} and matching the resulting expression to \re{3pt1}, we arrive at the algebra
\begin{align}\label{alg}
{[\mathcal J_{\omega,S} (n) , \mathcal J_{\omega',S'} (n') ]}_c=\delta^{(d-2)}(n,n')  \sum_{S''=2s}^{S+S'-2} C_{\ell\ell'}{}^{\ell''}(\omega,\omega') \mathcal J_{\omega+\omega',S''} (n) \,,
\end{align}
where $\ell_i=S_i-2s$ and the structure constants coincide with the expansion coefficients in \re{Q-sum}.
The minimal value of $S''$ in the sum \re{alg} equals $2s_\psi=1$ for fermion and $2s_F=2$ for gauge field flow operators. 

The relations \re{Q} and \re{Q-sum} lead to nontrivial conditions for the structure constants
\begin{align}\notag\label{C-pro}
& C_{\ell \ell'}{}^{\ell''}(\omega,\omega') = -C_{\ell' \ell}{}^{\ell''}(\omega',\omega)\,,
\\[2mm]\notag
& C_{\ell \ell'}{}^{\ell''}(\lambda\omega,\lambda\omega') = \lambda^{\ell+\ell'-\ell'' +2s-1} C_{\ell \ell'}{}^{\ell''}(\omega,\omega')\,, 
\\[2mm]
& C_{\ell \ell'}{}^{\ell''}(-\omega,-\omega') =-C_{\ell \ell'}{}^{\ell''}(\omega,\omega') \,.
\end{align}
Here the first relation follows from the antisymmetry of \re{Q} under the exchange of $(\omega,\ell)$ and $(\omega',\ell')$. 
To check the second relation, it is sufficient to compare the properties of both sides of \re{Q-sum} under rescaling  $\alpha\to \lambda \alpha$, $\omega\to \lambda\omega$ and $\omega'\to \lambda\omega'$. Moreover, matching \re{Q-sum} to \re{Q} we find that $C_{\ell \ell'}{}^{\ell''}(\omega,\omega')$ are homogenous polynomials in $\omega$ and $\omega'$ with real coefficients. It follows from the second relation in \re{C-pro} that their degree is $\ell+\ell'-\ell'' +2s-1=S+S'-S''-1$.
Finally, the last 
relation in \re{C-pro} can be derived by applying complex conjugation to both sides of \re{alg} and by taking into account the relation $(\mathcal J_{\omega,S} (n))^\dagger= \mathcal J_{-\omega,S} (n)$.

Substituting $\lambda=-1$ into the second relation in \re{C-pro} and comparing it with the third one, we find that the structure constants $C_{\ell \ell'}{}^{ \ell''}(\omega,\omega')$ vanish for even $\ell+\ell'-\ell'' +2s-1$, or equivalently for odd
$S+S'-S''$. This means that the sum in \re{alg} runs over $S''=S+S'-2p$ with $p=1,2,\dots$. 

 To summarize, the structure constants take the following general form 
\begin{align}\label{C-gen}
C_{\ell \ell'}{}^{\ell''}(\omega,\omega') = \frac{1+(-1)^{S+S'-S''}}{2}\sum_{k=0}^{S+S'-S''-1} c_{k}(\ell,\ell')\, \omega^k (\omega')^{S+S'-S''-1-k}\,,
\end{align}
where $S_i=\ell_i+2s$ and real-valued expansion coefficients depend on spins and satisfy $c_k(\ell,\ell') = - c_{S+S'-S''-1-k}(\ell',\ell)$.
Explicit expressions for the structure constants are derived in the next section. In what follows we do not display the signature factor and tacitly assume that $C_{\ell \ell'}{}^{\ell''}(\omega,\omega')$ vanish for odd $S+S'-S''$.

For the algebra \re{alg} to be consistent, the structure constants have to verify Jacobi identities. Indeed, it is straightforward to check using \re{Q} and \re{Q-sum} that $C_{\ell\ell'}{}^{\ell''}(\omega,\omega')$ satisfy 
\begin{align}\notag\label{Jacobi}
\sum_{\ell'}C_{\ell_1\ell_2}{}^{\ell'} (\omega_1,\omega_2)C_{\ell'\ell_3}{}^{\ell''} (\omega_1+\omega_2,\omega_3)&
\\\notag
+\sum_{\ell'}C_{\ell_2\ell_3}{}^{\ell'} (\omega_2,\omega_3)C_{\ell'\ell_1}{}^{\ell''} (\omega_2+\omega_3,\omega_1)&
\\
+\sum_{\ell'}C_{\ell_3\ell_1}{}^{\ell'} (\omega_3,\omega_1)C_{\ell'\ell_2}{}^{\ell''} (\omega_3+\omega_1,\omega_2)&=0\,.
\end{align}

The relation \re{alg} takes into account the contribution to \re{diff} of the connected part of the four-point function. We use \re{JJ} to obtain the disconnected contribution to \re{diff} as
\begin{align}\label{alg1}
\vev{\Phi(x_1) [\mathcal J_{\omega,S} (n), \mathcal J_{\omega',S'} (n')] \bar\Phi(x_2)}_d=
 -\delta_{SS'} \delta(\omega+\omega') \omega^{2S-1}\Omega_d\lr{n,n'} \vev{\Phi(x_1)\bar\Phi(x_2) } \,.
\end{align}
It defines the central extension of the algebra \re{alg}. Namely,
combining together \re{alg} and \re{alg1} we arrive at
\begin{align}\label{alg2}\notag
{[\mathcal J_{\omega,S} (n) , \mathcal J_{\omega',S'} (n') ]} &=\delta^{(d-2)}(n,n')  \sum_{S''=2s}^{S+S'-2} 
C_{\ell\ell'}{}^{\ell''}(\omega,\omega') \mathcal J_{\omega+\omega',S''} (n)  
\\
&
 -\delta_{SS'} \delta(\omega+\omega') \omega^{2S-1}\Omega_S\lr{n,n'} 
 \,,
\end{align}
where $\ell_i=S_i-2s$ and the sum runs over the spins $S''$ of the same parity as $S+S'$. Here
the structure constants take the form \re{C-gen} and
satisfy the relations \re{Q-sum} and \re{Q-phi-sum}. The central charge $\Omega_d\lr{n,n'}$ takes a finite value \re{Omega2} for $d=2$ but it diverges for $d>2$ as \re{Omegad}. For $S=2$ this agrees with the findings of Refs.~\cite{Wall:2011hj,Casini:2017roe,Huang:2020ycs,Belin:2020lsr,Besken:2020snx}.
 
\subsubsection*{Free scalar field}

Let us turn to the flow operators built out of the scalars. In this case, the four-point function is given by \re{4pt-Q} for $s=0$ and $j=(d-2)/4$. 
An important difference as compared to previous case is that \re{Q} is a meromorphic function and it is integrated in \re{4pt-Q} using the 
principal value prescription (see Eq.~\re{start2} in Appendix~\ref{app:scalar}). 

To separate the contribution of poles, it is convenient to rewrite $Q(\alpha_1)$ as
\begin{align}\label{Q-sum1}
Q(\alpha_1) = Q^{(\phi)}(\alpha_1) + 
\left[{P_{\ell}(\omega,0)P_{\ell'}(0,\omega')\over \alpha_1-\omega} -{P_{\ell'}(\omega',0)P_{\ell}(0,\omega)\over \alpha_1-\omega'}\right],
\end{align}
where $Q^{(\phi)}(\alpha_1)$ is a polynomial in $\alpha_1$ of degree $\ell+\ell'-2$
\begin{align}\notag\label{Q-phi}
Q^{(\phi)}(\alpha_1)=
& {P_{\ell}(\alpha_1,-\alpha_1+\omega)P_{\ell'}(\alpha_1-\omega,-\alpha_1+\omega'')-P_{\ell}(\omega,0)P_{\ell'}(0,\omega')\over \alpha_1-\omega} 
\\
{} -& { P_{\ell'}(\alpha_1,-\alpha_1+\omega')P_{\ell}(\alpha_1-\omega',-\alpha_1+\omega'')-P_{\ell'}(\omega',0)P_{\ell}(0,\omega)\over \alpha_1-\omega'}\,.
\end{align}
Substituting \re{Q-sum1} into  \re{4pt-Q} we obtain the sum of two terms. The first term contains $Q^{(\phi)} (\alpha_1)$.
In a close analogy with \re{Q-sum} we can write
\begin{align}\label{Q-phi-sum}
Q^{(\phi)} (\alpha_1) = \sum_{\ell''=0}^{\ell+\ell'-2} C^{(\phi)}_{\ell\ell'}{}^{\ell''} (\omega,\omega') P_{\ell''}(\alpha_1,-\alpha_1+\omega'')\,.
\end{align}
Going along the same lines as in the previous subsection, we find that \re{Q-phi-sum} produces the contribution to the commutator
$[\mathcal J_{\omega,S} (n) , \mathcal J_{\omega',S'} (n') ]$ that takes the same form as \re{alg} with the structure constants given by 
the expansion coefficients $C^{(\phi)}_{\ell\ell'}{}^{\ell''}$. The additional contribution to the commutator comes from the poles in \re{Q-sum1}. It reads
\begin{align}\label{alg-phi}\notag
{[\mathcal J^{(\phi)}_{\omega,S} (n) , \mathcal J^{(\phi)}_{\omega',S'} (n') ]} &=\delta^{(d-2)}(n,n') \left[ \sum_{S''=2s}^{S+S'-2} 
C^{(\phi)}_{\ell\ell'}{}^{\ell''}(\omega,\omega') \mathcal J^{(\phi)}_{\omega+\omega',S''} (n)  + 
\omega^S (\omega')^{S'}\mathcal O^{(\phi)}_{\omega,\omega'}(n) \right]
\\
&
 -\delta_{SS'} \delta(\omega+\omega') \omega^{2S-1}\Omega^{(\phi)}_S\lr{n,n'} 
 \,,
\end{align}
where  the structure constants $C^{(\phi)}_{\ell\ell'}{}^{\ell''}$ (with $\ell_i=S_i$) satisfy the relations \re{C-pro} and \re{Jacobi} for $j=(d-2)/4$ and $s=0$. 

The operator $\mathcal O^{(\phi)}_{\omega,\omega'}(n)$ satisfies the defining relation
\begin{align}\notag\label{nonlocal}
\vev{\phi(x_1)\mathcal O^{(\phi)}_{\omega,\omega'}(n)\bar\phi(x_2)}={1\over (2\pi)^{d-1}} \e^{-i\omega'' (n x_{2})}
\dashint_0^\infty d\alpha_1 \, (\alpha_1(\alpha_1-\omega''))^{d/2-2} \theta(\alpha_1-\omega'') 
\\
\qquad
\times   \e^{-i\alpha_1 (nx_{12})}
\omega^{-\ell} (\omega')^{-\ell'}\left[{P_{\ell}(\omega,0)P_{\ell'}(0,\omega')\over \alpha_1-\omega} -{P_{\ell'}(\omega',0)P_{\ell}(0,\omega)\over \alpha_1-\omega'}\right],
\end{align}
where $\omega''=\omega+\omega'$ and the integral is regularized in the vicinity of the poles $\alpha_1=\omega$ and $\alpha_1=\omega'$ using the principal value prescription. The operator $\mathcal O^{(\phi)}_{\omega,\omega'}(n)$ should be antisymmetric under the exchange of $(\omega,\ell)$ and $(\omega',\ell')$.  We expect that it should be linear in $\phi$ and $\bar\phi$.
With some work we obtain
\begin{align}\notag\label{bi}
\mathcal O^{(\phi)}_{\omega,\omega'}(n) &= -{i\over 2} \lim_{r\to\infty} (n\bar n)^2 r^{d-2}\int_{-\infty}^\infty dt_1 dt_2 \e^{-it_1 (n\bar n) \omega -it_2 (n\bar n) \omega'}  \left[\theta(t_1-t_2)-\theta(t_2-t_1)\right]
\\
& \times \omega^{-\ell} (\omega')^{-\ell'}  \left[ P_{\ell}(\omega,0)P_{\ell'}(0,\omega')\phi(x_1) \bar\phi(x_2)+ P_{\ell'}(\omega',0)P_{\ell}(0,\omega)\bar\phi(x_1)\phi(x_2)\right], 
\end{align}
where $x_1=r n+ t_1 \bar n$ and $x_2=r n+ t_2 \bar n$.

The scalar fields in \re{bi} are accompanied by polynomials defined in \re{Geg}.
It follows from \re{Geg} that 
\begin{align}
\label{eq:explicitP0}
\omega^{-\ell}P_{\ell}(\omega,0)= {\Gamma(d-3+\ell)\over \Gamma(d-3) \, \ell!}\,,\qqqquad P_{\ell}(0,\omega)=(-1)^\ell P_{\ell}(\omega,0)\,.
\end{align}
We recall that in $d=2$ dimensions, the polynomials $P_{\ell}$ have degree $\ell\ge 2$.  Then, we deduce from the last relation that $P_{\ell}(\omega,0)=P_{\ell}(0,\omega)=0$ in this case and, therefore, the nonlocal operator \re{bi} vanishes for $d=2$. The same is true at $d=3$ for $\ell\ge 1$ but this happens due to a particular choice of normalization of the polynomial \re{Geg} (see footnote~\ref{foot}).
Changing normalization by inserting the factor of $1/(d-3)$ on the right-hand side of \re{Geg} (see, e.g. \re{T}) one can obtain a nonvanishing result for \re{bi} at $d=3$.

Thus, the nonlocal operator \re{bi} is different from zero only for $d>2$. Substituting \re{eq:explicitP0} into \re{bi} we find that the resulting expression coincides with the nonlocal operator defined in \re{nonloc}
\begin{align}
\mathcal O^{(\phi)}_{\omega,\omega'}(n) = {1\over 4((d-1)(d-3))^2} {\cal O}^{\,\epsilon} _{\omega, \omega'}(n)\,,\qqqquad \epsilon=(-1)^{S+S'+1}\,.
\end{align}
We would like to emphasize that the commutation relations \re{alg2} hold for arbitrary $d\ge 2$. The dependence on $d$ enters \re{alg2} through the dependence of the structure constants on the conformal spin of fields, Eq.~\re{js}, and their Lorentz spins, $s_\phi=0$, $s_\psi=1/2$ and $s_F=1$.

For $d=2$ the leading twist operators that enter the definition of the flow operators \re{J-def} are known to possess an infinite-dimensional symmetry. Their commutation relations are described by (quantum) $W_\infty$ algebra which contains the Virasoro algebra and plays an important role in two-dimensional conformal field theory. Because the flow operators are given by integrated leading twist operators, we expect that for $d=2$ the commutation relations \re{alg2} (with $\delta^{(0)}(n,n')=1$) should be related to those of $W_\infty$ algebra. We establish this relation in the next section.

Because of the appearance of the nonlocal operator $\mathcal O^{(\phi)}_{\omega,\omega'}(n)$, a question arises whether the commutation relations \re{alg-phi} satisfy the Jacobi identity
\begin{align}\label{Jacobi1}
 [\mathcal J^{(\phi)}_{\omega_1,S_1}(n_1),[\mathcal J^{(\phi)}_{\omega_2,S_2}(n_2) ,\mathcal J^{(\phi)}_{\omega_3,S_3}(n_3) ]] + \text{cyclic} = 0\,.
\end{align}
To check this identity, we follow the same steps as in Section~\ref{sect:2}  and compute the connected part of the correlation function $\vev{\phi(x_1) \mathcal J^{(\phi)}_{\omega_1,S_1}(n_1)\mathcal J^{(\phi)}_{\omega_2,S_2}(n_2)\mathcal J^{(\phi)}_{\omega_3,S_3}(n_3) \bar\phi(x_2) }_c$. After an appropriate antisymmetrization this leads to the following result for a double commutator
\begin{align}\notag
& \langle \phi(x_1) [\mathcal J^{(\phi)}_{\omega_1,S_1}(n_1),[\mathcal J^{(\phi)}_{\omega_2,S_2}(n_2) ,\mathcal J^{(\phi)}_{\omega_3,S_3}(n_3) ]] \bar\phi(x_2) \rangle_c\sim 
\delta(n_1,n_2)\delta(n_2,n_3) 
\\[2mm]\notag
\times & \dashint_0^\infty d\alpha_1 \e^{-i\alpha_1 (n x_1) +i\alpha_2(n x_2)} \left[\Big( 
{P_{S_3}(\alpha_1,-\alpha_1
+\omega_3) P_{S_2}(\alpha_1-\omega_3,-\alpha_1+\omega_2+\omega_3)  \over \alpha_1-\omega_3} \right.
\\\notag
&\qquad -{P_{S_2}(\alpha_1,-\alpha_1+\omega_2) P_{S_3}(\alpha_1-\omega_2,-\alpha_1+\omega_2+\omega_3)   \over \alpha_1-\omega_2}   \Big) { P_{S_1}(\alpha_1-\omega_2-\omega_3,-\alpha_2) \over \alpha_1-\omega_2-\omega_3}
\\\notag
 & +\Big({  P_{S_2}(\alpha_1-\omega_1,-\alpha_1+\omega_1+\omega_2) P_{S_3}(\alpha_1-\omega_1-\omega_2,-\alpha_2) \over \alpha_1-\omega_1-\omega_2} 
\\ 
&\left. \qquad - {   P_{S_3}(\alpha_1-\omega_1,-\alpha_1+\omega_1+\omega_3) P_{S_2}(\alpha_1-\omega_1-\omega_3,-\alpha_2)  \over \alpha_1-\omega_1-\omega_3} \Big)
{P_{S_1}(\alpha_1,-\alpha_1+\omega_1) \over  \alpha_1-\omega_1}
\right],
\end{align} 
where $\alpha_2=\alpha_1-\omega_1-\omega_2-\omega_3$. Cyclically permuting the indices of the light-ray operators on both sides of this relation and adding up the resulting expressions, it is straightforward to verify that the identity \re{Jacobi1} is indeed verified.

\subsection{Structure constants}\label{sect:4}

According to \re{Q-sum} and \re{Q-phi-sum}, the structure constants of the algebra \re{alg2} are defined as coefficients of the expansion of \re{Q} and \re{Q-phi} over the polynomials $P_{\ell''}$ defined in \re{Geg}. 

We recall that the polynomials $P_\ell(p_1,p_2)$ 
are orthogonal to each other with respect to the inner product \re{ortho}. This allows us to find $C_{\ell\ell'}{}^{\ell''}(\omega,\omega')$ from  \re{Q-sum} and \re{Q-phi-sum} by projecting both sides of these relations onto $P_{\ell''}$. 
To this end, we replace $\alpha_1=\omega'' t$ in \re{Q-sum}, use the relation $P_\ell(\lambda p_1,\lambda p_2)=\lambda^\ell P_\ell(p_1,p_2)$ and integrate both sides of \re{Q-sum} with the measure \re{ortho} to find after some algebra
\begin{align}\notag\label{C-int}
C^{(s)}_{\ell \ell'}{}^{\ell''}& (\omega,\omega') = {2\over \mathcal N_{\ell''}}(\omega+\omega')^{S+S'-S''-1} 
\\[1.5mm]
& \times  \int_0^{1} dt\,(t (1-t))^{2j-1}   (t-\varepsilon)^{2s-1}\,P_{\ell}(t,\varepsilon-t)P_{\ell'}(t-\varepsilon ,1-t)P_{\ell''}(t,1-t) \,,
\end{align}
where the notation was introduced for $\varepsilon=\omega/(\omega+\omega')$ and the normalization factor $\mathcal N_{\ell''}$ was defined in \re{norm}. Here $S_i=\ell_i+2s$ 
and the superscript `$(s)$' refers to the spin of fields, $s_\psi=1/2$ and $s_F=1$. The polynomials $P_\ell$ are given by \re{Geg} with $j=s+(d-2)/4$.

Repeating the calculation for the scalar operators ($s=0$) we find from \re{Q-phi-sum}
\begin{align}\notag\label{C0-int}
C^{(0)}_{\ell \ell'}{}^{ \ell''}& (\omega,\omega') = {2\over \mathcal N_{\ell''}} (\omega+\omega')^{S+S'-S''-1} 
\\[1.5mm]
& \times  \int_0^{1} dt\,(t (1-t))^{d/2-2}  {P_{\ell}(t,\varepsilon-t)P_{\ell'}(t-\varepsilon ,1-t)-P_{\ell}(\varepsilon,0)P_{\ell'}(0,1-\varepsilon)\over t-\varepsilon}P_{\ell''}(t,1-t)\,,
\end{align}  
where $S=\ell$ and similar for $S'$ and $S''$. In this relation,  polynomials $P_\ell$ are given by \re{Geg} with $j=(d-2)/4$.

The relations \re{C-int} and \re{C0-int} hold for even  $S+S'-S''$. For odd $S+S'-S''$ the structure constants vanish.

\subsubsection*{Commutator of the energy flow operators and the Virasoro algebra}

Let us apply \re{alg2} to find the commutation relations of the energy flow operator. It is convenient to perform the conformal transformation  
\re{tr} to switch from \re{E-def} to the equivalent representation (see \re{J-z}) 
\begin{align}\label{calE}
\mathcal E_\omega(n) =\int_{-\infty}^\infty d z^- \e^{-i z^-\omega} T_{--}(0,z^-,\bit{z})\,.
\end{align}
We recall that the stress-energy tensor is given  by spin$-2$ operators \re{T}. As a consequence, the energy flow operator  
 takes the form
\begin{align}\label{E-J}\notag
 &\mathcal E^{(0)}_\omega(n) ={1\over 2(d-1)(d-3)} \mathcal J_{\omega,2}^{(0)}(n) \,,
\\\notag
 &\mathcal E^{(1/2)}_\omega(n) =  {1\over 2(d-1)} \mathcal J_{\omega,2}^{(1/2)}(n)  \,,
\\[2mm]
 &\mathcal E^{(1)}_\omega(n) =  \mathcal J_{\omega,2}^{(1)}(n) \,,
\end{align}
where $\mathcal J_{\omega,2}^{(s)}(n)$ is the spin$-2$ flow operator built out of fields of spin $s$. Here we inserted the subscript $(s)$ to indicate that these operators are defined in the free theory of scalars $(s=0)$, fermions $(s=1/2)$ and gauge fields $(s=1)$.

According to \re{alg2} and \re{alg-phi} the flow operators satisfy the relations
\begin{align}\label{EE} \notag
{[\mathcal J^{(s)}_{\omega,2} (n) , \mathcal J^{(s)}_{\omega',2} (n') ]} &=\delta^{(d-2)}(n,n')  \left[ 
C^{(s)}_{\ell\ell}{}^{\ell}(\omega,\omega') \mathcal J^{(s)}_{\omega'',2} (n) + \delta_{s,0} (\omega \omega')^2\mathcal O^{(\phi)}_{\omega,\omega'}(n) \right]
\\
& -  \omega^{3}\delta(\omega+\omega')\Omega_2^{(s)}\lr{n,n'}  
 \,,
\end{align}
where $C^{(s)}_{\ell\ell}{}^{\ell}(\omega,\omega')$ is given by \re{C-int} and \re{C0-int} for $S=S'=S''=2$ and $\ell=2-2s$. The second term inside the brackets only appears in the free scalar theory and it involves the nonlocal operator \re{bi}.
Being homogenous polynomials of degree $S+S'-S''-1$, the structure constants in this case look as
\begin{align}\label{C-lin}
C^{(s)}_{\ell\ell}{}^{\ell}(\omega,\omega')= (\omega'-\omega) c^{(s)}\,,
\end{align}
where relative sign inside the brackets is fixed by the first relation in \re{C-pro}. Applying \re{C-int} and \re{C0-int} we can find the normalization constant \begin{align}
c^{(0)}={2(d-1)(d-3)} \,,\qqquad c^{(1/2)}= 2(d-1)\,,\qqquad c^{(1)}=1\,.
\end{align}
Combining the above relations we arrive at the commutation relations of the flow operators in $d-$dimensions
\begin{align}\label{EE-com}\notag
[\mathcal E^{(s)}_\omega(n),\mathcal E^{(s)}_{\omega'}(n')] {} & =  \delta^{(d-2)}(n,n') \left[  (\omega'-\omega) \mathcal E^{(s)}_{\omega+\omega'}(n) + \delta_{s,0} (\omega   \omega')^{2} {\cal O}^-_{\omega, {\omega'}}(n)  \right]
\\[2mm]
 &- \omega^3 \delta(\omega+\omega')c^{(s)} (n,n')\,,
\end{align}
where the operator ${\cal O}^-_{\omega, {\omega'}}(n)$ is defined in \re{nonloc}. The central charge $c^{(s)} (n,n')$ is given by   $\Omega^{(s)}_{S=2}(n,n')$, Eq.~\re{Omega}, with the coefficients similar to those in \re{E-J}
\begin{align}\label{c-d} 
c^{(0)} = {\Omega^{(0)}_{2}(n,n')\over 4((d-1)(d-3))^2}   \,,\qqquad 
c^{(1/2)}  =   {\Omega^{(1/2)}_{2}(n,n')\over 4(d-1)^2}  \,,\qqquad
c^{(1)} = \Omega^{(1)}_{2}(n,n')\,.
\end{align}
For $d>2$ it vanishes for $n\neq n'$ and diverges otherwise as \re{Omegad}. For $d=2$ we have for free scalars and fermions, respectively, 
\begin{align}\label{c-charge}
c^{(0)} \big|_{d=2} = \frac16 \,, \qqqquad 
c^{(1/2)} \big|_{d=2} = \frac1{12} \,.
\end{align}

Let us examine the relation \re{EE-com} in $d=2$ dimension. In this case the stress-energy tensor  $T_{--}=T_{--}(x_1^-)$
satisfies the Virasoro commutation relations
\begin{align}\label{Vir}
[T_{--}(x_1^-),T_{--}(x_2^-)] = i \lr{\partial_{1-} -\partial_{2-}}\Big(T_{--}(x_1^-) \delta(x_1^--x_2^-)\Big) +{c\over 24\pi}  (i\partial_{1-})^3\delta(x_1^--x_2^-)\,,
\end{align}
where $\partial_{i-}=\partial/\partial {x_i^-}$ and $c$ is the central charge.
Using \re{calE} we find that the energy flow operator $\mathcal E_\omega$ in $d=2$ dimension satisfies the Virasoro algebra
\begin{align}\label{c}
[\mathcal E_\omega,\mathcal E_{\omega'}] = (\omega'-\omega) \mathcal E_{\omega+\omega'} - {c\over 12} \omega^3 \delta(\omega+\omega')\,.
\end{align}
We observe that the relation \re{EE-com} coincides with \re{c} (recall that the operator \re{nonloc} vanishes for $d=2$) and identify the central charge 
to be $c_\phi=2$ and $c_\psi=1$ for the free scalar and fermion, respectively.
These values of the central charge are twice larger than those of a real scalar and Majorana fermion in $d=2$. The reason for this is that we treated  from the start  the fields
$\Phi$ amd $\bar\Phi$ to be independent. In terms of the central charge this amounts to doubling the number of  fields. 
 
A distinguished feature of \re{EE} is that the expression on the right-hand side does not contain operators of lower spin. 
For the flow operators $\mathcal J^{(s)}_{\omega,S} (n)$ of spin $S>2$, the commutation relations \re{alg2} look as
\begin{align}\label{HS} 
{[\mathcal J^{(s)}_{\omega,S} (n) , \mathcal J^{(s)}_{\omega',S'} (n') ]} &=\delta^{(d-2)}(n,n')  
C^{(s)}_{\ell\ell'}{}^{\ell_{\rm max}}(\omega,\omega') \mathcal J^{(s)}_{\omega'',S+S'-2} (n)  +\dots\,,
\end{align}
where $\ell=S-2s$ and
$\ell_{\rm max}=S+S'-2-2s$. Here dots denote the contribution of operators with lower spin $S''=S+S'-2p$ with $p=2,3,\dots$. 
Similar to \re{C-lin},
the structure constants $C^{(s)}_{\ell\ell'}{}^{\ell_{\rm max}}(\omega,\omega')$ are linear functions of $\omega$ and $\omega'$ 
\begin{align}\label{HS-C}
C^{(s)}_{\ell\ell'}{}^{\ell_{\rm max}}(\omega,\omega') = ((S-1)\omega'-(S'-1)\omega)f^{(s)}_{S S'} \,.
\end{align}
 Using \re{C-int} and \re{C0-int} we can find the normalization constant
\begin{align}
f^{(s)}_{S S'}= \frac{2^{4(1- s)} \Gamma \left( S+\frac{d-3}{2} \right) \Gamma
   \left(S'+\frac{d-3}{2} \right) \Gamma (S+S'-2 s-1)}{\Gamma \left(2s+\frac{d-3}{2} \right) \Gamma (S+1-2
   s) \Gamma (S'+1-2 s) \Gamma \left(S+S'+\frac{d-7}{2} \right)}\,.
\end{align}
For $S=S'=2$ the relation \re{HS-C} coincides with \re{C-lin}.

\subsubsection*{$W-$algebra}

In $d=2$ dimensions, the spin$-2$ operators satisfy the Virasoro algebra \re{Vir}.
For the operators  $O_S$ with spin $S$ the commutation relations take a more complicated form of $W-$algebra~\cite{Pope:1989sr}
\begin{align}\notag\label{OO-W}
[O_{S_1}(x_1^-), O_{S_2}(x_2^-)] &=- \sum_{S_3=S_\text{min}}^{S_1+S_2-2}
 W_{S_1S_2}{}^{S_3}(i\partial_{1-},i\partial_{2-}) \Big(O_{S_3}(x_1^-) \delta(x_1^--x_2^-)\Big) 
\\
&+ \delta_{S_1,S_2} {\tilde c_{S_1}\over 24\pi}  (i\partial_{1-})^{2S_1+1}\delta(x_1^--x_2^-)\,,
\end{align}
where the sum runs over operators with spin $S_3$ of the same parity as
$S_1+S_2$. The minimal value of spin is $S_\text{min}=2$ for scalar operators and $S_\text{min}=1$ for fermionic operators. These two cases  correspond to 
$W_\infty$ and $W_{1+\infty}$ algebras, respectively. The latter algebra contains the additional operator of spin $1$. The derivatives on the right-hand side of \re{OO-W} are described by functions
 $W_{S_1S_2}{}^{S_3}(p_1,p_2)$ which are homogenous polynomials of degree $S_1+S_2-S_3-1$. As we show in Appendix~\ref{app:W} (see Eq.\,\re{C-sum-g}),  
the expansion coefficients of  $W_{S_1S_2}{}^{S_3}(p_1,p_2)$ are given by linear combinations of structure constants of the $W-$algebra. 
 The Virasoro algebra is a subalgebra of \re{OO-W}, it corresponds to $W_{22}{}^2(p_1,p_2)=p_2-p_1$. 
 
The $W-$algebra \re{OO-W} leads to the following commutation relations of the flow operators \re{J-z}
\begin{align} \label{J-W}
[\mathcal J_{\omega,S},\mathcal J_{\omega',S'}] &= \sum_{S''=S_\text{min}}^{S+S'-2}
 W_{SS'}{}^{S''}(\omega,\omega')  \mathcal J_{\omega+\omega',S''}
-\delta_{S,S'}  \delta(\omega+\omega') \omega^{2S+1}{\tilde c_S\over 12}  \,.
\end{align}
Comparing it with the analogous relation \re{alg2}  we observe that for $d=2$ the two relations take the same form.
Matching the expressions on the right-hand side of \re{J-W} and \re{alg2}, we establish the relation between the 
structure constants of the two algebras
\begin{align}\notag\label{W-C}
& C^{(0)}_{\ell_1\ell_2}{}^{\ell_3} (\omega,\omega')\Big|_{d=2}=W^{(0)}_{S_1S_2}{}^{S_3}(\omega,\omega') \,,&& 
\hspace*{-20mm} S_i=\ell_i\ge 2\,,
 \\[2mm]
&  C^{(1/2)}_{\ell_1\ell_2}{}^{\ell_3} (\omega,\omega')\Big|_{d=2}=W^{(1/2)}_{S_1S_2}{}^{S_3}(\omega,\omega') \,,&&\hspace*{-20mm}  S_i=\ell_i+1\ge 1\,,
\end{align}
where $W^{(0)}$ and $W^{(1/2)}$ are the structure constants of the $W_\infty$ and $W_{1+\infty}$ algebras, respectively.
In the similar manner, the relation between the central charges looks as 
\begin{align}
\tilde c^{(0)}_S =12\,\Omega_{S}^{(\phi)}\Big|_{d=2}\,, \qqqquad \tilde c^{(1/2)}_S = 12\,\Omega_{S}^{(\psi)}\Big|_{d=2}\,,
\end{align}
where $\Omega_{S}^{(\phi)}$ and $\Omega_{S}^{(\psi)}$ are defined in \re{Omega2}.

Substituting \re{C-int} and \re{C0-int} into \re{W-C} we obtain integral representation of the structure constants of $W-$algebras
\begin{align}\notag\label{W1-int}
& W^{(1/2)}_{{}\ SS'}{}^{S''}(\omega,\omega')=2(2S''-1) (\omega+\omega')^{S+S'-S''-1} 
\\[1.5mm]
&\qqqquad \times  \int_0^{1} dt\, P^{(1/2)}_{S-1}(t,\varepsilon-t)P^{(1/2)}_{S'-1}(t-\varepsilon ,1-t)P^{(1/2)}_{S''-1}(t,1-t) \,,
\\[2mm] \notag\label{W-int}
& W^{(0)}_{SS'}{}^{S''}(\omega,\omega')= - 
 {32(2S''-1) \over S(S-1) S'(S'-1)} (\omega+\omega')^{S+S'-S''-1} 
\\[1.5mm]
&\qqqquad \times  \int_0^{1} dt\,t(1-t) (t-\varepsilon) P^{(1)}_{S-2}(t,\varepsilon-t)P^{(1)}_{S'-2}(t-\varepsilon ,1-t)P^{(1)}_{S''-2}(t,1-t)\,,
\end{align}
where $\varepsilon=\omega/(\omega+\omega')$ and $P_\ell^{(j)}$ is given by \re{Geg} with the conformal spin $j$ replaced by its value \re{js}, $j_\psi=1/2$ and $j_\phi=0$ for $d=2$. In addition, in the last relation we applied the identity \re{P-sc} and replaced $P_S^{(0)}$ by its expression in terms of $P_{S-2}^{(1)}$. In agreement with \re{W-C} the minimal values of spins in \re{W1-int} and \re{W-int} are $S_{\rm min}=1$ and $S_{\rm min}=2$, respectively. 

The structure constants \re{W1-int} and \re{W-int} are homogenous polynomials in $\omega$ and $\omega'$ of degree $S+S'-S''-1$. Replacing the polynomials in \re{W1-int} and \re{W-int} by their explicit expressions \re{Geg} and carrying out integration, we can express $W^{(0)}_{SS'}{}^{S''}$ and $W^{(1/2)}_{SS'}{}^{S''}$ as finite nested sums. For our purposes however it is advantageous to use the integral representation \re{W1-int} and \re{W-int}. To best of our knowledge, the representation \re{W1-int} and \re{W-int} of the structure constants of the $W-$algebra
did not appear in the literature before.
 
We have shown in the previous subsection, that, aside from $\delta^{(d-2)}(n,n')$, the commutation relations \re{EE-com} and \re{c} of the energy flow operators look alike at $d=2$ and for $d>3$. The question arises whether similar relation holds between the algebras \re{alg2} of higher spin flow operators for different $d$. As we show in a moment, for $d=4$ the structure constants \re{C-int} and \re{C0-int} are related to the structure functions of the $W-$algebra \re{W1-int} and \re{W-int} in a nontrivial way.~\footnote{We would like to mention that this is not the first time that the $W$ algebra made its appearance in a context of $d=4$ dimensional CFT. It was noticed in Ref.~\cite{Huang:2021hye} that properly regularized integrated two-point function of the stress-energy tensors in $d=4$ is proportional to the central charge of the $W_3$ algebra.}  
 
We start with the structure constants \re{C-int} of the fermionic flow operators in $d=4$ dimension. Replacing $j_\psi=1$ and $s_\psi=1/2$ in \re{C-int}  we find
\begin{align}\notag\label{C-psi-4}
C^{(1/2)}_{\ell \ell'}{}^{\ell''}& (\omega,\omega') = {8(2S''+1) \over S''(S''+1)} (\omega+\omega')^{S+S'-S''-1} 
\\[1.5mm]
& \times  \int_0^{1} dt\, t (1-t) \,P^{(1)}_{S-1}(t,\varepsilon-t)P^{(1)}_{S'-1}(t-\varepsilon ,1-t)P^{(1)}_{S''-1}(t,1-t) \,,
\end{align}
where $S=\ell+1$ and similar for $S'$ and $S''$. The structure constants  \re{C0-int}  of the scalar  flow operators in $d=4$ dimension are given by (for $j_\phi=1/2$ and $s_\phi=0$)
\begin{align}\notag\label{C-phi-4}
C^{(0)}_{\ell \ell'}{}^{ \ell''}& (\omega,\omega') = {2(2S''+1)} (\omega+\omega')^{S+S'-S''-1} 
\\[1.5mm]
& \times  \int_0^{1} dt\, {P^{(1/2)}_{S}(t,\varepsilon-t)P^{(1/2)}_{S'}(t-\varepsilon ,1-t)-P^{(1/2)}_{S}(\varepsilon,0)P^{(1/2)}_{S'}(0,1-\varepsilon)\over t-\varepsilon}P_{S''}(t,1-t)\,,
\end{align}  
where $S_i=\ell_i$.

The above relations have a striking similarity to \re{W1-int} and \re{W-int}. Up to shift of spins, $S\to S-1$, the integrals in the these relations only differ by the factor of $(t-\varepsilon)$. Taking into account the properties of the Gegenbauer polynomials \re{Geg},
the integrals in \re{W1-int} and \re{W-int} can be expressed in terms of those in \re{C-psi-4} and \re{C-phi-4} with shifted indices. Namely, applying the identity \re{PP-id} we obtain 
\begin{align}\notag\label{rel1}
C^{(1/2)}_{\ell\ell'}{}^{\ell''}\Big|_{d=2}=
 {1\over 2 (1-\varepsilon )} & \bigg[
{1-2 \varepsilon\over 2 \ell+1}\lr{ \varepsilon^2 \omega''{}^2 \ell\e^{-\partial_\ell}+ (\ell+1) \e^{\partial_\ell}}
\\
&  
+ \varepsilon\lr{ {\ell''\over 2\ell''-1} \e^{-\partial_{\ell''}}+{\ell''+1\over 2\ell''+3}\omega''{}^2\e^{\partial_{\ell''}}}\bigg] C^{(0)}_{\ell\ell'}{}^{\ell''}\Big|_{d=4}\,,
\\[2mm]
\notag\label{rel2}
C^{(0)  \ \ell''+2}_{\ell+2,\ell'+2} \Big|_{d=2}=
 {1\over 2 (1-\varepsilon )} & \bigg[
 \frac{1-2 \varepsilon}{2 \ell+3}
   \lr{\varepsilon ^2\omega''{}^2 (\ell+2)
   \e^{-\partial_\ell} + (\ell+1)\e^{\partial_\ell} }
 \\
&  
+\varepsilon\lr{\frac{\ell'' }{2
   \ell''+1} \e^{-\partial_{\ell''}}+\frac{\ell''+3 
   }{2 \ell''+5}\omega''{}^2\e^{\partial_{\ell''}} }\bigg] C^{(1/2)}_{\ell\ell'}{}^{\ell''}\Big|_{d=4}\,,
\end{align}
where $\varepsilon=\omega/(\omega+\omega')$ and $\omega''=\omega+\omega'$. Here the operators $\e^{\pm\partial_\ell}$ shift of indices as $\ell\to\ell \pm 1$.   

The relations \re{rel1} and \re{rel2}  establish the correspondence between the structure constants in $d=2$ and $d=4$ dimensions. Similar (but algebraically more involved) relations can be derived for any even $d$ by applying \re{C-int} and expanding the polynomials \re{Geg} in terms of those in $d=2$ dimensions.
 
\subsubsection*{Conformal Ward identities}

Let us examine the commutation relation $ [\mathcal E_\omega(n), \mathcal J_{\omega',S'}(n')]$ containing the energy flow operator \re{calE}. 
Combining together \re{E-J} and \re{HS} we find 
\begin{align}\label{EJ-com}
[\mathcal E_\omega(n), \mathcal J_{\omega',S'}(n')] = -((S'-1)\omega-\omega') \delta^{(d-2)}(n,n')\mathcal J_{\omega+\omega',S'}(n') + \dots\,,
\end{align}
where dots denote the contribution of the flow operators $\mathcal J_{\omega+\omega',S''}(n')$ with spin $S''=S'-2p$ and $p\ge 1$. A close examination shows that it vanishes at small $\omega$ as $O(\omega^3)$ for $d=2$ and as $O(\omega^2)$ for $d>2$. This property is not obvious because the contribution of $\mathcal J_{\omega+\omega',S''}(n')$ to the commutator of generic flow operators is accompanied by the structure constants \re{C-gen}
which  do not vanish for $\omega\to 0$. The fact that the corrections to \re{EJ-com} vanish at small $\omega$ follows from conformal Ward identities. 

To show this we examine the integral of the energy flow operator over the sphere $\int d^{d-2} n \, \mathcal E_\omega(n) $. Its expansion at small $\omega$ yields global charges
\begin{align}\label{Ls}
\int d^{d-2} n \, \mathcal E_\omega(n) = Q^{(0)} - i \omega Q^{(1)} + {(i\omega)^2\over 2} Q^{(2)} + O(\omega^3)\,.
\end{align}
We find using \re{calE} that the expansion coefficients are given in the $z-$coordinates \re{tr} by
\begin{align}
Q^{(k)} =  \int d^{d-2} \bit z \int_{-\infty}^\infty dz^-  \,  (z^-)^{k}\,T_{--}(0,z^-,\bit z)\,,
\end{align}
with $k\ge 0$.
Integrating both sides of \re{EJ-com} over the sphere and taking into account \re{Ls} we find for arbitrary $d$
\begin{align}\notag\label{QQ}
& [Q^{(0)}, \mathcal J_{\omega',S'}(n')] = \omega' \mathcal J_{\omega',S'}(n')\,,
\\[2mm]
& [Q^{(1)}\,,  \mathcal J_{\omega',S'}(n')] = i \lr{\omega' \partial_{\omega'}-S'+1} \mathcal J_{\omega',S'}(n')\,.
\end{align}
In addition, for $d=2$ the vanishing of $O(\omega^2)$ corrections to \re{EJ-com} implies
\begin{align}\label{only}
 [Q^{(2)}\,,  \mathcal J_{\omega',S'}] =- \lr{\omega' \partial_{\omega'}^2-2(S'-1)\partial_{\omega'}}\mathcal J_{\omega',S'}\,.
\end{align}
For $d>2$ the right-hand side of this relation contains the additional terms involving the flow operators of smaller spin $\mathcal J_{\omega',S'-2p}(n')$ with $p\ge 1$.
 
The relations \re{QQ} follow from the observation that the operators $Q^{(0)}$ and $Q^{(1)}$ coincide with the total momentum 
and Lorentz boost operators, $Q^{(0)} = P_-$ and $Q^{(1)}=M_{+-}$. As a consequence, their commutation relations with the conformal operator $O_{S'}(0,z^-,\bit z)$ are
\begin{align}\notag\label{pro}
& i[Q^{(0)}, O_{S'}(0,z^-,\bit z) ] = \partial_{z^-} O_{S'}(0,z^-,\bit z)\,,
\\[2mm]
& i[Q^{(1)}, O_{S'}(0,z^-,\bit z) ] = (z^-\partial_{z^-} +S') O_{S'}(0,z^-,\bit z)\,.
\end{align}
Taking  the Fourier transform \re{J-z} on both sides we arrive at  \re{QQ}. 

To see why \re{only} only holds at $d=2$, we examine the generator of special conformal transformations
\begin{align}\notag
K^-  &=\int dz^- d^{d-2} \bit z \,(2 z^-z^\nu- z^2 g^{-\nu} ) T_{\nu -}(0,z^-,\bit z)
\\
&=\int dz^- d^{d-2} \bit z \, \left[ 2 (z^-)^2 T_{--} - 2 z^-z_\alpha T_{\alpha -} +z_\alpha ^2 T_{+-} \right] ,
\end{align}
where in the second relation we replaced $z^\nu=(z^+,z^-,\bit z)$ and  $\bit z= z_\alpha$ (with $\alpha=2,\dots,d-1$). For $d=2$ only the first term inside the brackets survives and $K^-/2$ coincides with the operator $Q^{(2)}$. As a consequence,  
\begin{align}\label{K}
i[Q^{(2)}, O_{S'}(0,z^- ) ] = ((z^-)^2\partial_{z^-} +2S'z^-) O_{S'}(0,z^- )\,,
\end{align}
leading to \re{only} after the Fourier transform \re{J-z}. For $d>2$ the operators $K^-/2$ and $Q^{(2)}$ differ by terms containing transverse components of
the stress-energy tensor which invalidate \re{K}. 

Because the relations \re{pro} follow from the Poincar\'e symmetry, they also hold in an interacting theory. The same is true for the relations \re{QQ}. Going back to \re{EJ-com} this suggests that the contribution of the flow operator $\mathcal J_{\omega+\omega',S'}(n')$ to \re{EJ-com} is protected from quantum corrections.

In a close analogy with \re{Ls}, we can define the global charges associated to spin $S$ flow operators
\begin{align}\label{Js}
\int d^{d-2} n \, \mathcal J_{\omega,S}(n) = Q^{(0)}_{S} - i \omega Q^{(1)}_{S} + {(i\omega)^2\over 2} Q^{(2)}_{S} + O(\omega^3)\,.
\end{align}
Then, it follows from \re{alg2} that  
\begin{align} \label{hs} 
{[ Q^{(0)}_{S}  , \mathcal J_{\omega',S'} (n') ]} &= \sum_{S''=2s}^{S+S'-2} 
(\omega')^{S+S'-S''-1}  f_{SS'}{}^{S''} \mathcal J_{\omega',S''} (n')   \,,
\end{align}
where the sum runs over spins $S''$ of the same parity as $S+S'$ and we took into account that $C_{\ell\ell'}{}^{\ell''}(0,\omega') = (\omega')^{S+S'-S''-1}  f_{SS'}{}^{S''}$. The explicit expressions for the normalization constants $f_{SS'}{}^{S''}$ can be found from \re{C-int} and \re{C0-int}. As compared to
the first relation in \re{QQ}, the right-hand side of \re{hs} contains the flow operators with lower spin. Applying \re{J-z} we can obtain from \re{hs} the action of the charges $Q^{(0)}_{S}$ on the leading twist operators $O_{S'}(0,z^-,\bit z)$. We verified that in $d=3$ dimensions, the resulting expressions for $[Q^{(0)}_{S}, O_{S'}(0,z^-,\bit z)]$ coincide with the analogous expressions derived in Refs.~\cite{Maldacena:2011jn,Maldacena:2012sf}.
 
\section{Commutator of the energy flow operators in interacting CFTs}
\label{sec:CFTlightrayalgebra}

In the previous sections we analyzed the algebra of the light-ray operators $\mathcal J_{\omega,S}$ built out of conserved currents in the free theory (and bilinear scalar operator of dimension $\Delta = d-2$ and spin $S=0$). In a generic interacting CFT the only conserved current is the stress-energy tensor and, possibly, the spin one conserved current. The remaining operators develop anomalous dimensions. In this section we address the question: which operators contribute to the commutator of the energy flow operators $[ {\cal E}_{\omega_1} (n_1) , {\cal E}_{\omega_2} (n_2)]$ in an interacting CFT?

To answer this question, we consider the three-point function
\be
 \label{eq:threepointalgebraEEC}
 \langle  0| [ {\cal E}_{\omega_1} (n_1) , {\cal E}_{\omega_2} (n_2) ] O_{\Delta, {\bf S}}(x) |0\rangle ,
 \ee
where $O_{\Delta, {\bf S}}(x)$ is a conformal primary operator of scaling dimension $\Delta$ belonging to an irreducible representation ${\bf S}$ of the Lorentz group in $d$ dimensions. Because the energy flow operator ${\cal E}_{\omega} (n)$ is given by the integrated stress-energy tensor, the correlation function \re{eq:threepointalgebraEEC} is different from zero provided that $O_{\Delta, {\bf S}}(x)$ can appear in the OPE of two stress-energy tensors.

To simplify the computation we choose $\omega_1<0$, $\omega_2>0$ and $\omega_1 + \omega_2 <0$. In this case,
we use the fact that  $\langle 0 | {\cal E}_{\omega_2} (n_2)  =0 $ for $\omega_2>0$, see Ref.~\cite{Korchemsky:2021okt}, to get
 \be
 \label{eq:actualcompute}
 \langle 0| [ {\cal E}_{\omega_1} (n_1) , {\cal E}_{\omega_2} (n_2) ]  O_{\Delta, {\bf S}} (x)|0\rangle = \langle 0| {\cal E}_{\omega_1} (n_1)  {\cal E}_{\omega_2} (n_2)  O_{\Delta, {\bf S}}(x)|0 \rangle .
 \ee 
Replacing the energy flow operators with \re{calE}, we can express the right-hand side of \re{eq:actualcompute} as  
the light-ray transform of the underlying three-point function
\begin{align}
\label{eq:three-pointfunctionC}
\lim_{\epsilon \to 0 }\int_{-\infty}^\infty d z_1^-d z_2^- \e^{-i z_1^-\omega_1 -i z_2^-\omega_2}  
\langle 0| T_{--}(\epsilon,z_1^-,\bit{z}_1)T_{--}(0,z_2^-,\bit{z}_2) O_{\Delta, {\bf S}}(x_3) |0\rangle\,,
\end{align}
where we wrote the light-ray operators in the null plane representation, see \eqref{J-z}.
The reason for introducing $\epsilon\to 0$ limit will be clear in a moment. 
The correlation function in \re{eq:three-pointfunctionC} is fixed by conformal symmetry up to a few constants.

To correctly compute the commutator \re{eq:actualcompute}, it is important to regularize the product of  the two stress-energy tensors  in  \eqref{eq:three-pointfunctionC} by separating them in the `$+$' direction by a non-zero $\eps$. Doing  the light transform and taking the limit $\epsilon \to 0$ we can identify   the distributional terms localized at $\bit{z}_1=\bit{z}_2$. 
Introducing non-zero $\eps$ introduces the regions in the light transform where the detectors are null (and timelike) separated. In this region the two detectors can communicate with each other, and this cross-talk produces a nontrivial commutator in the $\eps \to 0$ limit. By setting $\eps=0$ before doing the light transform this communication region is pushed to infinity (in variable $z_1^- - z_2^-$) and is not properly resolved. 

At the technical level, this can we understood as follows. The choice of the signs of $\omega_i$ dictates the direction in which the integration contour in \re{eq:three-pointfunctionC} can be deformed. For $\omega_1<0$ we can deform the integration contour for $z_1^-$ to the upper half-plane, for $\omega_2<0$ we can deform the integration contour for $z_2^-$ to the lower half-plane. The ordering of operators in \re{eq:three-pointfunctionC} dictates that the correlation function depends on $z_1^- - z_2^- - i 0$, $z_1^- - z_3^- - i 0$, and $z_2^- - z_3^- - i 0$. In particular, as we evaluate the integral over $z_2^-$ by deforming the contour into the lower half-plane only the singularity coming from $x_{12}^2 = 2 \epsilon (z_1^- - z_2^- - i 0) - (\bit{z}_1-\bit{z}_2)^2=0$ produces a nontrivial contribution. To correctly capture it is necessary to keep $\epsilon \neq 0$ when doing the light transform. 

We first consider the case when $O_{\Delta, {\bf S}}(x)$ is a traceless symmetric tensor in Lorentz indices of even spin $S$. We then show that the operators with twist $\tau =\Delta-S > d-2$ do not contribute to \re{eq:actualcompute}. This implies that the spinning operators (with $S>0$) which are not conserved currents cannot appear in the commutator of the energy flow operators $[ {\cal E}_{\omega_1} (n_1) , {\cal E}_{\omega_2} (n_2) ] $. On the other hand, conformal operators with zero spin $S=0$ and dimension $\Delta < d-2$ produce an infinite contribution.

At $d=3$ the three-point functions of conserved currents can contain additional conformal parity-odd structures involving the Levi-Civita tensor \cite{Giombi:2011rz,Costa:2011mg,Kravchuk:2016qvl}. This leads to the possibility that operators with odd spin  $S=5,7,9,\dots$ can appear in the OPE of the stress-energy tensors. We find that such operators do not contribute to the light-ray algebra.

For $d>3$ more general representations ${\bf S}$ of the Lorentz group appear in the OPE of two stress-energy tensors \cite{Costa:2016hju}. For simplicity we only consider $d=4$. In this case, the relevant representations are characterized by a Young tableaux with two rows of length $J$ and $j$ (with $J\ge j$). Following \cite{Chang:2020qpj}, we shall refer to $J$ and $j$ as    {\it spin}  and {\it transverse spin}, respectively. In the OPE of two stress-energy tensors only representations with $j \leq 4$ can appear. We show that the representations with $j\neq 0$ do not contribute to the light-ray algebra. We expect that the same should be true for $d>4$.
 
\subsection{Symmetric traceless tensors in $d$ dimensions}
\label{sec:symmtraceless}

A general expression for the three-point function $\langle T_{\mu \nu}(x_1)  T_{\rho \sigma}(x_2) O_{\mu_1 ... \mu_S}(x_3) \rangle$ can be found in 
Appendix A  of Ref.~\cite{Costa:2011mg},  we do not present it here.\footnote{This does not include parity odd structures in $d=3$ and $d=4$ which we consider separately in the following subsections.}

Let us start by considering the simplest case of the scalar primary operator. The relevant three-point function takes the form  
\be\nonumber
\label{eq:TTscalar}
&\langle T_{--}(x_1)  T_{--}(x_2) O_{\Delta}(x_3) \rangle = {\lambda_{TTO} \over  (x_{12}^2)^{d+2-{\Delta \over 2}} (x_{13}^2 x_{23}^2)^{{\Delta \over 2}}} \Big[ \Delta(\Delta+2)(d-2) V_1^2 V_2^2  \\
& - 2 (2+\Delta+d^2 - d (\Delta+1)) V_1 V_2 H_{12} + ((d-1) \Delta ^2-2 (d-1) d \Delta +(d-2) d (d+1)) H_{12}^2 \Big],  
\ee
where we follow the notations of Ref.~\cite{Costa:2011mg}. 
We next would like to compute \eqref{eq:three-pointfunctionC}. As a first step, we use differentiation by parts to rewrite the integral in \eqref{eq:three-pointfunctionC} as follows
\be
\label{eq:integralBasic}
{\cal {\hat D} } (\partial_{\om_1},\partial_{\om_2}) \int d z_1^- d z_2^-  {e^{- i \om_1 z_1^- - i \om_2 z_2^-} \over (x_{12}^2)^{d+2 - {\Delta \over 2} } (x_{13}^2 x_{23}^2)^{{\Delta \over 2} + 2}} \,,
\ee
where $x_1=(\epsilon,z_1^-,\bit{z}_1)$ and $x_2=(0,z_2^-,\bit{z}_2)$ in the light-cone coordinates and
the explicit form of the differential operator  ${\cal {\hat D} }$ can be found starting from \eqref{eq:TTscalar}.

We compute the light-ray transform \eqref{eq:integralBasic} in Appendix \ref{app:scalarintegral}. We then act with the differential operator and take the limit $\eps \to 0$.
The result is that the leading contribution takes the form
\be
\label{eq:scalaralgebra}
[{\cal E}_{\omega_1}(n_1) , {\cal E}_{\omega_2}(n_2)] &\sim   \epsilon^{{\Delta-(d-2)\over 2}} \delta^{(d-2)}(n_1,n_2) f_{d,\Delta}(\omega_1, \omega_2) {\cal J}_{\omega_1 + \omega_2,S=0}(n_1) ,
\ee
where $\epsilon \to 0$ and $f_{d,\Delta}(\omega_1, \omega_2)$ is a known function that can be written as a sum of ${}_2 F_1$ hypergeometric functions.
We see that scalar operators with $\Delta < d-2$ produce divergent contribution to the light-ray algebra, whereas scalars with $\Delta > d-2$ do not contribute. 
For scalar operators of dimension $\Delta = d-2$ we get  for $\omega_1 \omega_2 < 0$
\begin{align}
 f_{d,d-2}(\omega_1, \omega_2) &\sim 
\omega _1^2 \omega _2 \, _2F_1\left({1,\frac{d}{2}-1 \atop d-2} \Big|1+\frac{\omega _1}{\omega _2}\right).
\end{align}
One can check that $ f_{d,d-2}(\omega_1, \omega_2)  =  -f_{d,d-2}(\omega_2, \omega_1)$. 

Let us comment on the $ \epsilon^{{\Delta-(d-2)\over 2}}$ scaling in \eqref{eq:scalaralgebra}. It can be immediately understood  by examining
behaviour of the correlator \eqref{eq:TTscalar} in the limit $x_{12}^2=O(\epsilon)$ and $\epsilon=x_{12}^+\to 0$ before doing the light transform.
In this limit, the conformal-invariant tensor structures in \re{eq:TTscalar} scale as
$V_1, V_2 \sim O(\eps,x_{12}^2)$ and $H_{12} \sim O(\eps^2)$ leading to
\begin{align}
\label{eq:simpleasympt}
\langle T_{--}(x_1)  T_{--}(x_2) O_{\Delta}(x_3) \rangle \sim \sum_{i=0}^{4} c_i ^{(\Delta)}{\eps^i \over (x_{12}^2)^{d-2+i-{\Delta \over 2}}} .
\end{align}
where the coefficient functions $c_i ^{(\Delta)}$ depend on $x_{12}$ and $x_{23}$.
To probe the distributional term in \eqref{eq:scalaralgebra} we replace
$x_1=(\epsilon,z_1^-,\bit{z}_1)$ and $x_2=(0,z_2^-,\bit{z}_2)$ and
integrate the correlator  around the point $\bit{z}_1 = \bit{z}_2$. The relevant integral takes the schematic form 
\be
\label{eq:algebraintegral}
&   \int   {\epsilon^i \, d^{d-2} \bit x  \over (\bit x^2 - 2 \epsilon (z_1^- - z_2^- - i 0) )^{d-2+i-{\Delta \over 2}}}  \nn \\
&=  \epsilon^{{\Delta-(d-2)\over 2}} \left(    (z_1^- - z_2^- - i 0)^{{\Delta-(d-2) -2i \over 2}} {\Gamma({d \over 2}-1)\Gamma({d \over 2}+i-{\Delta \over 2} -1) \over \Gamma(d+i-{\Delta \over 2}-2)} + O(\epsilon) \right) ,
\ee
where $\bit x \equiv \bit{z}_1 - \bit{z}_2$. Note that the dominant contribution to the integral comes from $\bit x^2 \sim \eps$. 

This is how we proceed below. We first analyze the behavior of the three-point function to understand the expected scaling with $\eps$. In cases when we get a nontrivial contribution to the light-ray algebra, we do the light-ray transform to compute it.

Let us consider next spinning operators $O_{\Delta, S}(x,w) \equiv O_{\Delta}^{\mu_1 ... \mu_S}(x) w_{\mu_1} ... w_{\mu_S}$, where $S$ is an even positive integer. Unitarity bounds states that $\tau =\Delta-S \geq d-2$, with conserved currents saturating the bound $\tau = d-2$. 

In the same fashion we find that the three-point function scales for $x_{12}^2\to 0$ and $x_{12}^+=\epsilon\to 0$ as
\be
\label{eq:simplestr}
\langle T_{--}(x_1) T_{--}(x_2)  O_{\Delta, S}(x_3,w_3) \rangle =\sum_{i=0}^{4} c_i^{(\Delta, S)} {\eps^i \over (x_{12}^2)^{d-2+i-{\tau \over 2}}} + \dots ,
\ee
where dots denote less singular structures which will not play any role in our analysis. Here the coefficient functions $c_i^{(\Delta, S)}$ depend on $x_{13}$, $x_{23}$  and $w_3$. 
They can be straightforwardly computed using expression for the three-point function from Appendix A in \cite{Costa:2011mg}.
\footnote{ In particular, the relevant conformal tensor structures scale as
$V_1, V_2 = O(\eps,x_{12}^2)$, $V_3 = O(1/x_{12}^{2})$, $H_{12} = O(\eps^2)$ 
with $H_{13},H_{23}=O(1)$. 
}
From \eqref{eq:simplestr}, by integrating the correlator over $\bit{z}_1 - \bit{z}_2$,
we get
\be
\label{eq:symmtraceless}
[{\cal E}_{\omega_1}(n_1) , {\cal E}_{\omega_2}(n_2)] &\sim \epsilon^{{\tau-(d-2)\over 2}} \delta^{(d-2)}(n_1,n_2) {\cal J}_{\omega_1 + \omega_2,S}(n_1) ,
\ee
for $\epsilon \to 0$.
We immediately conclude from this relation that only conserved currents with $\tau = d-2$ contribute to the algebra! 

In the previous section we studied the case of free theories which contain infinitely many conserved currents that furnish a nontrivial light-ray algebra. In a generic interacting CFT the only conserved currents are spin-one and the stress energy tensor. Focusing on the commutator \eqref{eq:actualcompute} only the stress-energy tensor contributes to the algebra. To understand its contribution, recall that the three-point function in a generic interacting CFT takes the following form\footnote{In $d=3$ the vector structure is absent, and there is an additional parity odd structure. The latter however does not contribute to the algebra.}
\be
\label{eq:stresstensorthree}
\langle TTT\rangle_\text{CFT} = n_{\phi} \langle TTT\rangle_{\phi}+ n_{\psi} \langle TTT\rangle_{\psi}+ n_F \langle TTT\rangle_{F} ,
\ee
where $n_\phi, n_\psi, n_F$ are parameters of the theory and $ \langle TTT\rangle_{i=\phi,\psi,F}$ are conformal tensor structures computed in three free theories describing the complex scalar $(\phi)$,  the Dirac fermion $(\psi)$ and gauge field $(F)$, see Appendix B in Ref. \cite{Buchel:2009sk}.\footnote{This theory only exists in even $d$, but the resulting three-point function can be analytically continued to arbitrary $d$.} Their explicit form can be found for example in Appendix C of \cite{Zhiboedov:2012bm}. Conformal collider bounds \cite{Hofman:2008ar} together with the arguments \cite{Zhiboedov:2013opa,Meltzer:2017rtf} imply that in interacting CFTs
\be
n_\phi\,, n_\psi\,, n_F > 0 .
\ee
Using the conformal Ward identites we get from \eqref{eq:stresstensorthree} that the two-point function of the stress-energy tensors $\vev{TT}$ is proportional to a linear combination
\be
C_T = n_{\phi} C_T^{(\phi)} + n_{\psi}  C_T^{(\psi)} + n_F C_T^{(F)} \,,
\ee
where $C_T^{(i)}$ comes from the two-point functions $\vev{TT}_i$ in the corresponding free theories. For example, for the free complex scalar we have \cite{Osborn:1993cr}
\be
C_T^{(\phi)} = \frac{d\, \Gamma^2 \left(\frac{d}{2}\right)}{2 (d-1)\pi^d} \,. 
\ee

To compute the contribution of \eqref{eq:stresstensorthree} to the commutator $[{\cal E}_{\omega_1}(n_1) , {\cal E}_{\omega_2}(n_2)]$ we can use the results of the previous section for free theories. In this way we get
\be \label{eq:nphidefined}
[{\cal E}_{\omega_1}(n_1) , {\cal E}_{\omega_2}(n_2)] &=\delta^{(d-2)}(n_1,n_2) \Big(\omega_2 - \omega_1 + \tilde n_\phi\, \om_1^2 \om_2^2 f_\phi (\omega_1, \omega_2) \Big) {\cal E}_{\omega_1 + {\omega_2}}(n_1) +\dots\,,  
\ee
where dots denote the contribution of the remaining operators. 
Here $\tilde n_\phi\, \om_1^2 \om_2^2 f_\phi(\omega_1, \omega_2)$ is an extra contribution coming from $\langle TTT\rangle_{\phi}$ to the light-ray algebra. We compute it in Appendix \ref{sec:extralightray} with the result
\be\label{n-phi}
 \tilde n_\phi &= n_{\phi} {C_T^{(\phi)} \over C_T} \,,\qqqquad
 f_\phi (\omega_1,\omega_2) =- \omega_2^{-3}  {\,
   _2F_1\left(3,\frac{d+2}{2}, d+2, 1 + \frac{\omega _1}{\omega _2}\right)} \,.
\ee

\subsection{Parity odd structures in $d=3$}

In $d=3$ dimensions only symmetric traceless representations appear in the OPE of two stress-energy tensors. A special feature of $d=3$ however is that there are new parity odd conformal structures that can appear in the three-point functions. 
These go beyond the analysis done in the section above and therefore require a separate consideration.

As a simple example let us consider a parity odd structure in the three-point function involving the scalar primary operator
\be
\label{eq:scalarodd3}
\langle  T_{--}(x_1)  T_{--}(x_2) O_{\Delta}(x_3) \rangle_\text{odd} \sim {S_3 \Big( V_1 V_2 (\Delta+1) + H_{12} (\Delta -3) \Big) \over (x_{12}^2)^{{6-\Delta \over 2}} (x_{13}^2 x_{23}^2)^{{\Delta \over 2}} } ,
\ee
where the definition of $S_3$ can be found in Ref.~\cite{Giombi:2011rz}.
In the coincident detector limit, $x_{12}^2\to 0$ and $x_{12}^+=\epsilon\to 0$, the new parity odd tensor structure behaves as
\be
S_3 &\sim O\Big({\eps^2 \over |x_{12}|^3} ,  { \eps x \over |x_{12}|^3}  \Big) \ ,
\ee
where $x$ is the separation between detectors in the transverse direction $x \equiv \bit{z}_1 - \bit{z}_2$. Odd powers of $x$ do not contribute to the integral \eqref{eq:algebraintegral} and therefore do not contribute to the algebra.

Repeating the same analysis as above we get that the most singular structure generated by the three-point structure \eqref{eq:scalarodd3} takes the form
\be
\int  {\epsilon^4\, dx \over ( x^2 + \epsilon z_{12}^-)^{5-{\Delta+1\over 2}}} \sim \epsilon^{{\Delta \over 2}} .
\ee
As a result, in a unitary CFT with $\Delta > {1 \over 2}$ the odd structure \eqref{eq:scalarodd3} does not contribute to the algebra.

For $S=1$ the three-point function of interest does not exist. For $S=2$ there exists a parity-odd correction to the three-point function of the stress-energy tensors
\be
\label{eq:stressodd3}
\langle  T_{--}(x_1)  T_{--}(x_2) T(x_3,w_3) \rangle_\text{odd} \sim {S_1 \Big( 2 V_1^2 V_2 V_3 + H_{12} H_{13}  \Big) \over |x_{12}|^5  |x_{13}|^5 |x_{23}| } + {\rm cyclic} .
\ee
We found that this structure does not contribute to the light-ray algebra. 

For high spin operators   the  situation looks as follows. There exists a unique parity odd structure $\langle  T_{--}(x_1)  T_{--}(x_2)  O_{\Delta,S}(x_3, w_3) \rangle_\text{odd}$ for even spins $S \geq 2$ and for odd spins $S \geq 5$. We checked explicitly that it does not contribute to the algebra for a few lowest spins.  We expect the same pattern to continue for arbitrary $S$.

\subsection{Mixed tensors and parity odd structures in $d=4$}

In $d=4$ dimensions the OPE of two stress-energy tensors receives contribution from symmetric traceless even spin operators discussed in Section \ref{sec:symmtraceless}, as well as from two additional classes of operators. 

First, there are traceless symmetric odd spin operators which produce  parity odd structures in the correlation functions. Second, we have to include into consideration  more complicated representations of the Lorentz group, the so-called mixed-symmetry tensors \cite{Costa:2016hju}. In spinor notations they are described by tensors $O_{(\alpha_1 ... \alpha_k) (\dot \alpha_1 ... \dot \alpha_{\bar k})}$ which are completely symmetric in chiral and antichiral indices. They are conveniently characterised by the spin $J =(k + \bar k)/2$ and  transverse spin $j = |k - \bar k|/2$ which define the length of two rows in the corresponding Young tableaux \cite{Chang:2020qpj}.~\footnote{For $d>4$ more complicated representations are possible but we do not consider them here.
} The OPE of two stress-energy tensors involves operators $O_{\Delta,J,j}$ with the possible values of transverse spin $0\le j \leq 4 $. It is convenient to use index free notations for the operator and project its spinor indices onto auxiliary spinors $s^\alpha$ and $\bar s^{\dot\alpha}$
\begin{align}\label{O-dots}
O_{\Delta,J,j}(x,w) = O_{(\alpha_1 ... \alpha_k) (\dot \alpha_1 ... \dot \alpha_{\bar k})}(x)  s^{\alpha_1} \dots s^{\alpha_k} \bar s^{\dot \alpha_1} \dots \bar s^{\dot \alpha_{\bar k}}\,,
\end{align}
where $w^{\alpha\dot\alpha} = s^\alpha \bar s^{\dot\alpha}=w^\mu (\sigma_\mu)^{\alpha\dot\alpha}$ is a light-like vector in the spinor notations.

The general form of the three-point function involving \re{O-dots} and the stress-energy tensor was found  in \cite{Elkhidir:2014woa,Cuomo:2017wme}.  It takes the following form (for simplicity we assume that $k>\bar k$)
\be\label{sum-ki}
&\langle  T_{--}(x_1)  T_{--}(x_2) {O}_{\Delta,J,j}(x_3,w_3) \rangle = {1 \over (x_{12}^2)^{6-{\Delta+J \over 2}} (x_{13}^2 x_{23}^2)^{{\Delta + J \over 2}}} \sum_{\{ k_i \}} \lambda_{\{k_i \} } \left( \hat{\mathbb I}^{12} \right)^{k_1} \left( \hat{\mathbb I}^{21} \right)^{k_2} \left( \hat{\mathbb I}^{13} \right)^{k_3}  \  \nn 
\\
& \times\left( \hat{\mathbb I}^{31} \right)^{k_4} \left( \hat{\mathbb I}^{23} \right)^{k_5} \left( \hat{\mathbb I}^{32} \right)^{k_6} \left( \hat{\mathbb J}^{1}_{23} \right)^{k_7} \left( \hat{\mathbb J}^{2}_{13} \right)^{k_8}  \left(\hat{\mathbb J}^{3}_{12}  \right)^{k_9} \left(\hat{\mathbb K}_1^{23} \right)^{k_{10}} \left(\hat{\mathbb K}_2^{13} \right)^{k_{11}}\left( \hat{\mathbb K}_3^{12} \right)^{k_{12}},
\ee
where non-negative integers $k_i$ satisfy the relations
\be
k_1+ k_3 + k_7  &= 2 \,,  && k_2+ k_4 + k_7+ k_{11} + k_{12}  = 2 \,, \nn \\
k_2 + k_5+k_8 &=2\,, && k_1 + k_6+k_8 + k_{10}+ k_{12} =2\,, \nn \\
k_4+k_6+k_9 &= J-j\,, &&  k_3+k_5+k_9+k_{10}+k_{11} = J+j \,.
\ee
The definitions of various conformal tensor structures entering \re{sum-ki} can be found in Appendix D of \cite{Cuomo:2017wme}. In the spinor notations they look as
\be\label{spinor}
 \hat{\mathbb I}^{ij} &= [\bar s_i| x_{ij} |s_j\rangle , \nn \\[2mm]
\hat{\mathbb J}_{ij}^k &= {1\over x_{ij}^2}  [\bar s_k|  x_{ki}x_{ij}x_{jk} | s_k \rangle, \\
\hat{\mathbb K}_k^{ij} &= {|x_{ij}| \over |x_{ik}||x_{jk}|}\vev{s_i | x_{ik}x_{kj}| s_j}\,, \nn
\ee
where the auxiliary spinors satisfy $w_3^{\alpha\dot\alpha} = s_3^\alpha \bar s_3^{\dot\alpha}$ and $\bar n^{\alpha\dot\alpha} = s_i^\alpha \bar s_i^{\dot\alpha}$ (with $i=1,2$) for the particular choice of polarizations of operators in \re{sum-ki}.
The relation between the tensor structures \re{spinor} and analogous tensor structures used in section \ref{sec:symmtraceless} can be found in Appendix A of \cite{Karateev:2019pvw}.

In addition we need to impose permutation symmetry of \re{sum-ki} with respect to points $1$ and $2$, as well as the conservation condition for the stress-energy tensor. This leads to linear relations between $ \lambda_{\{k_i \} }$. These steps are implemented using the Mathematica package ${\rm CFTs4D}$ \cite{Cuomo:2017wme}. 

We then consider the limit $x_{12}^2\to 0$ and $x_{12}^+ \sim \eps \to 0$ relevant for the light-ray operator algebra. In this limit we have from \re{spinor}
\be
\label{eq:structureestimate4d}
& \hat{\mathbb I}^{12},  \hat{\mathbb I}^{21}, \hat{\mathbb J}^{2}_{13}, \hat{\mathbb J}^{1}_{23} \sim O(x_{12}^+) ,
&& \hat{\mathbb J}^{3}_{12}  \sim O(1/x_{12}^{2}), 
 \nn \\[2mm]
& \hat{\mathbb K}_1^{23},  \hat{\mathbb K}_2^{13} \sim O\Big( {x_{12}^+ \over |x_{12}|}  , {x_{12}^1 - i x_{12}^2 \over |x_{12}|} \Big)  \ , 
&& 
\hat{\mathbb K}_3^{12}\sim O\Big( x_{12}^+ |x_{12}| , (x_{12}^1 - i x_{12}^2) |x_{12}| \Big) , 
\ee
with the remaining tensor structures $\hat{\mathbb I}^{13},\hat{\mathbb I}^{23},\hat{\mathbb I}^{31},\hat{\mathbb I}^{32}=O(1)$. Notice that
 the $\hat{\mathbb K}-$structures depend on holomorphic component of the relative transverse coordinate,
 $[\bar s_3| x_{12} \ket{s_1}  = x_{12}^1 - i x_{12}^2$. To produce a nonvanishing contribution to the integral over the transverse coordinate $\bit x_{12}=(x_{12}^1,x_{12}^2)$ (see  \eqref{eq:algebraintegral}), they should be accompanied with the analogous structures depending on the holomorphic component, e.g.   $\hat{\mathbb I}^{13}-\hat{\mathbb I}^{23} \sim [\bar s_1|x_{12} \ket{s_3} = x_{12}^1 + i x_{12}^2$.
 
Using these formulas and going through the same steps as in the previous section,  we found that the contribution of the operators \re{O-dots} with   transverse spin $j$ to the light-ray algebra takes the following schematic form (for $\epsilon=x_{12}^+\to 0$)
\be
\label{eq:transversespin}
[{\cal E}_{\omega_1}(n_1) , {\cal E}_{\omega_2}(n_2)] &\sim  \epsilon^{{\tau + j - 2 \over 2}} \delta^{(2)}(n_1,n_2) 
 f_{\Delta,J,j}(\omega_1, \omega_2)
{\cal J}_{\omega_1 + \omega_2,J,j}(n_1)\,,
\ee
where 
 the light-ray operator ${\cal J}_{\omega_1 + \omega_2,J,j}(n_1)$ is defined similar to \re{J-def}. The explicit expression for the coefficient function $ f_{\Delta,J,j}(\omega_1, \omega_2)$ is not important for our purpose. The formula \re{eq:transversespin} generalizes \eqref{eq:symmtraceless} to the case $j \neq 0$ in $d=4$.

The expression on the right-hand side of \re{eq:transversespin} depends on  the twist $\tau=\Delta-J$ of the operator \re{O-dots}.
Unitarity bounds for such operators take the form \cite{Cordova:2017dhq}
\be
\label{eq:unitarityCordovaDiab}
\tau \geq {\rm max} \{ 2,j \} \, .
\ee
Combining this relation with \re{eq:transversespin} we observe that operators with $j>0$ do not contribute to  the light-ray algebra. 

It would be interesting to generalize the analysis of the present section to $d>4$. In this case more representations of the Lorentz group appear in the OPE of two stress-energy tensors, see \cite{Costa:2016hju} for  details. 

\subsection{Commutation relations and the Jacobi identities}  
\label{sec:jacobiInt}

Let us summarize the results of this section so far. We have observed that the commutator of the energy flow operators in a generic, interacting CFT is given by
\be
\label{eq:algebralight}
[{\cal E}_{\omega_1}(n_1) , {\cal E}_{\omega_2}(n_2)]  = \delta^{(d-2)}(n_1,n_2) \Big[  (\omega_2 - \omega_1 + \tilde n_{\phi}  f_\phi (\omega_1, \omega_2) ) {\cal E}_{\omega_1 + \omega_2}(n_1) - C \omega_1^3 \delta(\omega_1+\omega_2) + \dots  \Big] ,
\ee
where $ \tilde n_{\phi} f_\phi (\omega_1, \omega_2)$ arises from the free scalar structure in the expression  \re{eq:stresstensorthree} for the three-point function of the stress-energy tensors.~\footnote{Recall that in an interacting CFT we expect such structure to be always present in the three-point function of stress-energy tensors \cite{Zhiboedov:2013opa,Meltzer:2017rtf}.} The dots denote the contribution from light scalar primary operators of dimension ${d-2 \over 2} < \Delta \leq d-2$.
The operators with twist $\tau > d-2$ produce vanishing contribution to \re{eq:algebralight}.

Given the expression \eqref{eq:algebralight} it is tempting to ask if the commutator satisfies the Jacobi identity
\be
\label{eq:Jacobi}
[[{\cal E}_{\omega_1}(n_1) , {\cal E}_{\omega_2}(n_2)] , {\cal E}_{\omega_3}(n_3) ] + {\rm cyclic} \stackrel{?}{=} 0.
\ee
As opposed to the free theories, one can easily see that in a generic, interacting CFT the Jacobi identities \eqref{eq:Jacobi} are not satisfied. This sounds puzzling since we expect the Jacobi identity to hold as a consequence of crossing. Indeed, let us start with the four-point function $\langle T T T O_i\rangle$. We can ask if the operator algebra generated by the OPE satisfies the Jacobi identity at the level of the correlation function of local operators. This is of course nothing but the statement of crossing symmetry applied to the four-point function $\langle T T T O_i\rangle$. How is it possible then that the identity \eqref{eq:Jacobi} fails?

The resolution of the puzzle is that the commutator $[{\cal E}_{\omega_1}(n_1) , {\cal E}_{\omega_2}(n_2)]$ in $d>2$ CFTs requires a regularization. In this section we defined the commutator by first separating the light-ray operators 
in the `$+$' direction by distance $\eps$, and then taking $\eps \to 0$. Let us denote the commutator at finite $\eps$ as $[{\cal E}_{\omega_1}(n_1) , {\cal E}_{\omega_2}(n_2)]_{\eps}$. 
We then expect that the following sum rule holds 
\be
\label{eq:JacobiB}
\lim_{\eps \to 0} \Big( [[{\cal E}_{\omega_1}(n_1) , {\cal E}_{\omega_2}(n_2)]_{\eps} , {\cal E}_{\omega_3}(n_3)]_\eps + {\rm cyclic} \Big) = 0.
\ee

There is an important difference between \eqref{eq:Jacobi} and \eqref{eq:JacobiB}. 
According to \re{eq:symmtraceless},  
the contribution of the operator $O_{\Delta,S}$ with twist $\tau=\Delta-S$ to the commutator $[{\cal E}_{\omega_1}(n_1) , {\cal E}_{\omega_2}(n_2)]_\epsilon$ vanishes as $\epsilon^{{\tau-(d-2) \over 2}} {\cal J}_{\omega_1+\omega_2,S}$ for $\tau> d-2$. 
Performing the light-ray transform of the three-point function $\langle T T O_{\Delta,S}\rangle$, one can show that
$[{\cal J}_{\omega_1+\omega_2}(n_1) , {\cal E}_{\omega_3}(n_2)]_\epsilon$ receives a singular contribution $\epsilon^{{(d-2) - \tau \over 2}} {\cal E}_{\omega_1+\omega_2+\omega_3}$ which precisely compensates the factor $\epsilon^{{\tau-(d-2) \over 2}}$ coming from the commutator of the energy flow operators. Therefore, we conclude that the operators with twist $\tau > d-2$ do not contribute to \eqref{eq:Jacobi}, but they do contribute to \eqref{eq:JacobiB}.
 
The conclusion is that in a generic CFT we expect that the Jacobi identity \eqref{eq:Jacobi} receives contributions from infinitely many operators when inserted inside the correlation function, e.g. $$\lim_{\eps \to 0}  \langle \Big( [[{\cal E}_{\omega_1}(n_1) , {\cal E}_{\omega_2}(n_2)]_{\eps} , {\cal E}_{\omega_3}(n_3) ]_\eps + {\rm cyclic} \Big) T\rangle = 0 \ . $$ One might wonder what kind of sum rules can be obtained in this way, but this goes beyond the scope of the present paper. 

Instead, as we discuss in the concluding remarks, we can use \eqref{eq:algebralight} to write  simpler sum rules which generalize the superconvergence sum rules \cite{Kologlu:2019bco} and connect to the dispersive sum rules \cite{Caron-Huot:2020adz,Caron-Huot:2021enk}.
 
\section{Contact terms in Mellin space}
\label{sec:Mellin}

In this section, we apply the obtained results for the algebra of light-ray operators to understand the properties of (generalized) energy-energy correlation in ${\cal N}=4$ SYM 
\begin{align}\label{eec}
\text{EEC}(z,\omega_1,\omega_2) =\vev{0|\phi^\dagger \mathcal E_{\omega_1}(n_1)\mathcal E_{\omega_2}(n_2) \phi(q)|0}\,,
\end{align}
where $\phi$ and $\phi^\dagger$ are the half-BPS scalar primary operators in the representation $\mathbf{20'}$ of the $R-$symmetry group (see Ref.~\cite{Korchemsky:2021okt} for details).

The initial state $\phi(q)|0\rangle$ carries the total momentum $q^\mu$ and the energy flow operators $\mathcal E_{\omega_1}(n_1)$ and $\mathcal E_{\omega_2}(n_2)$ mimic two calorimeters detecting the particles propagating in the directions specified by null vectors $n_1$ and $n_2$. 
The function \re{eec} depends on dimensionless scaling variable
\begin{align}
z= {q^2 (n_1 n_2)\over 2(qn_1)(qn_2)}\,,
\end{align}
satisfying $0\le z\le 1$,
and two dimensionless frequencies $\hat\omega_i=2\omega_i (qn_i)/q^2$. To simplify notations, we will not display the additional factor in the expression for $\hat\omega_i$ and shall treat $\omega_i$ as dimensionless variables.

For $\omega_1=\omega_2=0$, the function \re{eec} coincides with the conventional energy-energy correlation.  A powerful approach to computing this function was developed in Refs.~\cite{Belitsky:2013xxa,Belitsky:2013bja}.
It was recently applied to the generalized energy-energy correlation \re{eec} with $\omega_i\neq 0$  in \cite{Korchemsky:2021okt}. 
The approach relies on the Mellin representation of the correlation functions and yields the following expression for the generalized energy-energy correlation  
\begin{align}\label{eec-sum}
\text{EEC}(z,\omega_1,\omega_2) =  \text{EEC}^{\text{tree}}(z, \omega_1, \omega_2) + \text{EEC}^{\text{Mellin}}(z, \omega_1, \omega_2)\ , 
\end{align}
where $\text{EEC}^{\text{tree}}$ describes the tree-level (zero coupling) contribution and $ \text{EEC}^{\text{Mellin}}$ is given by a convolution of the Mellin amplitude of the correlator and a kinematical kernel depending on dimensionless variables $z$ and $\omega_i$    
\be\label{Mellin}
\text{EEC}^{\text{Mellin}}(z,\omega_1,\omega_2) = \int {d j_1 d j_2 \over (2 \pi i)^2} K_{\text{EEC}}(j_1,j_2 | z, \omega_1, \omega_2) M(j_1,j_2) \,.
\ee
The explicit form of $K_{\text{EEC}}(j_1,j_2 | z, \omega_1, \omega_2)$ can be found in the ancillary file contained in the submission of Ref. \cite{Korchemsky:2021okt}.

The results obtained in Refs.~\cite{Belitsky:2013xxa,Belitsky:2013bja,Korchemsky:2021okt} apply when the detectors are located at separated points on the celestial sphere $n_1 \neq n_2$. In this case,
the energy flow operators commute and the function $\text{EEC}(z,\omega_1,\omega_2)$ is symmetric under the exchange of the detectors. 
The commutation relations  \re{eq:algebrahigherd} imply that this property does not hold if the detectors are oriented along the same direction $n_1=n_2$, or equivalently $z=0$. It follows from \re{eq:algebrahigherd} that $\text{EEC}(z,\omega_1,\omega_2)-\text{EEC}(z,\omega_2,\omega_1)\sim \delta(z)$.\footnote{The locus $z=0$ corresponds to $n_1 = n_2$, as such $\delta^{d-2}(n_1,n_2)$ that we used in the previous sections becomes $\delta(z)$ up to a trivial constant pre-factor.} It is natural to ask to what extent the analysis in Mellin space can be generalized to also cover such contact terms. This is the subject of the present section.
  
Let us start with an important subtlety which we already discussed at the beginning of Section \ref{sec:CFTlightrayalgebra}. In doing the computations in the papers mentioned above, the detector operators were first placed to future null infinity, and only then the integral over working time of the detector, or the light transform, was performed. It is possible to check that if we were to do this in the previous sections of the present paper we would miss important contact terms that produce nontrivial contributions to the commutator. As a result, if we want to correctly capture the commutator the analysis of \cite{Belitsky:2013xxa,Korchemsky:2021okt} is not obviously applicable.

At the technical level, the subtlety arises from the treatment of the separation between detectors $(x_{12}^2)^{- \delta_{12}}$ in the definition of the Mellin amplitude. If we do the light transform first, we can potentially pick the poles from $x_{12}^2=0$ which are sensitive to the light-ray commutators. If we, on the other hand, take the detectors to null infinity first, the $x_{12}^2=0$ poles never contribute to the light transform.

There is another possibility for poles $x_{12}^2=0$ not to contribute even if we want to discuss the locus $n_1 = n_2$. If the light-ray operators are ordered as ${\cal E}_{\omega_1 \geq 0}(n_1){\cal E}_{\omega_2 \leq 0}(n_2)$ then by evaluating the light transform we will not get the contribution from the $x_{12}^2=0$ poles due to the $i \epsilon$ prescription, which takes the form $t_1 - t_2 - i \epsilon$, where we used the detector time from the definition \eqref{J-def}. For $\omega_1 \geq 0$ we can deform the $t_1$ contour to the lower half-plane, and given that $\omega_2 \leq 0$, the $t_2$ contour to the upper half-plane, so that the pole at ${\rm Im}[t_1 - t_2] = \epsilon$ never contributes. 
 
In principle, it should be possible to generalize the analysis of \cite{Belitsky:2013xxa,Korchemsky:2021okt} to include such contributions, but here for simplicity we only restrict our consideration to the particular ordering ${\cal E}_{\omega_1 \geq 0}(n_1){\cal E}_{\omega_2 \leq 0}(n_1)$ for which the contact terms in the event shapes can be simply computed using the formulas of \cite{Belitsky:2013xxa,Korchemsky:2021okt} as we explain below.

\subsection{Zero frequencies}

Let us first consider the conventional energy-energy correlation corresponding to $\omega_1 =\omega_2=0$. 
In this case, we have (see Refs.~\cite{Belitsky:2013xxa,Korchemsky:2021okt})
\begin{align}\notag\label{eec-tree}
& \text{EEC}^{\text{tree}}(z, 0, 0) = {1 \over 4} \Big( \delta(z) + \delta(1-z) \Big)\,,
\\
& K_{\text{EEC}}(j_1,j_2 |0,0, z ) =  {\pi \over 2\sin \pi (j_1+j_2)} z^{-2-j_1 - j_2}\left(1 - z  \right)^{j_1 + j_2 - 1} \,.
\end{align} 
In order to reveal the origin of contact term $\delta(z)$ in the Mellin integral  \re{Mellin}, we change integration variable to $j=j_1+j_2$ and rewrite 
\re{Mellin} as
\begin{align}\label{int-f}
\text{EEC}^{\text{Mellin}}(z,0,0) = \int_{- \epsilon - i \infty}^{- \epsilon + i \infty} {d j \over 2 \pi i} {1 \over z^{1 + (1+j)}} f(j,z) \,,
\end{align}
where the notation was introduced for
\begin{align}\label{f-fun}
f(j,z) = {2\pi \over \sin \pi j}  \left(1 - z  \right)^{j - 1}  \int {d j_1   \over  2 \pi i }   M(j_1,j-j_1)  \,.
\end{align}

To find the leading behaviour of \re{int-f} at small $z$, we deform the integration contour to the left and pick up a residue at the rightmost pole of $f(j,z)$. Imagine that it is located at $j=-1$ and the function $f(j,z)$ has the following expansion around it
\be\label{f-res}
f(j,z) = {f_{-1}(z) \over 1+j} + f_0(z) + \dots \ .
\ee
Then, we expand ${1 / z^{1 + (1+j)}}$ around $j=-1$ using the identity
\be
{1 \over z^{1+(1+j)}} = - {1 \over 1+ j} \delta(z) + \Big[ {1 \over z} \Big]_+ + \dots \  ,
\ee
where the plus distribution is defined as $\int_0^1 d z  \, \phi(z) \big[ {1 \over z} \big]_+ =\int_0^1 d z (\phi(z)-\phi(0))/z$ for an arbitrary test function. In this way, closing the contour in \re{int-f} to the left we get
\be\label{eec-d}
\text{EEC}^{\text{Mellin}}(z,0,0) = - f_0(0) \delta(z) + f_{-1}(0) \Big[ {1 \over z} \Big]_+ 
+ \dots \,.
\ee

Let us apply the relations \re{f-fun}  -- \re{eec-d}  to the leading expressions for the Mellin amplitude $M(j_1,j_2)$  at weak and strong coupling
\begin{align}\notag
& M^\text{weak}(j_1,j_2) = -{a\over 4} {(j_1+j_2)^2\over (j_1j_2)^2}\,,
\\ 
& M^\text{strong}(j_1,j_2) = -{1\over 2} {(j_1+j_2)^2(1+j_1+j_2)\over j_1j_2}\,,
\end{align}
where $a=g_\text{YM}^2 N/(4\pi^2)$ is the 't Hooft coupling constant. Substituting these expressions into \re{f-fun} and \re{f-res} we get
\be\notag
f_{-1}^{\text{weak}}(0)&={a \over 4}, && \hspace*{-15mm} f_{0}^{\text{weak}}(0)={a \over 4}\,,
\\
f_{-1}^{\text{strong}}(0)&=0, && \hspace*{-15mm} f_{0}^{\text{strong}}(0)={1 \over 4} \,.
\ee
Combining these relations with \re{eec-d} we reproduce the known expressions for the energy-energy correlation at small $z$.

For example, at strong coupling we get from \re{eec-sum} and \re{eec-d}
\be
\text{EEC}^{\text{strong}}(z,0,0) = {1 \over 4} \Big( \delta(z) + \delta(1-z) \Big) - {1 \over 4} \Big( \delta(z) + \delta(1-z)-2 \Big) = {1 \over 2}\, ,
\ee
where the first term comes from $\text{EEC}^{\text{tree}}$ and the second one from $\text{EEC}^{\text{Mellin}}$. Notice that the contact terms cancel in the sum of two terms as expected (see Ref.~\cite{Hofman:2008ar}).
The origin of $- {1 \over 4} \delta(z)$ term in $\text{EEC}^{\text{Mellin}}$ was explained above. The presence of $- {1 \over 4}(\delta(1-z)-2)$ can be equally detected by computing the moments $\int_0^1 d z z^N \text{EEC}^{\text{Mellin}}(z,0,0)$. 

\subsection{Finite frequencies}

Let us now consider generalized energy-energy correlation \re{eec-sum} and \re{Mellin} for $\omega_i \neq 0$. 
For the sake of simplicity we restrict our consideration to $\omega_1 \geq 0$ and $\omega_2\le 0$ and denote the corresponding function \re{eec-sum} as $\text{EEC}_{+-}(z, \omega_1, \omega_2)$. 
 
In this case, we have in $\mathcal N=4$ SYM theory~\footnote{Here we ignore the contribution of the disconnected correlator which is localized at $z=0$ and in addition is proportional to $\delta(\omega_1+\omega_2)$. We also write down the tree-level expression that is valid for any signs of $\omega_i$.}
\be
\label{eq:treeEECN4}
\text{EEC}^{\text{tree}}(z, \omega_1, \omega_2) &=  \frac{(\omega _2^2+6 \omega _2+6) (\omega _1^2+6 (\omega
   _2+1) \omega _1+6 (\omega _2+1)^2)}{144 (\omega
   _2+1)} \theta(1+\omega_2) \delta(z) \nn  \\
   &-  \frac{(\omega _1^2+6 \omega _1+6) (\omega _2^2+6 (\omega
   _1+1) \omega _2+6 (\omega _1+1)^2)}{144 (\omega
   _1+1)} \theta(-1-\omega_1) \delta(z) \nn \\
   &+   \frac{(\omega _1^2+6 \omega_1+6) (\omega _2^2 +6 \omega _2+6)}{144}  \theta(1+\omega_1)  \theta(1+\omega_2) \delta(1-z) \,.
\ee
It is easy to check that for $\omega_1=\omega_2=0$ this relation coincides with the first relation in \re{eec-tree}.

As in the previous section, the contact term $\delta(z)$ in $\text{EEC}_{+-}^{\text{Mellin}}(z, \omega_1, \omega_2)$ arises from small $z$ behaviour of the kernel $K_{\text{EEC}}(j_1,j_2 | z, \omega_1, \omega_2)$ in \re{Mellin}. Using the explicit form of the kernel we find 
\be
K_{\text{EEC}}(j_1,j_2 | z, \omega_1, \omega_2) \sim {1 \over z^{1+(1+j_1+j_2)}} \, . 
\ee
Going through the same steps as in the previous section, we obtain the following result for $\text{EEC}_{+-}^{\text{Mellin}}$ at strong coupling for $\omega_1 \geq0$ and $-1<\omega_2\leq0$
\be
\label{eq:strongcouplingcontact}
\text{EEC}_{+-}^{\text{Mellin}}(z, \omega_1, \omega_2)  &=\frac{(\omega _2 ^2+6\omega _2+6)(6  (\omega
   _2+1)^2+6 \omega _1 (\omega _2+1) (\omega _2
   z+1)+\omega _1^2(\omega _2 z+1)^2 )}{72 (\omega
   _2 z+1){}^3} \nn \\
   &- \frac{(\omega _2^2+6 \omega _2+6) (\omega _1^2+6 (\omega
   _2+1) \omega _1+6 (\omega _2+1)^2)}{144 (\omega
   _2+1)} \delta(z) \nn \\
   &- \frac{(\omega _1^2+6 \omega_1+6) (\omega _2^2+6
   \omega _2+6)}{144}  \delta(1-z) \, .
\ee
Substituting \re{eq:treeEECN4} and \re{eq:strongcouplingcontact} into \re{eec-sum} we
observe that the contact terms cancel in the sum of two expressions leading to
\be
\text{EEC}^\text{strong}_{+-}(z, \omega_1, \omega_2)  &=\frac{ (\omega _2^2 +6 \omega _2+6) (6 (\omega
   _2+1)^2+6 \omega _1 (\omega _2+1) (\omega _2
   z+1)+\omega _1^2(\omega _2 z+1)^2)}{72 (\omega
   _2 z+1)^3}  \, .
\ee

Thus, the generalized energy-energy correlation does not contain any distributional terms at strong coupling. 
This is indeed consistent with the following simple observation. 
Taking into account that the energy flow operator $\mathcal E_\omega(n)$ annihilates the left (or right) vacuum for $\omega>0$ (or $\omega<0$) we can write for $\omega_1 \geq 0$ and $\omega_2\le 0$
\be
\text{EEC}_{+-}(z, \omega_1, \omega_2)  = \langle 0|  [\phi^\dagger, {\cal E}_{\omega_1 }(n_1)] [{\cal E}_{\omega_2 }(n_2), \phi] |0 \rangle\,.
\ee
Following  \cite{Kologlu:2019bco}, we can insert a complete state of states between the two commutators to get the OPE representation of $\text{EEC}_{+-}(z, \omega_1, \omega_2)$. At strong coupling only $\phi(x)$ itself contributes at leading order in ${1 / N_c^2}$, due to the familiar suppression of the double trace operators in the double commutator. We do not expect to get any distributional terms from the exchange of $\phi(x)$, hence we get a regular result in agreement with \eqref{eq:strongcouplingcontact}. This is a very nontrivial check of consistency of our approach.

If we, on the other hand, exchange the flow operators in \re{eec} the same argument does not apply because  for $\omega_1 \geq 0$ and $\omega_2\le 0$ the function $\text{EEC}_{-+}(z, \omega_2, \omega_1)$
cannot be writen as  a double commutator. In particular, the double trace operators are not suppressed  in this case and they eventually generate the nontrivial contact term $\delta(z)$ in  $\text{EEC}_{-+}(z, \omega_2, \omega_1)$.

\subsection{Asymmetry}

As was mentioned above, the commutator of the energy flow operator controls the asymmetry of the energy-energy correlation, $\text{EEC}(z,\omega_1,\omega_2)-\text{EEC}(z,\omega_2,\omega_1)\sim \delta(z)$. 

Let us first consider the limit of zero coupling in $\mathcal N=4$ SYM when  the Mellin contribution to \re{eec-sum} is absent. It follows from the algebra \eqref{eq:freescalar} that
\be
\label{eq:treelevelalgebra}
\text{EEC}^{\text{tree}}(z, \omega_1, \omega_2) - \text{EEC}^{\text{tree}}(z, \omega_2, \omega_1) = {1 \over 4} \delta(z) \langle  (\omega_2 - \omega_1) {\cal E}_{\omega_1 + \omega_2}(n_1) + \omega_1^{2}  \omega_2^{2} {\cal O}^{-}_{\omega_1, \omega_2}(n_1)  \rangle_\text{tree} \,,
\ee
where an average on the right-hand side is taken with respect to the state $\ket{\phi(q)}$ containing in the free theory two on-shell scalar particles with the total momentum $q$, in the representation $\mathbf{20'}$ of the $R-$symmetry group (see Ref.~\cite{Korchemsky:2021okt} for details).
The relevant one-point functions take the form
\be\label{mat}
 \langle {\cal E}_{\omega}(n)  \rangle_\text{tree} &= 1+\omega+{\omega^2 \over 6} \,, \nn \\
  \langle  {\cal O}^{-}_{\omega_1, \omega_2}(n)\rangle_\text{tree} &= {\omega_1-\omega_2 \over 36(1+\omega_1)(1+\omega_2)}\,.
\ee 
We can use \eqref{eq:treeEECN4} to check that the light-ray algebra identity \eqref{eq:treelevelalgebra} is indeed satisfied.
Notice that the first relation in \re{mat} is protected from corrections in $\mathcal N=4$ SYM.

The situation becomes more interesting at finite coupling. We show in Appendix~\ref{sec:extralightray} that the operator ${\cal O}^{(\phi)}_{\omega_1, \omega_2}$ is given by the sum of light-ray operators \re{J-def} of an arbitrary spin $S$. In $\mathcal N=4$ SYM the operators with $S\neq 2$ acquire anomalous dimensions.  Their twist satisfies $\tau>2$ and, as a consequence, they do not contribute to the light-ray algebra. The only operator that contributes is the stress-energy tensor. This leads to the algebra \re{eq:energyalgebra}. As a consequence, the energy-energy correlation satisfies
\be
\text{EEC} (z, \omega_1, \omega_2) - \text{EEC} (z, \omega_2, \omega_1) = {1 \over 4} \delta(z) (\omega_2 - \omega_1+ {1 \over 10} f_\phi (\omega_1,\omega_2)) \langle    {\cal E}_{\omega_1 + \omega_2}(n_1)  \rangle \,.
\ee 
Combining this relation with \re{eec-sum} and \re{eq:treelevelalgebra} 
we expect that  the following identity holds in $\mathcal N=4$ SYM at finite coupling
\be\notag
\label{eq:Melling}
& \text{EEC}^{\text{Mellin}}(z, \omega_1, \omega_2) - \text{EEC}^{\text{Mellin}}(z, \omega_2, \omega_1) 
\\
& ={1 \over 4} \delta(z) \left[ {1 \over 10} f_\phi (\omega_1,\omega_2) \langle  {\cal E}_{\omega_1+ \omega_2}(n_1)  \rangle -  \omega_1^{2}  \omega_2^{2} \langle {\cal O}^{-}_{\omega_1, \omega_2}(n_1)  \rangle \right],
\ee 
where the one-point functions on the right-hand side are given by \re{mat}.
It would be interesting to derive this relation by an explicit computation.  
 
\section{Concluding remarks}
\label{sec:conclusions}

In this paper we explored commutation relations of the light-ray operators \re{J-def} built out of local operators in a $d-$dimensional CFT. 
The commutator of these operators $[\mathcal J_{\omega,S} (n), \mathcal J_{\omega',S'} (n')]$ is localized at $n=n'$, when both operators are pointed in the same direction on the celestial sphere. We argued that it has different form in free and interacting CFTs.

We demonstrated that in a free theory in $d$ dimensions, the flow operators form a closed algebra and derived the explicit expressions for the corresponding structure constants and the central charges. 
In $d=2$ the light-ray operators are closely related to generators of infinite-dimensional symmetries of CFT${}_2$. More precisely, 
the light-ray operators of spin $S=2$, or equivalently the energy flow operators, are generating functions of the generators of Virasoro algebra. In the similar manner, for $d=2$,  the light-ray operators of higher spin are related to the generators of the $W_\infty$ algebra. We found that in the free theory of 
complex scalar and fermion in $d=4$ dimensions  (and, more generally, for any even $d$), the structure constants of the algebra of the light-ray operators
are related to those of the Virasoro and $W_\infty$ algebras by a linear finite-difference relation.

Defining the light-ray operators \re{J-def} in free CFTs, we have chosen the conformal operators to be built from two fundamental fields of the same type (scalars, fermions and gauge fields). In a similar manner, we can consider a free CFT describing one fermion, one complex scalar and gauge field, define ``mixed''  operators (built out of fields of different type) and construct the corresponding flow operators. In $d=2$ dimensions, this theory respects $\mathcal N=2$ supersymmetry and, as a consequence, such flow operators form a $\mathcal N=2$ supermultiplet. Their algebra coincides with super$-W_\infty$ algebra~\cite{Bergshoeff:1988uc,Bergshoeff:1990cz,Bergshoeff:1991dz}. In $d=4$ dimensions, all leading twist operators belong to the same $\mathcal N=4$ supermultiplet~\cite{Belitsky:2003sh}. This suggests that the flow operators generate an $\mathcal N=4$ superalgebra whose bosonic sector is given by \re{comm} and \re{eq:freescalar}. It would be interesting to compute the structure constants of this superalgebra and to establish their relation to those of super$-W_\infty$ algebra.
 
We found that in an interacting CFT in $d>2$ dimensions, the commutator of the energy flow operators only receives contribution from the stress-energy tensor itself and light scalar primary operators (which appear in the OPE of two stress-energy tensors) with dimension $\Delta \le d-2$.
The latter contribution is divergent for $\Delta < d-2$. 

As opposed to the free theories considered in Section \ref{sec:freeCFTs}, the light-ray operators do not form an algebra in interacting CFTs. 
As explained in Section \ref{sec:jacobiInt},  this is related to the careful, regularized definitions of the commutators.
It would be interesting to understand if there is a sense in which the light-ray algebra that we found in the free CFTs is weakly broken at weak coupling, see e.g. Refs.~\cite{Maldacena:2012sf,Gerasimenko:2021sxj}.

We believe that the significance of the commutation relations considered in the present paper is two-fold. First, the algebra of the light-ray operators is connected to the symmetries of the theory, be it global conformal invariance in interacting CFTs or higher spin symmetry in the free CFTs. For example, the only spinning operators that contribute to the commutator of the energy flow operators are conserved currents out of which the symmetry generators are constructed. The result \re{eq:algebrahigherd} contains a universal term 
 $[\mathcal E_{\omega_1}(n_1),\mathcal E_{\omega_2}(n_2)] = \delta^{(d-2)}(n_1,n_2)(\omega_2 - \omega_1) {\cal E}_{\omega_1 + \omega_2}(n_1) + \dots$ which comes from the OPE $T \times T \sim T + \dots$ and is closely related to the conformal Ward identities. In a free theory, in a close analogy with the $W$ symmetry in CFT$_2$, the algebra gets extended to include the light-ray operators built out of higher spin currents.
 
Second, the commutator of light-ray operators naturally leads to a family of sum rules
\be
\label{eq:sumrulecomplex}
\int \prod_{i=1}^{2} d \omega_i d \vec n_i\, \delta(\vec n_i^2-1)  g (n_i, \omega_i)   \langle \phi_i  | (\text{light-ray commutator}) |  \phi_j \rangle  = 0,
\ee
where $g (n_i,\omega_i)$ is a priori an arbitrary function. The sum rules are obtained by inserting a complete set of states between the light-ray operators in the commutator.
For example, let us consider a generic interacting CFT, where the light-ray algebra takes the form \eqref{eq:algebrahigherd}. The presence of the divergent contributions in the algebra due to light scalar primary operators, $\Delta < d-2$,
signify the necessity of subtractions to get well-defined CFT sum rules. For the identity operator subtraction boils down to considering the connected correlator on the left-hand side of \eqref{eq:sumrulecomplex}. For other light scalar primary operators 
of dimension ${d-2\over 2}\le  \Delta < d-2$, 
that appear in the OPE of stress-energy tensors, we can for example subtract the corresponding scalar exchange Witten diagram which effectively removes light scalars from the $[{\cal E}_{\omega_1}(n_1) , {\cal E}_{\omega_2}(n_2)]$. The sum rules \eqref{eq:sumrulecomplex} generalize the superconvergence relations of Ref.~\cite{Kologlu:2019bco} and we can refer to the sum rules that follow from the light-ray algebra as {\it generalized superconvergence relations}. We do not attempt studying the consequences of these sum rules for the OPE data of CFTs in the present paper. 

Let us note in this context that the generalized superconvergence relations, analogous to \eqref{eq:sumrulecomplex}, were shown to be equivalent to the dispersive sum rules in some interesting cases  \cite{Caron-Huot:2020adz,Caron-Huot:2021enk}. It would be very interesting to explore this connection in a greater depth. For example, by making $g (n_i,\omega_i)$ vanishing in the limit $\omega_1 - \omega_2 \to 0$ with $\omega_1 + \omega_2$  fixed,  one damps the contribution of the Regge limit to the sum rule. Functionals with this property are related to subtractions in the CFT dispersion relations \cite{Carmi:2019cub,Mazac:2019shk,Penedones:2019tng}.
 
The light-ray algebra has important application to the generalized event shapes introduced in Ref.~\cite{Korchemsky:2021okt}. It allows one to control the contribution to these observables of distributions localized at the coinciding points on the celestial sphere.  Already for the conventional event shapes, for $\omega_i=0$, understanding of such distributional terms was important in order to verify the conformal Ward identities \cite{Kologlu:2019mfz,Korchemsky:2019nzm,Dixon:2019uzg}, which serve as a nontrivial check of the computations. The same applies to the generalized event shapes for $\omega_i \neq 0$. In this context, we have explained how the distributional terms naturally emerge in Mellin space in ${\cal N}=4$ SYM and presented expected
expression for the commutator \eqref{eq:Melling}. It would be interesting to derive this relation from the first principles. 

Relatedly, generalized event shapes satisfy the following positivity condition
\be
\label{eq:normposgen}
\int \prod_{i=1}^{2} d \omega_i d \vec n_i\, \delta(\vec n_i^2-1) \, g^{*}(n_1, \omega_1) g(n_2, \omega_2) \langle \Psi | \mathcal E_{-\omega_1}(n_1) \mathcal E_{\omega_2}(n_2) | \Psi \rangle \geq 0 \,.
\ee
It follows from the fact that $\mathcal E_{\omega_1}^\dagger (n_1) = \mathcal E_{-\omega_1}(n_1)$ and therefore \eqref{eq:normposgen} can be interpreted as a norm, or the total cross-section of the scattering $\ket{\Psi}$ off the state created by the integrated energy flow operator. Computing \eqref{eq:normposgen} necessarily involves evaluation of the generalized event shapes at coincident point $n_1 = n_2$, which we considered in Mellin space in Section \ref{sec:Mellin}. 
 
\section*{Acknowledgments}

We are grateful to Alex Belin,  Andrei Belitsky, Diego Hofman, Denis Karateev, Marco Meineri, Emery Sokatchev, Eric Perlmutter, Slava Rychkov, and Matthew Walters for useful discussions. This project has received funding from the European Research Council (ERC) under the European Union's Horizon 2020 research and innovation programme (grant agreement number 949077). The work of GK was supported by the French National Agency for Research grant ANR-17-CE31-0001-01.

\appendix

\section{Conventions}\label{App:conv}

For an arbitrary $d-$dimensional vector $x^\mu=(x^0,\dots,x^{d-1})$ the light-like coordinates are defined as
\begin{align}\label{lc}
x^+={1\over \sqrt{2}}(x^0+x^{d-1})\,,\qqquad x^-={1\over \sqrt{2}}(x^0-x^{d-1})\,,\qqquad \bit{x}=(x^1,\dots,x^{d-2}) \,.
\end{align}
The metric in these coordinates is
\begin{align}
dx^2 = 2dx^+dx^-- d\bit{x} d\bit{x}\,,
\end{align}
and the scalar product looks as
\begin{align}
(x y) = x^+ y^-+x^-y^+-\bit x \cdot \bit y\,.
\end{align}
The definition of the flow operator \re{J-def} involves an auxiliary light-like vector $\bar n^\mu$ that we choose to be along the `$-$' direction, 
 $\bar n^-=1$ and $\bar n^+=\bar{\bit{n}}=0$. In $d=2$ dimensions, the light-like vectors satisfying $(n\bar n)\neq 0$ take the form $n^\mu \sim (1,1)$,
or equivalently  $n^-=0$. 

To obtain representation of the light-ray operator \re{J-z}, we
apply a conformal transformation   ~\cite{Cornalba:2007fs,Hofman:2008ar} 
\begin{align}\label{tr}
z^+=-{1\over 2x^+}\,,\qqqquad z^-=x^--{\bit x^2\over 2x^+}\,,\qqqquad \bit z={\bit x\over \sqrt{2} x^+}\,,
\end{align}
where $dz^2 = dx^2/(2(x^+)^2)$.  It maps a light ray at null infinity $x^\mu= n r + \bar n t$ with $r\to\infty$ and $-\infty < t<\infty$ to a line
$-\infty< z^-<\infty$ with $z^+=0$,  $n^\mu=(1,\bit z^2,\sqrt{2}\bit z)$ and $\bar n^\mu=(0,1,\bit 0)$ in the light-cone coordinates

The commutation relations of the flow operators \re{alg2} involve the delta-function on the sphere $S^{d-2}$. For two light-like vectors $n^\mu=(1,\vec n)$ and $n'{}^\mu=(1,\vec n')$ it is defined as
\begin{align}\label{delta}
\delta^{(d-2)}(n,n') = \lim_{\epsilon\to 0} {1\over (2\pi i\epsilon)^{d/2-1}} \e^{i(nn')/\epsilon}\,.
\end{align}
It is straightforward to verify that it satisfies
\begin{align}
\int d \vec  n\, \delta(\vec n^2-1) f(n) \delta^{(d-2)}(n,n') = f(n')\,,
\end{align}
where $\int d \vec  n = \int d n_1 \dots dn_{d-1}$ and $f(n)$ is a test function. 
For $d=2$, it follows from $(\bar nn)\neq 0$ and $(\bar nn')\neq 0$ that $n\sim n'$ and, therefore, $\delta^{(0)}(n,n') =1$.

Taking into account \re{delta} we obtain  
\begin{align}\notag\label{limit}
\lim_{\epsilon\to 0}{1 \over (2\pi i \epsilon)^{d/2-1}}  {\Gamma(2j)(-i)^{2j} \over (-(nn')/\epsilon +t-i0)^{2j}}
&
 =\int_0^\infty  d\alpha\, \alpha^{2j-d/2} \e^{-i\alpha t} \lim_{
\epsilon\to 0}{\e^{is (nn')/\epsilon} \over (2\pi i \epsilon/s)^{d/2-1}}   
\\
&= \delta^{(d-2)}(n,n')\int_0^\infty  d\alpha\, \alpha^{2j-d/2} \e^{-i\alpha t}\,,
\end{align}
where in the first relation we applied \re{Sch}.

\section{Gegenbauer polynomials}\label{App:Geg}

The leading twist operators \re{O-P} and \re{Geg} are expressed in terms of Gegenbauer polynomials $C_n^{(\lambda )}(z)$. Their  index $\lambda=2j-1/2$ is related to the conformal spin of the field \re{js}. In particular,  $j=0$ for a scalar field in $d=2$ dimensions. In this case, we have
 \begin{align}
C_n^{(-1/2)}(x) ={(1-x^2)\over n(n-1)} C_{n-2}^{(3/2)}(x)
\end{align}
for $n\ge 2$,
or equivalently in terms of polynomials \re{Geg}
\begin{align}\label{P-sc}
P_n^{(0)}(p_1,p_2) = {4p_1 p_2\over n(n-1)} P_{n-2}^{(1)}(p_1,p_2)\,,
\end{align}
where the superscript refers to the conformal spin $j$. 

The Gegenbauer polynomials satisfy a recurrence relation
\begin{align}
z\, C_{n}^{(\lambda )}(z)=
{(2 \lambda +n-1)\over 2 (\lambda +n)}C_{n-1}^{(\lambda )}(z)+{(n+1)\over 2 (\lambda +n)} C_{n+1}^{(\lambda )}(z)  \,.
\end{align}
One can use it to verify the following identity
\begin{align}\notag\label{PP-id}
 (t-\varepsilon) & P_{\ell}(t ,-t +\varepsilon)P_{\ell''}(t,1-t) 
\\\notag
& =
\frac{\varepsilon  }{2 (1-\varepsilon )}\Bigg[\frac{(4 j+{\ell''}-2)
}{(4 j+2 {\ell''}-1)}\e^{-\partial_{\ell''}}
+\frac{({\ell''}+1)}{(4 j+2 {\ell''}-1)}\e^{\partial_{\ell''}}
\\   
&+\frac{\varepsilon(1-2 \varepsilon) (4 j+\ell -2)}{(4 j+2 \ell -1) } \e^{-\partial_\ell} +\frac{(1-2 \varepsilon)(\ell +1)}{\varepsilon(4 j+2 \ell -1) }\e^{\partial_\ell} \Bigg]   P_{\ell}(t ,-t +\varepsilon)P_{\ell''}(t,1-t) \,,
\end{align}
where $P_\ell$ is given by \re{Geg}. 

The Gegenbauer polynomials are orthogonal with respect to inner product
\begin{align}\label{CC}
\int_{-1}^1 dx\, (1-x^2)^{\lambda-\frac12}C_n^{(\lambda)} (x) C_m^{(\lambda)} (x) = \delta_{nm} {\pi 2^{1-2\lambda} \Gamma(n+2\lambda)
\over n!(n+\lambda)\Gamma^2(\lambda)}\,.
\end{align}
This relation ensures that the two-point correlation function of conformal operators \re{OO} is diagonal in spins. To show this, we consider the correlation function $\vev{O_{S_1}(x_1) O_{S_2}(x_2)}$ for $x_i=(x_i^+,x_i^-,\bit{0})$.
Using \re{O-P} one obtains 
\begin{align} \label{OO-pre}
\vev{O_{S_1}(x_1) O_{S_2}(x_2)}= 
\vev{\Phi(x_1) \bar\Phi(x_2)}
P_{\ell_1} (\stackrel{\leftarrow}{i\partial_{1-}},\stackrel{\rightarrow}{i\partial_{1-}})
P_{\ell_2} (\stackrel{\rightarrow}{i\partial_{2-}},\stackrel{\leftarrow}{i\partial_{2-}})
\vev{\bar\Phi(x_1)\Phi(x_2)}\,,
\end{align}
where $\partial_{i-}\equiv  \partial_{x_i^-}$ and the propagator is given by \re{phiOphi}. Introducing Schwinger parametrization for $x^--$dependent part of the propagator \re{phiOphi}
\begin{align}\label{Sch}
{\Gamma(2j) \over (x_{12}^--i0)^{2j} } = i^{2j} \int_0^\infty dp\, p^{2j-1} \e^{-ip x_{12}^-}\,,
\end{align}
one finds from \re{OO-pre}
\begin{align}
\vev{O_{S_1}(x_1) O_{S_2}(x_2)}\sim \int_0^\infty dp_1 dp_2 (p_1p_2)^{2j-1} \e^{-i(p_1+p_2)x_{12}^-}
P_{\ell_1}(p_1,p_2) P_{\ell_2}(p_1,p_2)\,.
\end{align}
After change of the integration variables, $p_1=t p $ and $p_2=(1-t) p$, with $0\le x\le 1$ and $0\le p <\infty$, the integral over $t$ takes the form 
\re{ortho} and leads to \re{OO}.

\section{Commutation relation for scalar operators}\label{app:scalar}

In this appendix, we examine the correlation function $\vev{\phi(x_1) \mathcal J_{\omega,S}(n)\mathcal J_{\omega',S'}(n')\bar\phi(x_2)}_c$
involving scalar flow operators and derive their commutation relations. For fermions and gauge fields, the corresponding correlation function is given by
\re{4pt} and \re{I-gen-j}. Substituting $s=0$ into there relations we find that the integral in \re{I-gen-j} diverges logarithmically. 
 
To elucidate its origin, we repeat the calculation of diagrams shown in Figure~\ref{four}. Introducing notation for $\epsilon=(n\bar n)/r'-(n'\bar n)/r$ we find for $r,r'\to\infty$
\begin{align} \notag\label{start}
& \vev{\Phi(x_1) \mathcal J_{\omega,S}(n)\mathcal J_{\omega',S'}(n')\bar\Phi(x_2)}_c= {1\over (2\pi)^{d-1}}   \lim_{\epsilon\to 0} {1\over (2\pi i\epsilon)^{d/2-1}}  \,  \int_0^\infty d\alpha_1 \, (\alpha_1 \alpha_2)^{2j-1}  \theta(\alpha_2)
\\[2mm]\notag
& \times\Big[ \e^{ i (\alpha_1-\omega) ( n n')/\epsilon}
 \e^{-i\alpha_1 (n_1x_1)+i \alpha_2(n_2x_2)}\theta(\alpha_1-\omega)
 (\alpha_1-\omega)^{2j-1} P_{\ell}(\alpha_1,-\alpha_1+\omega)P_{\ell'}(\alpha_1-\omega,-\alpha_2)
\\[2mm]
&{} +(-1)^{2s} \e^{ i (\omega'-\alpha_1) ( n n')/\epsilon}\e^{-i\alpha_1 (n_2x_1)+i \alpha_2(n_1x_2)}\theta(\omega'-\alpha_1)
(\omega'-\alpha_1)^{2j-1} P_{\ell'}(\alpha_1,-\alpha_1+\omega')P_{\ell}(\alpha_1-\omega',-\alpha_2)\Big]\,,
\end{align}
where integration goes over the energy of the outgoing particle and $\alpha_2=\alpha_1-\omega''$ (with $\omega''=\omega+\omega'$) is the energy of the incoming particle. The two terms inside the brackets correspond to the diagrams shown in Figure~\ref{four}(a) and (b), respectively.
 
Applying the identity \re{delta}, one is tempting to replace the first exponential factor on the second line of \re{start} with $\delta^{(d-2)}(n,n')/(\alpha_1-\omega)$ and similar on the last line. This would lead to relations \re{4pt} and \re{I-gen-j}. Such transformations are only legitimate for $\alpha_1-\omega\neq 0$ and $\omega'-\alpha_1\neq 0$. The boundary values $\alpha_1=\omega$ and $\alpha_1=\omega'$ correspond to the vanishing energy of the particle exchanged between the detectors, see Figure~\ref{four}. Assuming that $(nn')\neq 0$ and integrating in the vicinity of the boundary values, we get from \re{start}
\begin{align} \notag 
& \vev{\Phi(x_1) \mathcal J_{\omega,S}(n)\mathcal J_{\omega',S'}(n')\bar\Phi(x_2)}_c={\Gamma(2j)  \over (2\pi)^{3d/2-2}} \, (-\omega \omega')^{2j-1} 
\\[2mm]\notag
& \times {(i\epsilon)^{2s}\over (nn')^{2j}}\Big[ P_{\ell}(\omega,0)P_{\ell'}(0,\omega')\theta(\omega)\theta(-\omega')
  +(-1)^{2s}  P_{\ell'}(\omega',0)P_{\ell}(0,\omega)\theta(-\omega)\theta(\omega')\Big] +\dots\,,
\end{align}
where dots denote terms localized at $n=n'$ and we took into account that $j=s+(d-2)/4$. We observe that for $s>0$ this expression vanishes whereas for $s=0$ it leads to \re{4pt-s}. 

Thus,  for scalar fields the four-point function \re{start} receives an additional contribution that does not vanish for $n\neq n'$ and scales as $1/(nn')^{(d-2)/2}$. It is symmetric under the exchange of the detectors and does not contribute to the difference of the correlation functions \re{diff}. Taking into account \re{start} we obtain from \re{diff}
\begin{align} \notag\label{start1}
& \vev{\phi(x_1) [\mathcal J_{\omega,S}(n),\mathcal J_{\omega',S'}(n')]\bar\phi(x_2)}_c= {1\over (2\pi)^{d-1}}   \lim_{\epsilon\to 0} {1\over (2\pi i\epsilon)^{d/2-1}}  \,  \int_0^\infty d\alpha_1 \, (\alpha_1 \alpha_2)^{d/2-2}  \theta(\alpha_2)
\\[2mm]\notag
& \times\Big[ \e^{ i (\alpha_1-\omega) ( n n')/\epsilon}
 \e^{-i\alpha_1 (n_1x_1)+i \alpha_2(n_2x_2)} 
 (\alpha_1-\omega)^{d/2-2} P_{\ell}(\alpha_1,-\alpha_1+\omega)P_{\ell'}(\alpha_1-\omega,-\alpha_2)
\\[2mm]
&{} +\e^{ i (\omega'-\alpha_1) ( n n')/\epsilon}\e^{-i\alpha_1 (n_2x_1)+i \alpha_2(n_1x_2)} 
(\omega'-\alpha_1)^{d/2-2} P_{\ell'}(\alpha_1,-\alpha_1+\omega')P_{\ell}(\alpha_1-\omega',-\alpha_2)\Big],
\end{align}
where  $\alpha_2=\alpha_1-\omega''$ and we took into account that $\epsilon$ changes the sign under the exchange of the detectors. As follows from the above discussion, the contribution from $\alpha_1=\omega$ and $\alpha_1=\omega'$ cancels on the right-hand side of \re{start1}. This allows us to deform the integration contour in the vicinity of the two points. Then, applying the identity \re{delta} we arrive at
\begin{align} \notag\label{start2}
 \vev{\phi(x_1) [\mathcal J_{\omega,S}(n),\mathcal J_{\omega',S'}(n')]\bar\phi(x_2)}_c= {\delta^{(d-2)}(n,n')\over (2\pi)^{d-1}}  \dashint_0^\infty d\alpha_1 \, (\alpha_1 \alpha_2)^{d/2-2}  \theta(\alpha_2)\e^{-i\alpha_1 (nx_1)+i \alpha_2(nx_2)} 
\\[2mm] 
 \times\Big[  {P_{\ell}(\alpha_1,-\alpha_1+\omega)P_{\ell'}(\alpha_1-\omega,-\alpha_2)\over \alpha_1-\omega} - 
 {  P_{\ell'}(\alpha_1,-\alpha_1+\omega')P_{\ell}(\alpha_1-\omega',-\alpha_2)\over \alpha_1-\omega'}\Big],
\end{align}
where dash denotes the principal value prescription. The expression in the second line of \eqref{start2} coincides with \re{Q-sum1}.

\section{Relation to the Virasoro and $W$ algebras}\label{app:W}
    
In this appendix we show that the algebra of the flow operators \re{alg2} in $d=2$ dimensions is equivalent to the $W-$algebra in two-dimensional Euclidean CFT~\cite{Bouwknegt:1992wg}. 

The commutation relations in the $W$ algebra are~\cite{Pope:1989ew,Pope:1989sr,Pope:1991ig,Korybut:2020ncv}
 \begin{align}\label{VV}
[V_m^i,V_n^j] = \sum_{p\ge 0} g_{2p}^{ij}(m,n) V_{m+n}^{i+j-2p} + c_i(m) \delta^{ij} \delta_{m+n,0}\,,
\end{align}
where the operators $V_m^i$ are modes in the Laurent expansion of the currents $O_S(z)$ of spin $S=i+2$ on two-dimensional Euclidean plane
\begin{align}
O_S(z)  = \sum_{m=-\infty}^\infty V_m^i\, z^{-m-i-2}\,,
\end{align}
where  $z=x^1+i x^2$.
For the $W_\infty$ and $W_{1+\infty}$ algebras the spins satisfy condition $S\ge 2$ and $S\ge 1$, respectively.

For $S=2$ the current $O_{S=2}(z)$ coincides with the stress-energy tensor and the corresponding modes $L_n=V_n^0$ satisfy the Virasoro algebra
\begin{align}\label{L}
[L_m,L_n] = (m-n)L_{m+n} + {c\over 12} m(m-1)(m+1)\delta_{m+n,0}\,.
\end{align}
The structure constants and the central charges in \re{VV} can be determined by requiring the relations \re{VV} to be consistent with the Jacobi identities. In particular, the central charges take the form
\begin{align}\label{ci} 
& c_i(m) = m(m^2-1)(m^2-4) \dots (m^2-(i+1)^2) c_i\,,
\end{align}
where $c_i$ is proportional to the central charge $c$.
 
To compare \re{VV} with the algebra \re{alg2} one has to extend the above relations to Lorentzian signature.  In two-dimensional Euclidean CFT, the modes $V_n^j$ are given by
\begin{align}\label{oint}
V_n^j = \oint {dz\over 2\pi i}  z^{n+j+1} O_{S=j+2}(z)  \,,
\end{align} 
where the integration contour encircles the origin. After Wick rotation $x^2\to i x^0$, the complex variable $z=x^1+ix^2$ turns into the light-cone coordinate $z= -\sqrt{2} x^-$. 
Then, one may try to generalize the relation \re{oint} as $V_n^j \sim \int dx^- \,(x^-)^{n+j+1} O_{S=j+2}(x^-)$. However, as was noticed in Refs.~\cite{Belin:2020lsr,Besken:2020snx}, such operator is not well-defined because inserted inside a correlation function it leads to integrals that are divergent as $x^-\to\infty$.~\footnote{These papers deal with spin-two operators, the same arguments apply to operators of high spin.} 

Following Ref.~\cite{Besken:2020snx} we can generalize \re{oint} to Lorentzian signature by applying a (complexified) conformal transformation 
$z\to w=i (1+iz)/(1-iz)$ that maps a circle of unit radius $|z|=1$ to a line $-\infty < w<\infty$. Taking into account transformation properties of the currents $O_{S}(z)$, we obtain from \re{oint}
\begin{align}\label{hat-V}
  V_n^j= {1\over 2^{S-1}}\int_{-\infty}^\infty dw\, (1-iw)^{S-1-n} (1+iw)^{S-1+n}O_S(w) \Big|_{S=j+2}  \,,
\end{align}
where the additional factor comes from $(\partial w/\partial z)^{S-1}$. These operators satisfy the same commutation relations as the operators \re{oint} but they do not suffer from divergences since the large $w$ asymptotics of the integrand in \re{hat-V} does not depend on $n$. 

To express \re{hat-V} in terms of the flow operators \re{J-z}, it is convenient to introduce the weight function
\begin{align} \label{muS}
\mu_{n,S}(\omega) & =  {1\over 2^{S-1}} \int_{-\infty}^\infty {dx\over 2\pi} \e^{ i\omega x}(1-ix)^{S-1-n}(1+ix)^{S-1+n}\,.
\end{align}
Then, the operators \re{hat-V} can be written as weighted integrals of the flow operators
\begin{align}\label{conv}
V_n^j= \int_{-\infty}^\infty d\omega\,\mu_{n,S}(\omega)\, \mathcal J_{\omega,S}\Big|_{S=j+2}\,.
\end{align}
This relation establishes the correspondence between the $W-$algebra \re{VV} and the algebra of the flow operators \re{alg2} in $d=2$ dimensions.

The weight function \re{muS} verifies relations
\begin{align}\label{mu-rel}
\mu_{n,S}(\omega) = \mu_{-n,S}(-\omega) = \lr{\mu_{n,S}(\omega) }^*\,.
\end{align}
It has different properties depending on the value of integer $n$. 

For $-S<n<S$, the expression on the right-hand side of \re{muS} is given by a linear combination of $\delta(w)$ and its derivatives
\begin{align}
\mu_{n,S}(\omega) &= {1\over 2^{S-1}}  (1-\partial_\omega)^{S-1-n}(1+\partial_\omega)^{S-1+n} \delta(\omega)\,.
\end{align}
Its substitution to \re{conv}  yields expansion of $\mathcal J_{\omega,S}$ at small $\omega$
\begin{align}\label{sum-der}
V_n^j = {1\over 2^{S-1}}  (1+\partial_\omega)^{S-1-n}(1-\partial_\omega)^{S-1+n}\mathcal J_{\omega,S}\Big|_{\omega=0}\,,
\end{align}
where $j=S-2$. For $S=2$ this relation looks as 
\begin{align}
L_{-1} = {1\over 2} (1+\partial_\omega)^2 \mathcal E_{\omega=0} \,,\qqquad
L_{0} =  {1\over 2} (1-\partial_\omega ^2) \mathcal E_{\omega=0} \,,\qqquad
L_{1} = {1\over 2} (1-\partial_\omega)^2 \mathcal E_{\omega=0} \,,
\end{align}
where we replaced $\mathcal J_{\omega,S=2}=\mathcal E_\omega$ and $V_n^0=L_n$. It is straightforward to verify that the algebra \re{c} of the energy flow operators translates to the commutation relations \re{L} for the modes $L_{-1}$, $L_0$ and $L_1$.
Notice that the expression on the right-hand side of \re{sum-der} is given by a linear combination of the first $(2S-1)$ terms of the small $\omega$ expansion of the flow operator $\mathcal J_{\omega,S}$. This is in agreement with the observation made in Section~\ref{sect:2} that the flow operators have a well-defined expansion at small $\omega$ up to order $O(\omega^{2S-1})$.  
 
For $n \ge S$ the function \re{muS} can be expressed in terms of (generalized) Laguerre polynomial
\begin{align} \label{muS1}
\mu_{n,S}(\omega) &=  (-1)^{n-S} 2^S \e^{\omega} L_{n-S}^{(2S-1)}(-2\omega)\theta(-\omega)\,.
\end{align}
For $n\le -S$ the function $\mu_{n,S}(\omega)$ can be found using the first relation in \re{mu-rel}. 
Taking into account the properties of the Laguerre polynomials, we find that $\mu_{n,S}(\omega)$ satisfies the orthogonality condition
\begin{align}\label{mu-ortho}
\int_{-\infty}^\infty d\omega\, \omega^{2S-1} \mu_{m,S}(\omega)\mu_{-n,S}(\omega) = - {\Gamma(m + S)\over \Gamma(m-S+1)}  \delta_{n+m,0} \,.
\end{align}

Applying \re{conv} we obtain
\begin{align}\label{VV-com}
[V_m^i,V_n^j] = \int_{-\infty}^\infty d\omega d\omega'  \,\mu_{m,S}(\omega)\mu_{n,S'}(\omega')\,[\mathcal J_{\omega,S},\mathcal J_{\omega',S'}]\,,
\end{align}     
where $S=i+2$ and $S'=j+2$.
Replacing the commutator of the flow operators with \re{alg2} we can reproduce the $W-$algebra \re{VV}. 

The contribution of the second term in \re{alg2} to \re{VV-com} yields the central charge
\begin{align}\label{c-W}
&-\delta_{SS'}  \Omega_S  \int_{-\infty}^\infty d\omega \,\mu_{m,S}(\omega)\mu_{n,S}(-\omega)  \omega^{2S-1} 
=\delta_{ij}\delta_{n+m,0}  \Omega_{i+2} m( m^2-1) \dots (m^2-(i+1)^2),
\end{align}  
where we applied \re{mu-rel} and \re{mu-ortho} and replaced $S=i+2$ and $S'=j+2$. This relation coincides with the expression for the central charge \re{ci} of the $W-$algebra \re{VV}. The central charge of Virasoro algebra \re{L} is given by $c=12\Omega_2$, see Eq.\re{Omega2}.
For the central charge of the $W-$algebra we get $c_i=\Omega_{i+2}$. Notice that the central charge \re{c-W} vanishes for $-(i+1)\le m \le i+1$. We recall that the corresponding operators $V_m^i$ form a finite-dimensional representation of the $SL(2;\mathbb R)$ subgroup  the conformal group in $d=2$ dimensions and their commutation relations do not contain central extension.

The contribution of the first term in \re{alg2} to \re{VV-com} can be written as (for $\ell_i=S_i$)
\begin{align} \label{aux1}
& \int d\omega d\omega' \mu_{m,S} (\omega)\mu_{n,S'} (\omega')C_{S S'}{}^{S''}(\omega,\omega') \mathcal J_{\omega+\omega',S''} = \int dx \, O_{S''}(x) f(x)\,,
\end{align}
where we replaced $\mathcal J_{\omega,S}=\int dx \e^{-i\omega x} O_S(x)$ and introduced notation for 
\begin{align}\label{f1}
f(x) =  \int d\omega d\omega' \mu_{m,S} (\omega)\mu_{n,S'} (\omega')C_{S S'}{}^{S''}(\omega,\omega') \e^{-ix(\omega+\omega')} \,.
\end{align}
As we will see in a moment, this function is given by
\begin{align}\label{f2}
f(x) = g_{2p}^{ij}(m,n) \int d\omega'' \mu_{n+m,S''}(\omega'')  \e^{-ix\omega''} \,,
\end{align}
where $S=2+i$, $S'=2+j$ and $S''=S+S'-2(p+1)$. Then, 
replacing $f(x)$ in \re{aux1} with this expression and applying \re{conv}, we can express the right-hand side of \re{aux1} as $g_{2p}^{ij}(m,n)V_{m+n}^{i+j-2p}$, thus reproducing the first term in \re{VV}.

To find the relation between the structure constants $g_{2p}^{ij}(m,n)$ and $C_{S S'}{}^{S''}(\omega,\omega')$, we apply \re{muS} and
match the expression on the right-hand side of \re{f1} and \re{f2}
\begin{align}
 {(1+ix)^{m+S-1}\over  (1-ix)^{m-S+1}}
C_{S S'}{}^{S''} (i\!\!\stackrel{\leftarrow}{\partial_x},i\!\!\stackrel{\rightarrow}{\partial_x}) 
 {(1+ix)^{n+S'-1}\over  (1-ix)^{n-S'+1}}=2^{2p+1}
g_{2p}^{ij}(m,n) {(1+ix)^{m+n+S''-1}\over  (1-ix)^{m+n-S''+1}}\,,
\end{align}
where $i=S-2$, $j=S'-2$ and $2p=S+S'-S''-2$. This relation holds for arbitrary $x$. Choosing $x=i(1+\epsilon)$ and 
taking the limit $\epsilon\to 0$ we get
\begin{align}\label{g-C}
g_{2p}^{ij}(m,n)   =-
{1\over \epsilon^{m+n+S''-1} }
\lr{\epsilon^{m+S-1} 
C_{S S'}{}^{S''} (\stackrel{\leftarrow}{\partial_\epsilon},\stackrel{\rightarrow}{\partial_\epsilon}) \,
 \epsilon^{n+S'-1}} + O(\epsilon) \,.
\end{align}
We recall that $C_{S S'}{}^{S''}(\omega,\omega')$ is a homogenous polynomial of degree $S+S'-S''-1$. It is easy to 
check that the first term in \re{g-C} approaches a finite value as $\epsilon\to 0$. For instance, for $p=0$, or equivalently
$S''=S+S'-2$, we apply the relation \re{HS-C} to find from \re{g-C}
\begin{align} 
g_0^{ij}(m,n) 
& =c^{(s)}_{S S'} \lr{m(S'-1)-n(S-1)}\big|_{S=i+2,\,S'=j+2}\,,
\end{align}
in agreement with the known result of Ref.~\cite{Pope:1991ig}.

The structure constants \re{g-C} should be independent on $\epsilon$.~\footnote{One can show that this condition fixes $C_{S S'}{}^{S''}(\omega,\omega')$ up to a normalization constant.} This imposes nontrivial restrictions on $C_{S S'}{}^{S''}(\omega,\omega')$.
One can show that vanishing of the $O(\epsilon)$ term in \re{g-C} follows from invariance of  the commutation relations \re{OO-W} under the $SL(2)$ conformal transformations. We can invert the relation  \re{g-C} to get
\begin{align}\label{C-sum-g}
C_{S S'}{}^{S''} (\omega_1,\omega_2) = -\sum_{m, n \ge 0\atop m+n=2p+1}{\omega^m \over m!}{ (\omega')^n\, \over n!} 
g_{2p}^{ij}(m-S+1,n-S'+1) \,,
\end{align}
where $S=i+2$, $S'=2+j$ and $S''=S+S'-2(p+1)$. This provides the relation between the structure constants of the algebra of flow operators in $d=2$ dimensions and the $W-$algebra. 

The structure constants in \re{VV} are sensitive to the normalization of the operators $V_m^i$. Replacing $V_m^i \to f_i V_m^i$ in \re{VV} one finds that the structure constants transform as
\begin{align}
g_{2p} ^{ij}(m,n) \to {f_i f_{j}\over f_{i+j-2p}} g_{2p} ^{ij}(m,n)\,.
\end{align}
Applying this transformation to \re{C-sum-g} with $f_i= -2^{1-2i} (2i+2)!/((i+1)!(i+2)!)$ and replacing $C_{S S'}{}^{S''} (\omega_1,\omega_2)$ with its expression \re{W-int}, we find that $g_{2p} ^{ij}(m,n)$ coincide with the known expressions for the structure constants of the $W_\infty$ algebra, see Eqs.~(3.4) -- (3.6)  
in Ref.~\cite{Pope:1991ig}.

\section{Scalar integral} \label{app:scalarintegral}

In this appendix, we compute the integral \eqref{eq:integralBasic} that appeared in the discussion of the light-ray algebra in interacting CFTs. 
It takes the following general form
\be\label{I-def}
I_{\alpha \beta}(\omega_1, \omega_2) = \int_{-\infty}^\infty d x_1^- d x_2^- e^{- i \om_1 x_1^- - i \om_2 x_2^ -} {1 \over (x_{12}^2)^{\alpha } (x_{13}^2 x_{23}^2)^{\beta}} \,,
\ee
where $x_i=(x_i^+,x_i^-,\bit x_i)$ in the light-cone coordinates and singularities of the denominator at $x_{12}^2=0$ are resolved as
\be
{1\over (x_{12}^2)^\alpha }
\equiv {1 \over (2x_{12}^+ (x_{12}^- - i \eps) - \bit x_{12}^2)^\alpha}   = {1 \over (2x_{12}^+)^\alpha} {i^\alpha \over \Gamma(\alpha)}   \int_0^\infty d s s^{\alpha - 1} e^{ -i s \left(x_{12}^- - {\boldsymbol{x}_{12}^2 \over 2x_{12}^+}\right)} ,
\ee
and similar for the two remaining factors in \re{I-def}.
Here the `$-i\epsilon$' prescription reflects the ordering of the operators in the correlation function \re{eq:three-pointfunctionC}.

Applying the last relation and performing integration in \re{I-def}, we encounter  integrals of the form
\be
\int_{- \infty}^\infty   {d x_1^-\,e^{- i \omega x_1^-} \over (2x_{13}^+ (x_{13}^- - i \epsilon) - \bit x_{13}^2 )^\beta}  = {2 \pi \theta(-\omega) \over (2x_{13}^+)^\beta} {i^\beta  \over \Gamma(\beta)} (- \om)^{\beta - 1} e^{i \omega \left(x_3^- + {\boldsymbol x_{13}^2 \over 2x_{13}^+} \right)} .
\ee
Combining various factors together we obtain in the limit $x_{12}^+\to 0$
\begin{align}\notag
I_{\alpha \beta}(\omega_1, \omega_2) = & (2\pi)^2 \theta(-\omega_1)\theta(-\omega_1-\omega_2) {i^{\alpha+2\beta}\over \Gamma(\alpha)\Gamma^2(\beta) } 
\\
 \times & (2x_{12}^+)^{-\alpha} (4 x_{13}^+ x_{23}^+)^{-\beta} \int_{\omega_2} ^{-\omega_1} ds\, s^{\alpha-1}((-s-\omega_1)(s-\omega_2))^{\beta-1} e^{is{\boldsymbol x_{12}^2\over 2x_{12}^+}}\,.
\end{align}
We deduce from this relation that for $\boldsymbol x_{12}^2\neq 0$ the integral decreases exponentially fast as $x_{12}^+\to 0$. At the same time, for $\boldsymbol x_{12}^2= 0$ it scales as $(x_{12}^+)^{-\alpha}$. This suggests that $I_{\alpha \beta}(\omega_1, \omega_2)$ is proportional to a distribution localized at $\boldsymbol x_{12}=0$. To identify its form we integrate both sides of the last relation over
$\bit x= \boldsymbol x_{12}$
\begin{align}
\int d^{d-2} \bit x \, I_{\alpha \beta}(\omega_1, \omega_2) \sim (x_{12}^+)^{-\alpha+(d-2)/2} \,,
\end{align}
leading to $I_{\alpha \beta}(\omega_1, \omega_2)\sim (x_{12}^+)^{-\alpha+(d-2)/2}\delta^{(d-2)}(\bit x_{12})$.
 
\section{Extra light-ray operator in the free scalar theory}
\label{sec:extralightray}

In this Appendix, we examine the properties of nonlocal operators $\mathcal O^\pm _{\omega_1,\omega_2}(n)$ that enters the commutation relations \re{eq:freescalar} and \re{odd} of the light-ray operators in a free scalar theory. Using \re{nonloc} it is convenient to write $\mathcal O^\pm _{\omega_1,\omega_2}(n)$ as
\begin{align}\label{Opm}
{\cal O}^\pm _{\omega_1, \omega_2}(n) \sim {\cal O}_{\omega_1, \omega_2}(n) \pm  {\cal O}_{\omega_2, \omega_1}(n)\,,
\end{align}
where the notation was introduced for 
\begin{align}\label{app-O}
{\cal O}_{\omega_1, \omega_2}(n) =i (n\bar n)^{2} \lim_{r\to\infty} r^{d-2}  \int_{-\infty}^\infty dt_1 dt_2  \,  \e^{-i( t_1\omega_1+ t_2 \omega_2)(n \bar n)}
{\rm sign}(t_1-t_2) \phi( r n + t_1 \bar n)\bar\phi( r n + t_2\bar n) .
\end{align}
In the embedding space, the light-ray operator \eqref{app-O} takes the form ${\cal O}_{\omega_1, \omega_2}(n) \equiv {1 \over 4} \mathbb{O}_{{\om_1 \over 2} , {\om_2 \over 2} }(X_{\infty},Z_\infty)$, see \cite{Korchemsky:2021okt} for the details,
\be
\label{eq:lightraynew}
\mathbb{O}_{\om_1 , \om_2}(X,Z) =i \int_{- \infty}^{\infty} d \alpha_1 \int_{- \infty}^{\infty} d \alpha_2 e^{- i \om_1 \alpha_1} e^{- i \om_2 \alpha_2} {\rm sign}(\alpha_1-\alpha_2) \phi(Z - \alpha_1 X)  \bar \phi(Z - \alpha_2 X) .
\ee

The product of scalar fields on the right-hand side of \eqref{app-O} can be expanded over local conformal operators $O_S(x) \equiv O_{\mu_1\dots \mu_S}(x) \bar n^{\mu_1} \dots \bar n^{\mu_S}$ and their descendants (see e.g. \cite{Braun:2003rp})
\begin{align}\label{app2}
\phi( r n + t_1 \bar n)\bar\phi( r n + t_2\bar n)=\sum_{S\ge 0} c_S {(i t_{12})^S\over S!} \int_0^1 du \, u^{S+\nu} (1-u)^{S+\nu} O_S(rn +(t_1 - u t_{12}) \bar n) \,,
\end{align}
where $t_{12}=t_1-t_2$ and $\nu=2j_\phi-1=(d-4)/2$ is related to the conformal spin of the scalar field.\footnote{In writing \eqref{app2} we implicitly subtracted the contribution of the identity operator $\langle \phi( r n + t_1 \bar n)\bar\phi( r n + t_2\bar n) \rangle$ on the right-hand side.} The integral over $u$ takes into account the contribution of descendants. The OPE coefficient $c_S$ depends on the normalization of the local operator $O_S$. At $S=2$, for the stress-energy tensor,  we have $c_2={\Gamma(6+2\nu)/ \Gamma^2(3+\nu)}$.

Combining the relations \re{app-O} and \re{app2} we get for $(n\bar n)=1$
\begin{align} \label{O-f}
{\cal O}_{\omega_1, \omega_2}(n) &=\sum_{S\ge 0}  {c_S\over S!} \lim_{r\to\infty} r^{d-2} \int_{-\infty}^\infty dt\,O_S(rn +t\bar n)
\e^{-it(\omega_1+\omega_2)}  \left[ I_S(\omega_1,\omega_2)-(-1)^S I_S(\omega_2,\omega_1)\right]
\end{align}
where the notation was introduced for
\begin{align}\notag
I_S(\omega_1,\omega_2)= i^{S+1}\int_0^1 du \, (u (1-u))^{S+\nu} \int_{-\infty}^\infty dt_1 dt_2  \,  \e^{-i( (t_1-t)\omega_1+ (t_2-t) \omega_2)}
\\
\times
\theta(t_1-t_2) 
(t_1-t_2)^S    \delta(t_1-t - u (t_1-t_2))
\end{align}
Shifting the integration variables, $t_i\to t_i+t$, it is easy to see that $I_S(\omega_1,\omega_2)$ does not depend on $t$. Going through calculation we obtain
\begin{align}\notag
I_S(\omega_1,\omega_2) &=\int_0^1 {dx\, (x(1-x))^{S+\nu} \over [x\omega_1-(1-x)\omega_2-i0]^{S+1}}
=
{\Gamma^2 (S+\nu +1)\over \Gamma (2 S+2 \nu +2)}f_S   (\omega_1,\omega_2) 
  \,,
\\
f_S   (\omega_1,\omega_2) &={(-\omega_2)^{-S-1}  \, _2F_1\left({S+1,S+\nu +1\atop 2 (S+\nu
   +1)}\Big| 1+\frac{\omega_1 - i 0}{\omega_2}\right)}\,,
\end{align}
where $\nu=(d-4)/2$.
Notice that ${\rm Re} [I_S(\omega_1,\omega_2) ] = - (-1)^{S} {\rm Re} [I_S(\omega_2,\omega_1) ]$, and ${\rm Im} [I_S(\omega_1,\omega_2) ] = (-1)^{S} {\rm Im} [I_S(\omega_2,\omega_1) ]$.

Taking into account \re{J-def}, we can express \re{O-f} as the sum over light-ray operators of arbitrary spin
\begin{align} \label{O-sum}
{\cal O}_{\omega_1, \omega_2}(n) &=\sum_{S\ge 0}  {c_S\over S!}  \left[ I_S(\omega_1,\omega_2)-(-1)^S I_S(\omega_2,\omega_1)\right]
 \mathcal J_{\omega_1+\omega_2,S} (n)\, \nn \\
 &=2 \sum_{S\ge 0}  {c_S\over S!}  {\rm Re} [I_S(\omega_1,\omega_2) ]  \mathcal J_{\omega_1+\omega_2,S} (n) \ . 
\end{align}
Substituting this relation into \re{Opm}, we find that ${\cal O}^\pm _{\omega_1, \omega_2}(n)$ is given by the sum over odd and even spins, respectively. 

For $S=2$  the contribution of the energy flow operator $\mathcal E_{\omega_1+\omega_2} (n)=\mathcal J_{\omega_1+\omega_2,S=2} (n)$ to the right-hand side of \re{O-sum} takes the form
\begin{align} \label{O-sum1}\notag
{\cal O}_{\omega_1, \omega_2}(n) &= f_{\phi}   (\omega_1,\omega_2)
 \mathcal E_{\omega_1+\omega_2} (n) + \dots\,,
 \\
 f_\phi(\omega_1,\omega_2)&=-{\rm Re}\Big[\omega_2^{-3}  {\,
   _2F_1\left({3,\frac{d}{2}+1\atop d+2}\Big|1+\frac{\omega_1- i 0}{\omega_2}\right)} \Big] ,
\end{align}
where $f_\phi=f_{S=2}   (\omega_1,\omega_2)$ and dots denote the contribution of the spins $S\neq 2$. Its explicit expression for different values of $d$ looks as 
\begin{align}\notag
& f_\phi \big|_{d=3}=\frac{16 \left(4 (\omega _2 - \omega_1)+ 
   \left( \omega _1^2-6 \omega _1\omega _2+\omega _2^2\right) (-\omega_1 \omega_2)^{-1/2} [\theta(\om_2) - \theta(\om_1) ] \right)}{\left(\omega
   _1+\omega _2\right){}^4} \ , 
\\
 & f_\phi \big|_{d=4}=\frac{30 \left(3 (\omega _2^2 - \omega_1^2)+\left(\omega _1^2-4 \omega _2 \omega
   _1+\omega _2^2\right) \log |\frac{\omega _1}{\omega
   _2}|\right)}{\left(\omega _1+\omega _2\right){}^5} .
\end{align}
We recall that for $d=2$ the operator \re{nonloc} vanishes and does not contribute to the algebra of the light-ray operators.

\bibliographystyle{JHEP} 



\providecommand{\href}[2]{#2}\begingroup\raggedright\endgroup

\end{document}